\documentclass[11pt,a4paper]{article}
\pdfoutput=1 
\DeclareMathAlphabet{\scr}{U}{rsfs}{m}{n}

\usepackage{latexsym}
\usepackage{epsfig}
\usepackage[mathscr]{eucal}
\usepackage{amsfonts}
\usepackage{amscd}
\usepackage{cite}
\usepackage{array}   
\usepackage{amssymb}
\usepackage{colordvi}
\usepackage[centertags]{amsmath}
\usepackage{enumerate}
\usepackage{graphicx}
\usepackage{booktabs}
\usepackage{theorem}
\usepackage{multirow}
\usepackage[footnotesize]{caption}
\usepackage{bbm}
\usepackage[labelfont=bf,textfont={sl,bf}]{subfig}
 \usepackage{slashed}
\usepackage[obeyFinal]{todonotes}
\usepackage{color}
\usepackage{ulem}
\let\newabs=\abs
\let\newnorm=\norm
\let\newbraket=\braket
\let\newpb=\pb
\usepackage{physics}
\usepackage{float}

\let\abs=\newabs
\let\norm=\newnorm
\let\braket=\newbraket
\let\pb=\newpb

\usepackage{mathtools}

\DeclarePairedDelimiter\abs{\lvert}{\rvert}%
\DeclarePairedDelimiter\norm{\lVert}{\rVert}%

\newcommand{\newc}{\newcommand}
\newc{\be}{\begin{equation}}
\newc{\ee}{\end{equation}}
\newc{\bea}{\begin{eqnarray}}
\newc{\eea}{\end{eqnarray}}
\newc{\ol}{\overline}
\newc{\wt}{\widetilde}
\newc{\bs}{\boldsymbol}
\newc{\m}{\mathcal}
\newc{\lan}{\langle}
\newc{\ra}{\rangle}
\newc{\pa}{\partial}

\newcommand{\sbeta}{{s_{\beta}}}
\newcommand{\cbeta}{{c_{\beta}}}
\newcommand{\tbeta}{{t_{\beta}}}
\newcommand{\mueff}{{\mu_{\text{eff}}}}

\newcommand{\crn}{\nonumber \\}
\newcommand{\beq}{\begin{eqnarray}} 
\newcommand{\eeq}{\end{eqnarray}} 
\newcommand{\bpmatrix}{\begin{pmatrix}}
\newcommand{\epmatrix}{\end{pmatrix}}
\newcommand{\ba}{\begin{array}}
\newcommand{\ea}{\end{array}}
\newcommand{\braket}[1]{\left(#1\right)}

\newcommand{\fr}{\frac}

\newcommand{\diag}{\text{diag}}

\newcommand{\calM}{{\cal M}}
\newcommand{\calA}{{\cal A}}

\newcommand{\tb}{\tan\beta}  

\newcommand{\calL}{{\cal L}}

\newcommand{\DRbar}{\overline{\text{DR}}}

\newcommand{\figref}[1]{Fig.~\ref{#1}}
\renewcommand{\eqref}[1]{Eq.~(\ref{#1})}

\newcommand{\tab}[1]{Table~\ref{#1}}
\newcommand{\sect}[1]{Section~\ref{#1}}
\newcommand{\ssect}[1]{Subsection~\ref{#1}}

\newcommand{\appen}[1]{Appendix~\ref{#1}}

\newcommand{\DRb}{\overline{\text{DR}}}

\newcommand{\OS}{\text{OS}}

\renewcommand{\Re}{\text{Re}\,}
\renewcommand{\Im}{\text{Im}\,}

\newcommand{\ie}{{\it i.e.\;}}

\newcommand{\bc}{\begin{center}}
\newcommand{\ec}{\end{center}}

\newcommand{\gev}{~\text{GeV}}

\newcommand{\EV}{~\text{eV}}
\newcommand{\tev}{~\text{TeV}}
\newcommand{\pb}{{~\text{pb}}}

\newcommand{\ti}{\tilde}

\newcommand{\ISS}{\text{ISS}}

\newcommand{\expeta}{e^{i\varphi_u }}
\newcommand{\expetam}{e^{-i\varphi_u }}
\newcommand{\expetas}{e^{i\varphi_s }}

\newcommand{\lsim}{\raisebox{-0.13cm}{~\shortstack{$<$ \\[-0.07cm]
      $\sim$}}~}
\newcommand{\s}{\newline \vspace*{-3.5mm}}

\usepackage{bm}
\allowdisplaybreaks

\newcommand{\Br}{\text{BR}}
\newcommand{\baru}{\bar{u}}

\newcommand{\DRbt}{$ \DRb $ }

\newcommand{\vp}{\varphi}

\newcommand{\absl}{\qty|\lambda|}
\newcommand{\absk}{\qty|\kappa|}
\newcommand{\reAl}{{\Re A_\lambda}}
\newcommand{\reAk}{{\Re A_\kappa}}
\newcommand{\imAl}{{\Im A_\lambda}}
\newcommand{\imAk}{{\Im A_\kappa}}
\newcommand{\half}{\dfrac{1}{2}}
\newcommand{\calR}{\mathcal{R}}
\newcommand{\divt}{\text{div}}
\newcommand{\Usn}{{U_{\tilde{\nu}}}}
\newcommand{\tn}{{\tilde{\nu}}}
\newcommand{\tN}{{\tilde{N}}}
\newcommand{\tX}{{\tilde{X}}}

\usepackage{hyperref,cleveref}

\begin{document}
\title{
\vspace*{-3cm}
\phantom{h} \hfill\mbox{\small IFIRSE-TH-2021-3} \\[-0.2cm]
\phantom{h} \hfill\mbox{\small KA-TP-15-2021}
\\[1cm]
Loop-corrected Higgs Masses in the NMSSM with Inverse Seesaw Mechanism\\[4mm]}

\author{
Thi Nhung Dao$^{1}$\footnote{E-mail: \texttt{dtnhung@ifirse.icise.vn}}\;,
Margarete M\"{u}hlleitner$^{2}$\footnote{E-mail: \texttt{margarete.muehlleitner@kit.edu}}\;,
Anh Vu Phan$^{1,3,4}$\footnote{E-mail: \texttt{phananhvu1609@gmail.com}}
\\[9mm]
{\small\it
$^1$Institute For Interdisciplinary Research in Science and Education, ICISE,}\\
{\small\it 590000, Quy Nhon, Vietnam.}\\[3mm]
{\small\it $^2$Institute for Theoretical Physics, Karlsruhe Institute of Technology,} \\
{\small\it Wolfgang-Gaede-Str. 1, 76131 Karlsruhe, Germany.}\\[3mm]
{\small\it $^3$University of Science, Ho Chi Minh City, Vietnam.}
\\[3mm]
{\small\it $^4$Vietnam National University, Ho Chi Minh City, Vietnam.}
}
\maketitle

\begin{abstract}
In this study, we work in the framework of the Next-to-Minimal
extension of the Standard Model (NMSSM) extended by six singlet
leptonic superfields. Through the mixing with the three doublet
leptonic superfields,
the non-zero tiny neutrino masses can be generated through the inverse
seesaw mechanism. While $R$-parity is conserved in this model lepton
number is explicitly violated. We quantify the impact of the extended
neutrino sector on the NMSSM Higgs sector by computing the complete
one-loop corrections with full momentum dependence to the Higgs boson
masses in a mixed on-shell-$\overline{\mbox{DR}}$ renormalization
scheme, with and without the inclusion of CP violation. The results are
consistently combined with the dominant two-loop corrections at ${\cal
O}(\alpha_t(\alpha_s+\alpha_t))$ to improve the predictions for the Higgs
mixing and the loop-corrected masses. In our numerical study we
include the constraints 
from the Higgs data,  the neutrino oscillation data, the charged lepton
flavor-violating decays $l_i \to l_j + \gamma$, and the new physics constraints
from the oblique parameters $S,T,U$. We present in this context the
one-loop decay width for $l_i \to l_j + 
\gamma$. The loop-corrected Higgs boson masses are included in the
Fortran code {\tt NMSSMCALC-nuSS}.
\end{abstract}
\thispagestyle{empty}
\vfill
\newpage

\section{Introduction}
Both cosmological and neutrino oscillation data have indicated the existence
of three neutrino flavors, non-zero neutrino masses, and neutrino mixing. The three
observed neutrinos are called active neutrinos. Since their masses are tiny, 
the absolute values have not been measured so far and 
their Yukawa interactions with the Higgs boson are extremely small. 
Therefore the neutrino effects on the Higgs sector are
negligible. However, in models with an extended neutrino sector that
allows incorporating the tiny neutrino masses, there exists an
arbitrary number of sterile neutrinos. Their masses can be light, 
heavy, or extremely heavy. Current experiments have not observed them
yet but still allow for a small mixing between sterile and active
neutrinos. For precise investigations and meaningful interpretations
of both the Higgs and the neutrino sector, it is therefore worthwhile and
mandatory to consider the effects of these sterile neutrinos on the
Higgs sector. With the increasing amount of the experimental LHC data on
the Higgs mass, couplings, production, and decay processes, one can
expect stronger constraints on new physics affecting directly and/or indirectly the
Higgs sector. \s

In this study, we consider the impact on the Higgs boson masses in 
a supersymmetric theory. More specifically, we work in the framework of
the  Next-to-Minimal Supersymmetric extension of the Standard Model
(NMSSM)
\cite{Fayet:1974pd,Barbieri:1982eh,Dine:1981rt,Nilles:1982dy,Frere:1983ag,Derendinger:1983bz,Ellis:1988er,Drees:1988fc,Ellwanger:1993xa,Ellwanger:1995ru,Ellwanger:1996gw,Elliott:1994ht,King:1995vk,Franke:1995tc,Maniatis:2009re,Ellwanger:2009dp}
with a Higgs sector consisting of two complex Higgs doublets and a
complex Higgs singlet. After electroweak symmetry breaking (EWSB), the NMSSM Higgs sector features seven Higgs bosons, five neutral and two charged Higgs bosons. 
One of the neutral Higgs states is identified with the Standard Model
(SM) Higgs boson. Its tree-level mass is light and can receive large
loop corrections to explain the 125 GeV mass of the observed Higgs
boson by ATLAS \cite{Aad:2012tfa} and CMS
\cite{Chatrchyan:2012ufa}. The neutrino sector in this model is
extended to include six singlet leptonic superfields. $R-$parity is
conserved while the lepton number is explicitly violated by an interaction
term between two singlet neutrino superfields and a singlet Higgs
superfield. This type of model was first discussed in
\cite{Gogoladze:2008wz}. The six singlet neutrinos mix with the three
doublet ones to generate nine neutrino mass eigenstates. Three of them
have very light masses, that can be explained through the inverse
seesaw mechanism \cite{Mohapatra:1986aw,PhysRevD.34.1642,Bernabeu:1987gr}.
The six remaining neutrinos can have masses of order TeV which may
be observable in collider experiments.  \s
   
In the literature, there exist many studies on the effects of the 
(s)neutrinos on the loop-corrected Higgs boson masses in the context of a supersymmetric theory with the type I or inverse seesaw mechanism. We briefly review here
those studies that are close to our subject. 
The impact of the extended neutrino and sneutrino sector on the
lightest CP-even Higgs mass in the NMSSM with the inverse seesaw mechanism
(ISS), was presented in \cite{Gogoladze:2012jp} using an approximate
one-loop correction neglecting the effects from external momentum
dependence and mixings between Higgs bosons.   
The authors of \cite{Wang:2013jya} have computed the one-loop corrections
stemming solely from the neutrino/sneutrino sector to
the lightest CP-even Higgs boson in the NMSSM
extended by a right-handed neutrino superfield with $R-$parity
conservation. The full one-loop 
corrections to neutral Higgs boson masses were presented in a mixed
on-shell (OS)-$\DRb$ scheme for the $\mu\nu\text{SSM}$ model with only one
generation in \cite{Biekotter:2017xmf} and for three generations of
right-handed neutrinos in \cite{Biekotter:2019gtq}. In the
$\mu\nu\text{SSM}$, the Higgs sector contains two Higgs doublets while
the neutrino sector is extended to include singlet right-handed
neutrino superfields. Lepton number and $R-$parity are not protected
in the $\mu\nu\text{SSM}$ so that the superpartners of the singlet
right-handed neutrinos can develop vacuum expectation values
(VEVs). In the Minimal Supersymmetric Standard Model (MSSM) extended
by the type I seesaw mechanism, the full one-loop corrections
with full momentum dependence combined with the dominant two-loop
corrections to the Higgs boson masses were presented in
\cite{Heinemeyer:2010eg}, which showed a non-decoupling effect for a
large right-handed neutrino scale. The authors of
\cite{Draper:2013ava} have shown that the decoupling property is
preserved with a suitable renormalization scheme of the parameter
$\tan\beta$, which denotes the ratio of the two vacuum
expectation values of the two Higgs doublets in the MSSM. With the
inverse seesaw mechanism incorporated in the MSSM, the one-loop
corrections of (s)neutrinos  have been studied in
\cite{Guo:2013sna,Chun:2014tfa} by using the one-loop effective
potential approach. \s

Our goal is to present here the
complete one-loop corrections with full momentum dependence to the
Higgs boson masses  in a mixed OS-$\DRb$ renormalization scheme using
the Feynman diagrammatic approach. The calculation has been done both
in the real and the complex NMSSM. We consistently  combine our result
with the dominant two-loop corrections  
of ${\cal O}(\alpha_s\alpha_t)$ \cite{Muhlleitner:2014vsa} and ${\cal
  O}(\alpha_t^2)$ \cite{Dao:2019qaz} computed by our group. In order to investigate
the impact of our newly computed corrections, we perform a numerical
study where we apply constraints from the Higgs data, the neutrino
oscillation data, the charged lepton flavor-violating decays $l_i\to l_j+\gamma$,
and the constraints on new physics from the oblique parameters
$S,T,U$. The explicit computation of the one-loop decay width for
$l_i\to l_j+\gamma$ is also presented in this study. 
We furthermore provide the Fortran code, dubbed {\tt
  NMSSMCALC-nuSS}, for the
computation of the loop-corrected Higgs 
boson masses and Higgs boson decay branching ratios incorporating
higher-order corrections. This code is adapted from the code {\tt
  NMSSMCALC} \cite{Baglio:2013vya} published by our
group.\footnote{Our recently published code {\tt
      NMSSMCALCEW} also includes the supersymmetric (SUSY) electroweak
    \cite{Baglio:2019nlc} and SUSY-QCD corrections to the NMSSM Higgs boson decay
    widths and branching ratios \cite{Dao:2020dfb}.} \s

The paper is organized as follows. In \sect{sec:tree-levelspectrum}, we
describe the model and the masses and mixings of each sector at 
tree level. In \sect{sec:HMass}, we present details of our calculation
of the loop-corrected Higgs boson masses and mixing. We also discuss
our renormalization scheme for parameters and fields needed to obtain
finite renormalized Higgs self-energies. In \sect{sec:constraints} 
we present all constraints related to the Higgs data, the neutrino oscillation
data, the
oblique parameters  $S,T,U$, and the charged lepton flavor-violating decays $l_i\to l_j+\gamma$ that we apply in our phenomenological
study. \sect{sec:analysis} is dedicated to the
numerical analysis. We present the size of the loop corrections and
their dependencies on the parameters of the neutrino and sneutrino sectors. We
furthermore discuss the effects of different constraints on the neutrino
sector parameters. Finally, we present our conclusions in \sect{sec:conclusions}. 

\section{The NMSSM with Inverse Seesaw Mechanism}
\label{sec:tree-levelspectrum}
The NMSSM realization of the seesaw mechanism through the
$\mathbb{Z}_3$ discrete symmetry with a unit charge of $\omega=
e^{i2\pi/3}$ has been introduced in
\cite{Gogoladze:2008wz,Gogoladze:2012jp}. Depending on the
$\mathbb{Z}_3$ charge assignment for the lepton doublet superfields and
the two Higgs doublet superfields, the neutrino masses may arise from
effective dimension five, six or seven operators. We consider in this
paper the case of the dimension six operator. Tiny neutrino masses are
obtained through the well-known inverse seesaw mechanism. This is 
an interesting case because the dimension six operators generating
neutrino masses are not present in the non-supersymmetric seesaw
models\footnote{In non-supersymmetric seesaw models, effective
  dimension five operators give masses for light neutrinos while
  dimension six operators affect their kinetic terms.}. Another reason
that makes this case more interesting is that the new appearing
particles need not to be too heavy in order to obtain tiny masses for
the observed neutrinos. They can be at the TeV scale and hence in the
reach of present and future colliders. We consider the simple case
where  we introduce six gauge singlet chiral superfields, $\hat N_i,
\hat X_i$ ($i=1,2,3$). These superfields carry lepton number. The $\mathbb{Z}_3$ charge assignment\footnote{Other
  assignments of the 
  $\mathbb{Z}_3$ charge were also discussed in
  \cite{Gogoladze:2008wz}.} used in this paper  for the relevant NMSSM superfields is
given in \tab{tab:chargeassignment}. \s
\begin{table}[h]
	\begin{center}
		\begin{tabular}{l||c|c|c| c}
			& $SU(2)_L$ & $U(1)_Y$ & $Z_3$ & $L$\\
			\hline
            $\hat Q_i$ & 2 & 1/3 & 1& 0\\
             $\hat U_i^c$ & 1 & -4/3 & $\omega^2$& 0\\ 
              $\hat D_i^c$ & 1 & 2/3 & $\omega^2$& 0\\
			$\hat L_i$ & 2 & -1 & 1& 1\\
			$\hat E_i^c$& 1& 2 & $\omega^2$& -1\\
			$\hat H_u$ & 2 & 1 &$\omega$& 0\\
			$\hat H_d$ & 2 & -1 &$\omega$&0\\
			$\hat S$ & 1 & 0& $\omega$& 0 \\
			\hline
			$\hat N_i^c$ & 1 & 0 & $\omega^2 $& -1\\
			$\hat X_i$ & 1 &0 &$\omega$& 1 \\
		\end{tabular}
		\caption{Quantum numbers associated with the gauge
                  symmetry group $SU(2)_L, U(1)_Y$, the $\mathbb{Z}_3$
                  charge ($\omega=e^{i2\pi/3}$) and the lepton number
                  $L$ for the NMSSM superfields. The generation index
                  $i$ runs from one to
                  three.} 
		\label{tab:chargeassignment}
	\end{center}
\end{table}

The NMSSM superpotential including the new superfields is given by
\begin{align} 
W_{\text{NMSSM}} = W_{\text{MSSM}} - \epsilon_{ab} \lambda \hat{S} \hat{H}_d^a
\hat{H}_u^b +  \frac{1}{3} \kappa \hat{S}^3- y_\nu \epsilon_{ab}  \hat{H}_u^a \hat{L}^b \hat{N}^c +  \lambda_X  \hat S \hat X
\hat X + \mu_X  \hat X  \hat N^c \,, \label{eq:wnmssm}
\end{align}
where  $\epsilon_{ab}$ is the totally antisymmetric tensor
with $\epsilon_{12}=\epsilon^{12}=1$, $\hat{H}_u,\hat{H}_d$ denote
the two complex Higgs doublet superfields and $\hat{S}$ the complex singlet
superfield. The MSSM superpotential reads 
\begin{align}
W_{\text{MSSM}} = - \epsilon_{ab} \bigl( y_u \hat{H}_u^a \hat{Q}^b \hat{U}^c -
y_d \hat{H}_d^a \hat{Q}^b \hat{D}^c - y_e \hat{H}_d^a \hat{L}^b
\hat{E}^c\bigr)
\; ,
\end{align}
in terms of the left-handed quark and lepton superfield doublets
$\hat{Q}$ and $\hat{L}$ and the right-handed up-type, down-type and
electron-type superfield singlets $\hat{U}$, $\hat{D}$ and $\hat{E}$,
respectively. Charge conjugation is denoted by the superscript $c$,
and color and generation indices have been omitted. The NMSSM
superpotential contains the coupling $\kappa$ of the self-interaction
of the singlet superfield $\hat S$, the coupling $\lambda$ for the
$\hat{S}$ interaction with the two Higgs doublet superfields, and the
coupling $\lambda_X$ for the interaction of the  Higgs singlet with the
two singlets $\hat{X}$. In general, the coupling
$\lambda_X$ is a $3\times 
3$ matrix.
This is the only term in the superpotential that violates
the lepton number. The $3\times 3$ matrix $\mu_X$ is the only parameter
with the dimension of mass in the superpotential so that it can be of the
order of the SUSY conserving mass scale and is naturally large. 
This is essential for the seesaw mechanism. The quark and
lepton Yukawa couplings $y_d,$ $y_u, y_e,y_\nu$ and the couplings
$\lambda,\kappa,\lambda_X, \mu_X$ are in general complex. In the numerical analysis,
we chose $y_d,y_u, y_e, \lambda_X, \mu_X$ to be diagonal. However in the code $\lambda_X, \mu_X$
can be chosen to be non-diagonal.
The soft SUSY breaking NMSSM Lagrangian respecting the gauge symmetry
and the $\mathbb{Z}_3$ symmetry reads 
\begin{align}
\mathcal L_{\text{NMSSM}}^{\text{soft}} =& \mathcal
L_{\text{MSSM}}^{\text{soft}} - \ti m_S^2 |S|^2 
+ 
(\epsilon_{ab} A_{\lambda} \lambda S H_d^a H_u^b - \frac{1}{3}
A_{\kappa}\kappa S^3 
+ \epsilon_{ab} y_\nu  A_{\nu}  H_u^a \ti{L}^b\ti N^*  \crn
& + \lambda_X A_X \ti S\ti X \ti X
+\mu_X B_{\mu_X}  \ti X \ti N^* + h.c.)   - \ti m^2_X \abs{\ti X}^2 - \ti m^2_N \abs{\ti N}^2   \; ,
\end{align}
and contains the soft SUSY breaking trilinear couplings $A_\lambda, A_\kappa, A_\nu$
and $A_X$, the soft SUSY breaking masses $\ti m_S^2,\ti m_X^2,\ti
m_N^2$ and the soft SUSY breaking bilinear mass
$B_{\mu_X}$. In general, $A_\lambda, A_\kappa, A_\nu,A_X$
and $B_{\mu_X}$ are complex parameters.
For simplicity, in our numerical analysis we chose $\tilde{m}_X, \tilde{m}_N,  A_X,$ and $B_{\mu_X}$ to be diagonal. 
The SM-type and SUSY fields corresponding to a superfield
(denoted with a hat) are represented by a letter without and with a
tilde, respectively.
The soft SUSY breaking MSSM contribution can be cast into
the form
\begin{align}\nonumber
\mathcal L_{\text{MSSM}}^{\text{soft}} &= - \ti m_{H_d}^2 |H_d|^2 - \ti  m_{H_u}^2 
|H_u|^2  
- \ti  m_{Q}^2 |\tilde{Q}|^2  - \ti  m_{U}^2 |\tilde{u}_R|^2  - \ti  m_{D}^2 |\tilde{d}_R|^2 
- \ti m_{L}^2 |\tilde{L}|^2 -\ti  m_{E}^2 |\tilde{e}_R|^2 \\& \quad \nonumber
+ \epsilon_{ab} ( y_u A_u H_u^a \tilde{Q}^b \tilde{u}^*_R 
-  y_d A_d H_d^a \tilde{Q}^b \tilde{d}^*_R 
-  y_e A_e H_d^a \tilde{Q}^b \tilde{e}^*_R + h.c.)\\& \quad
- \frac{1}{2} (M_1  \tilde{B} \tilde{B} +   M_2 \tilde{W}_i \tilde{W}_i + M_3
\tilde{G} \tilde{G} + h.c) \; .
\end{align}
The indices of the soft SUSY breaking masses,
$Q$ ($L$), stand for the left-handed doublet of the three quark
(lepton) generations, and $U,D,E$ are the indices for the right-handed
up-type and down-type quarks and charged leptons, respectively. In
the trilinear coupling parameters, the indices $u,d,e$ represent the
up-type and down-type quarks and charged leptons. 
While the trilinear couplings $A_u$, $A_d$ and $A_e$
are complex, the soft SUSY breaking mass terms $\ti m_x^2$
($x=S,H_u,H_d,Q,U,D,L,E$) are real. The soft SUSY breaking mass
parameters of the gauginos, $M_1$, $M_2$, $M_3$, for the bino, the winos and
the gluinos, $\tilde{B}$, $\tilde{W}_i$ ($i=1,2,3$) and $\tilde{G}$,
corresponding to the weak hypercharge $U(1)_Y$, the weak isospin $SU(2)_L$
and the colour $SU(3)_C$ symmetry, are in general complex. 
In this paper  we are working in complex NMSSM where the parameters
are kept complex. Furthermore, we apply flavor conservation in the
charged (s)lepton and (s)quark sectors so that all matrices including 
soft mass matrices $ \ti m_L^2, \ti m_E^2, \ti m_Q^2, \ti m_U^2, \ti m_D^2$, the
trilinear couplings $A_{e,d,u}$ and the Yukawa matrices $Y_{e,d,u}$ are diagonal
in any basis. Flavor mixing occurring in our model arises solely from
the neutrino and sneutrino sectors. 
The Lagrangian contains two lepton number violating terms, namely
  $\lambda_X \hat{S} \hat{X} \hat{X}$ and $\lambda_X A_X \tilde{S}
  \tilde{X} \tilde{X}$. \s

The Higgs, neutralino, chargino, (s)quark and charged (s)lepton
sectors are the same as in the usual NMSSM without seesaw. For
completeness, we recall briefly these sectors here to introduce our
notation. The sectors that receive significant changes are the neutrino
and sneutrino ones. We will present them in detail later on. Expanding
the scalar Higgs fields about their vacuum expectation 
values (VEVs) $v_u$, $v_d$, and $v_s$,\footnote{The six new scalar
  fields carrying lepton numbers do not develop VEVs, since we do not
  want to break lepton number spontaneously.} we have 
\begin{align}
H_d = \begin{pmatrix}\frac{1}{\sqrt{2}}(v_d + h_d + i a_d) \\ 
h_d^- \end{pmatrix}~, \;\;\; 
H_u =  e^{i\varphi_u}\begin{pmatrix} h_u^+ \\
\frac{1}{\sqrt{2}}(v_u + h_u + i a_u) \end{pmatrix}~, \;\;\; 
S = e^{i\varphi_s}\frac{1}{\sqrt{2}} (v_s + h_s + i a_s)~,
\label{Higgsdecomp}
\end{align}
where two additional complex phases, $\varphi_u,\varphi_s$, have been introduced.
The fields $h_i$ and  $a_i$ with 
$i = d, u, s$  correspond to the CP-even and CP-odd part, respectively, of the
neutral entries of $H_u$, $H_d$ and $S$. The charged
components are denoted by $h_{d,u}^\pm$. \s

After EWSB, there are mixings between
the three CP-even and the three CP-odd Higgs interaction states. In the
basis $\phi=(h_d, h_u, h_s,a_d,a_u,a_s)$, the mass term is given by 
\be 
{\cal L}= \fr 12 \phi^T M_{\phi\phi}\phi\;.
\ee
The explicit expression of the mass matrix $M_{\phi\phi}$ can be found
in \cite{Dao:2019qaz}.  The transformation into mass eigenstates
at tree-level can be performed in two steps. First, $\mathcal R^G$ is used  
to single out the Goldstone boson whose mass  is equal to the
$Z$ boson mass in the 't\,Hooft-Feynman gauge, 
\begin{align}
M_{hh}^{(6)}&=\mathcal R^G  M_{\phi\phi}\braket{\mathcal R^G }^T,\\
(h_d,h_u,h_s,a,a_s,G)^T& =   \mathcal R^G (h_d, h_u, h_s,a_d,a_u,a_s)^T.
\end{align}
Here one can remove the Goldstone state from the rest by crossing out
the sixth row and column of $M_{hh}^{(6)}$, so that
it becomes a $5\times 5$ mass matrix in the basis $(h_d,h_u,h_s,a,a_s)$.
In the second step, we diagonalize the thus obtained
  $5\times 5$ matrix $M_{hh}$ with an orthogonal matrix $\mathcal R $
\begin{align}
{\text{diag}}(m_{h_1}^2,
  m_{h_2}^2,m_{h_3}^2,m_{h_4}^2,m_{h_5}^2)&=\mathcal R M_{hh} 
\mathcal R^T, \\
(h_1,h_2,h_3,h_4,h_5)^T& =  \mathcal R   (h_d, h_u, h_s,a,a_s)^T \;.
\end{align}
  The tree-level Higgs mass eigenstates are denoted
by the small letter $h$. The masses are ordered as $m_{h_1} \leq
m_{h_2} \leq m_{h_3} \leq m_{h_4} \leq m_{h_5}$. \s

The mass matrix in the 't\,Hooft-Feynman gauge for the charged
components of the Higgs doublets, 
\be  
\bpmatrix h_d^+, h_u^+ \epmatrix M_{h^+h^+} \bpmatrix h_d^-\\ h_u^-
\epmatrix,
\ee
is given by 
\begin{align} 
M_{h^+h^+}= \fr12 \bpmatrix \tbeta & 1 \\
1 & 1/\tbeta \epmatrix \bigg[& M_W^2 s_{2\beta} +
\fr{\abs{\lambda}v_s}{\cos( \varphi_\lambda+\varphi_u+ \varphi_s) }
\braket{\sqrt{2} \,\Re A_\lambda+\abs{\kappa}v_s \cos(\varphi_\kappa+3\varphi_s) } \crn  
& -\fr{2 |\lambda|^2 M_W^2 s_{\theta_W}^2}{e^2}s_{2\beta} \bigg]+ M_W^2
\bpmatrix \cbeta^2 & -\cbeta\sbeta \\
                                -\cbeta\sbeta & \sbeta^2 \epmatrix, 
\end{align}
where $M_W$ is the mass of the $W$ boson, $\theta_W$ the
electroweak mixing angle, $e$ the electric charge and
$\varphi_\lambda,$ $\varphi_\kappa$ the complex phases of $\lambda$ and
$\kappa$, respectively. The angle $\beta$ is defined as
\beq
\tan \beta =\fr{ v_u}{v_d} \;.
\eeq
Here and in the following we use the short hand notation $c_x = \cos
x$, $s_x = \sin x$ and $t_x = \tan x$. The mass matrix, $ M_{h^+h^+}$, can be
diagonalized by a rotation matrix with the angle $\beta_c=\beta$
leading to the charged Higgs mass given by
\be 
M_{H^\pm}^2 = M_W^2 +\fr{\abs{\lambda}v_s}{s_{2\beta}\cos(
	\varphi_\lambda+\varphi_u+ \varphi_s) } \braket{\sqrt{2} \Re
	A_\lambda+\abs{\kappa}v_s \cos( \varphi_\kappa+ 3\varphi_s)
}-\fr{2 |\lambda|^2 M_W^2 s_{\theta_W}^2}{e^2}.  
\ee  
The mass of the charged Goldstone boson $G^\pm$ is equal to $M_W$. \s

The fermionic superpartners of the neutral Higgs bosons,
$\tilde{H}_d^0$, $\tilde{H}_u^0$, $\tilde{S}$, and of the neutral gauge
bosons, $\tilde{B}$, $\tilde{W}_3$, mix, and in the Weyl spinor basis
$\psi^0 =  (\tilde{B},\tilde{W}_3, \tilde{H}^0_d,\tilde{H}^0_u,  \tilde{S})^T$ 
the neutralino mass matrix $M_N$ is given by
\begin{align}
&M_N = \nonumber \\& \begin{pmatrix} 
M_1               & 0        & - c_\beta M_Z s_{\theta_W} &   
M_Z s_\beta s_{\theta_W} e^{-i \varphi_u}
& 0\\
0                 & M_2      &    c_\beta M_W    & - M_W s_\beta e^{-i \varphi_u} 
& 0\\
- c_\beta M_Z s_{\theta_W} & c_\beta M_W & 0                &
- \lambda \frac{v_s}{\sqrt{2}} e^{i \varphi_s}   & 
- \frac{\sqrt{2} M_W s_\beta s_{\theta_W} \lambda e^{i\varphi_u}}{e}
\\
M_Z s_\beta s_{\theta_W} e^{-i \varphi_u} &  - M_W s_\beta e^{-i \varphi_u} & 
- \lambda \frac{v_s}{\sqrt{2}}e^{i \varphi_s} & 0 &
-  \frac{\sqrt{2} M_W c_\beta s_{\theta_W} \lambda}{e} 
\\
0   & 0     & -  \frac{\sqrt{2}M_W s_\beta s_{\theta_W}\lambda e^{i \varphi_u}}{e}  & 
- \frac{\sqrt{2} M_W c_\beta s_{\theta_W}\lambda}{e} &
\sqrt{2} \kappa v_s e^{i \varphi_s} 
\end{pmatrix} \label{eq:neuMass}       
\end{align}
after EWSB, where  $M_Z$ is the $Z$ boson mass.   
The neutralino mass matrix is symmetric and can be
diagonalized by a $5 \times 5$ matrix $N$, yielding   
$\text{diag}(m_{\tilde{\chi}^0_1},
m_{\tilde{\chi}^0_2},m_{\tilde{\chi}^0_3},m_{\tilde{\chi}^0_4},
m_{\tilde{\chi}^0_5}) = N^* M_N N^\dagger$,  
where the mass values are ordered as $m_{\tilde{\chi}^0_1}\leq ... 
\leq m_{\tilde{\chi}^0_5}$. The neutralino mass eigenstates $\tilde{\chi}^0_i$, expressed 
as a Majorana spinor, are then obtained by 
\begin{align}\label{eq:neuspinor}
\tilde{\chi}_i^0 = \left( \begin{array}{c} \chi_i^0
\\ \overline{\chi_i^0} \end{array}
\right) \quad\ \text{with}\quad\ \chi^0_i = N_{ij} \psi^0_j,\quad\  i,\,j 
= 1,\dots,5~,
\end{align}
where
\beq
\overline{\chi_i^0} = i \sigma_2 \chi_i^{0*}
\eeq
in terms of the Pauli matrix $\sigma_2$. \s

The fermionic superpartners of the charged Higgs and gauge bosons are
given in terms of the Weyl spinors $\tilde{H}_d^\pm$,
$\tilde{H}_u^\pm$, $\tilde{W}^-$, and $\tilde{W}^+$. With
\beq
\psi^-_R = \begin{pmatrix} \tilde{W}^- \\ \tilde{H}_d^- \end{pmatrix}
\quad \mbox{and} \quad
\psi^+_L = \begin{pmatrix} \tilde{W}^+ \\ \tilde{H}_u^+ \end{pmatrix},
\eeq
the mass term for these spinors reads 
\be 
{\cal L}= (\psi^-_R)^T M_C \psi^+_L + h.c.  \;,
\ee
where
\begin{align}
M_C = \begin{pmatrix} M_2 & \sqrt{2}  s_\beta M_W e^{-i \varphi_u}\\
\sqrt{2} c_\beta M_W & \lambda \frac{v_s}{\sqrt{2}} e^{i \varphi_s}
\end{pmatrix}~. \label{eq:chaMass} 
\end{align}
The chargino mass matrix $M_C$ can be diagonalized with the help of two unitary 
$2 \times 2$ matrices, $U$ and $V$, resulting in 
\beq
\text{diag}(m_{\tilde{\chi}^\pm_1}, m_{\tilde{\chi}^\pm_2}) = U^* M_C
V^\dagger \;,
\eeq 
with $m_{\tilde{\chi}^\pm_1} \leq  m_{\tilde{\chi}^\pm_2}$. The
left-handed and the right-handed part of the mass eigenstates are 
\begin{align}
\tilde{\chi}^+_L = V \psi^+_L \quad \mbox{and} \quad \tilde{\chi}^-_R = U \psi^-_R~,
\end{align}
respectively, with the mass eigenstates ($i=1,2$)
\beq
\tilde{\chi}^+_i = \left( \begin{array}{c} 
	\tilde{\chi}^+_{L_i} \\ \overline{\tilde{\chi}^-_{R_i}}^T \end{array} \right)
\eeq
written as Dirac spinors. \s

The scalar partners of the left- and right-handed quarks are denoted as
$\ti q_L$ and $\ti q_R$, respectively. Assuming no generation mixing
in the squark sector the mass matrix for the 
top squark in the interaction basis $(\ti t_L,\ti t_R)$  reads
\be 
M_{\ti t} = \bpmatrix m_{\ti Q_3}^2 + m_t^2 + M_Z^2c_{ 2\beta}(\fr12 -
\fr23 s^2_{\theta_W}) & 
m_t\braket{A_t^* \expetam - \mueff/\tbeta} \\
m_t\braket{A_t \expeta - \mueff^*/\tbeta} & m_t^2 + m_{\ti t_R}^2
+\fr23 M_Z^2 c_{ 2\beta} s^2_{\theta_W}   \epmatrix,
\label{eq:mstopmat}
\ee 
while the bottom squark mass matrix is given by
\be 
M_{\ti b} = \bpmatrix m_{\ti Q_3}^2 + m_b^2 + M_Z^2c_{ 2\beta}(-\fr12
+ \fr13 s^2_{\theta_W}) & 
m_b\braket{A_b^*  - \expeta\mueff \tbeta} \\
m_b\braket{A_b  -\expetam \mueff^* \tbeta} & m_b^2 + m_{\ti b_R}^2
-\fr13 M_Z^2 c_{ 2\beta}s^2_{\theta_W}   \epmatrix,
\label{eq:msbotmat}
\ee 
where
\be 
\mueff = \fr{\lambda v_s \expetas}{\sqrt{2}}.
\ee
The mass eigenstates are obtained by diagonalizing these squark
matrices with the unitary transformations
\be   
\diag(m_{\ti q_1}^2,m_{\ti q_2}^2 ) = U_{\ti q}  M_{\ti q}  U_{\ti
	q}^\dagger, \quad \bpmatrix 
\ti q_1\\ \ti q_2\epmatrix =  U_{\ti q} \bpmatrix
\ti q_L\\ \ti q_R\epmatrix, \quad q=t,b,
\ee
with the usual convention $m_{\ti q_1}\le m_{\ti q_2}$. \s

For the charged leptonic sector, we use the same assumption of no
generation mixing as in the squark sector. In each generation, the
left- and right-handed sleptons mix. For example, the mixing matrix
for the third generation, {\it i.e.}~for the left- and right-handed stau, is given by  
\be
M_{\ti \tau} = \bpmatrix m_{\ti L_3}^2 + m_\tau^2 + M_Z^2c_{ 2\beta}(-\fr12
+  s^2_{\theta_W}) & 
m_\tau\braket{A_\tau^*  - \expeta\mueff \tbeta} \\
m_\tau\braket{A_\tau  -\expetam \mueff^* \tbeta} & m_\tau^2 + m_{\ti \tau_R}^2
- M_Z^2 c_{ 2\beta}s^2_{\theta_W}   \epmatrix,
\label{eq:mstaumat}
\ee 
Upon diagonalization we obtain the mass eigenstates $\ti \tau_1$ and
$\ti\tau_2$ whose masses are ordered as $m_{\ti\tau_1}\leq m_{\ti\tau_2}$.\s 

In the neutral leptonic sector, the three left-handed neutrinos,
$\nu_{L_i}$, mix with the six leptonic component fields of the six singlet 
superfields $ \hat N_i^c, \hat X_i$, $i=1,2,3$, and the mass term in
the Lagrangian reads 
\be 
\calL_{\text{mass}}^\nu =-\fr{1}{2} \bpmatrix \nu_L & N^c & X
\epmatrix  M_\ISS^\nu  \bpmatrix \nu_L\\ N^c\\ X \epmatrix 
\ee
where the mixing mass matrix is given by
\be M_{\ISS}^\nu = \bpmatrix  0 & M_D & 0 \\
M_D^T & 0 & \mu_X \\
0  & \mu_X^T & M_X \epmatrix. \label{eq:neumassmatrix}\ee
Note that $\nu_{L_i}, N_i^c, X_i$ are left-handed Weyl spinors and the
products of them are defined in such a way that they are Lorentz
invariant. For example, $ \nu_{L_i} N^c_j=
\epsilon^{ab}\nu_{L_i,a}N^c_{j,b} $, where the spinor indices are
denoted by $a,b=1,2$, and the generation indices by $i,j=1,2,3$.  
The blocks $M_D, \mu_X$ and $M_X$ are $3\times 3$ matrices with $
\mu_X $ defined in Eq.~(\ref{eq:wnmssm}) and 
\be 
M_D =\fr{v_u e^{i\varphi_u}}{\sqrt{2}} y_\nu, \quad   M_X =
\fr{v_se^{i\varphi_s}}{\sqrt{2}} (\lambda_X + \lambda_X^T). \label{eq:lambdaxinmasses}
\ee
 The mass matrix $M^{\nu}_{\ISS}$ can be diagonalized by a $9\times 9$ unitary matrix as
\be 
U_\nu^* M_{\ISS}^\nu U^\dagger_\nu = \diag(m_{n_1},\cdots,m_{n_9}).  \label{eq:fulldia}
\ee 
The diagonalization process is done numerically in our code. It can be
performed, however, by using an expansion approximation
\cite{Gonzalez-Garcia:1988okv,Grimus:2000vj} to separate the $3\times
3$ light neutrino mass matrix from the $6\times 6 $ heavy states
exploiting the fact that all matrix elements 
of $M_D,M_X$ are much smaller than the eigenvalues of $\mu_X$. In
particular, at the lowest order, the $3\times 3$ light neutrino mass
matrix can be expressed as 
\be 
M_{\text{light}} = M_D M_N^{-1} M_D^T\,, \; \mbox{ with } \; M_N= \mu_X M_X^{-1}
\mu_X^T\,.\label{eq:mlight}
\ee
One then defines Majorana neutrino fields as  
\begin{align}
 n_i = \left( \begin{array}{c} \nu_i
\\ \overline{\nu_i} \end{array}
\right) \quad\ \text{with}\quad\ \nu_i = (U_\nu)_{ik} \nu_{L,k}+ (U_\nu)_{i(k+3)} N^c_{k}
+ (U_\nu)_{i(k+6)} X_{k},
\end{align}
where $i=1,\cdots,9,\quad k=1,2,3,$ and 
\beq
\overline{\nu_i} = i \sigma_2 \nu_i^{*}.
\eeq
The neutrino spectrum should contain three light active neutrinos
$n_i$ ($i=1,2,3$) with masses of order eV and six heavy neutrinos $n_{I}$,
$I=4,\cdots,9$. Their masses can be  of order TeV. The diagonalization
process can lead to negative mass eigenvalues, $m_{n_x}$, which
in the version of the code for the CP-violating NMSSM
we make positive by multiplying the corresponding $x_{\text{th}}$ rotation matrix row
with the imaginary unit $i$.
 So in our convention, the neutrino masses are all positive. In principle, one
gives arbitrary inputs for $M_D, \mu_X, 
M_X$ and then obtains the corresponding neutrino masses and their rotation matrix.  
However, the obtained masses and mixing angles must satisfy the experimental data of the three active neutrinos. The chance to get parameter points passing 
these constraints starting from arbitrary input values is very low. A
way out of this technical difficulty is to use a parameterization of
$M_D$ in terms of $\mu_X, M_X$ and the light neutrino masses and mixing angles. We
follow the  Casas-Ibarra  parameterization \cite{Casas:2001sr},  that
makes use of the leading order  relation for the light neutrino mass
matrix 
\bea
  U_{\text{PMNS}}^* M_{\text{light}} U_{\text{PMNS}}^\dagger = m_\nu \;, \label{eq:LOexpansion}
\eea  
with
\bea
m_\nu= \diag({m_{\nu_1}},{m_{\nu_2}},{m_{\nu_3}}) \;. \label{eq:mnui}
\eea
Using the expression of $M_{\text{light}}$ given in \eqref{eq:mlight}
one then gets
\bea 
M_D =  U_{\text{PMNS}}^T \sqrt{m_\nu}  R  \sqrt{{\calM}_N} V, \label{eq:MDdefinition}  
\eea
where 
\be 
{\calM}_N=\diag({M_{N_1}},{M_{N_2}},{M_{N_3}}) \; . 
\ee 
The Pontecorvo-Maki-Nakagawa-Sakata (PMNS) matrix $U_{\text{PMNS}}$ and the
active neutrino masses $m_{\nu_i}$ (i=1,2,3) are input values based on
the available experimental data, while $M_{N_1},M_{N_2},M_{N_3}$ are the
positive roots of $M_N$ and $V$ is a unitary matrix diagonalizing $M_N$ as 
\be 
\diag(M_{N_1},M_{N_2},M_{N_3}) = V^* M_N V^\dagger,
\ee
and $R$ is  a complex orthogonal matrix that can be written in terms
of three complex angles $\theta_i$ ($i=1,2,3$) 
\be 
R=\bpmatrix c_2s_3& -c_1s_3-s_1s_2c_3& s_1s_3-c_1s_2c_3\\
c_2s_3 & c_1c_3-s_1s_2s_3 & -s_1c_3 -c_1s_2s_3 \\
s_2 & s_1c_2& c_1c_2  \epmatrix,  \label{eq:rdefmat}
\ee
with $c_i=\cos\theta_i,\, s_i=\sin\theta_i$. In this study we set the
three angles $\theta_{1,2,3}$  to be real. Using this Casas-Ibarra
parameterization, the light neutrino masses denoted $m_{n_i}$
($i=1,2,3$) in \eqref{eq:fulldia} are  approximately the input neutrino
masses $m_{\nu_i}$ in \eqref{eq:LOexpansion}. If the relative difference, defined as the maximum of
 $\abs{(m_{n_i}-m_{\nu_i})/m_{\nu_i}}$ ($i=1,2,3$),
is  more than one percent, the code will print out a warning about the breakdown of the 
Casas-Ibarra  parameterization.  \s

With the introduction of the new superfields, the sneutrino sector is
also changed. To incorporate CP violation, each sneutrino field is
separated into its CP-even and CP-odd components as 
\begin{align}
\tn &= \dfrac{1}{\sqrt{2}} \qty(\tn_+ + i \tn_-)\\
\tN^* &= \dfrac{1}{\sqrt{2}} \qty(\tN_+ + i \tN_-)\\
\tX &= \dfrac{1}{\sqrt{2}} \qty(\tX_+ + i \tX_-).
\end{align}
The mass term in the basis $ \psi = (\tn_+,\tN_+,\tX_+,\tn_-,\tN_-,\tX_-)^T $ (generation indices are suppressed) is given by
\be 
{\cal L}= \fr 12 \psi^T M_{\tn}\psi\;,
\ee
where the mass matrix $ M_\tn $ is an $18\times 18$ symmetric matrix
that can be found in \appen{appen:sneumass}.  An orthogonal matrix $
\Usn $ can be used to find the masses of the sneutrinos as follows, 
\be
\diag \qty(m^2_{\tilde{n}_1},\cdots,m^2_{\tilde{n}_{18}}) = \Usn M_\tn \Usn^T,
\ee
where their mass values are ordered as $ m^2_{\tilde{n}_1} \leq \cdots \leq m^2_{\tilde{n}_{18}} $.

\section{Calculation of the Neutral Higgs Boson Masses and Mixings}
\label{sec:HMass}
In this section, we describe in detail our computation of the complete
one-loop contribution to the loop-corrected neutral Higgs boson masses
including the full momentum dependence  and give a detailed
description of the renormalization procedure.  We apply dimensional
reduction (DRED) \cite{Siegel:1979wq,Stockinger:2005gx} to
regularize the UV-divergences, which has been proven to conserve SUSY
at one-loop order. We have used several programs to compute the 
one-loop self-energies. To generate the Feynman diagrams and self-energies
we use \texttt{FeynArts}~\cite{Kublbeck:1990xc,Hahn:2000kx} together with a model
file created by
\texttt{SARAH}~\cite{Staub:2009bi,Staub:2010jh,Staub:2012pb,Staub:2013tta}. The
output self-energies were further processed using
\texttt{FeynCalc}~\cite{FeynCalc,Shtabovenko:2016sxi} for the
simplification of the Dirac matrices and for the tensor reduction. The
one-loop one- and two-point integrals were evaluated with a modified
loop library of \texttt{NMSSMCALC}~\cite{Baglio:2013iia}, in
particular we used quadruple precision and complex external momentum
squared for the two-point integrals to increase the convergence and
stability of the code.\footnote{It is also possible to use double precision
for the evaluation of the loop-corrected Higgs boson masses. This can
be set in the makefile. However, for some parameter points, the
convergence of the iterative method is not good compared to the usage
of quadruple precision. This does not happen in
the NMSSM without the seesaw mechanism.} The new model and the calculation of the 
loop-corrected Higgs boson masses and mixings have been implemented in
the code, called \texttt{NMSSMCALC-nuSS}.  The code can be
  downloaded from the url:
\begin{center}
https://www.itp.kit.edu/$\sim$maggie/NMSSMCALC-nuSS/
\end{center}

\subsection{Loop-corrected Higgs Boson Masses and Mixings}
The loop-corrected Higgs boson masses can be obtained from the real parts of
the complex mass eigenvalues of the $5\times 5$ Higgs mass matrix with
its elements 
\begin{align}
 \mathcal{M}_{h_i h_j}= m_{h_i}^2 \delta_{h_i h_j} - \hat{\Sigma}_{h_i
  h_j} (p^2), \quad i,j=1,\cdots,5,  
\end{align}
where $ \hat{\Sigma}_{h_i h_j} (p^2) $ is the renormalized self-energy
of the transition $h_i\to h_j$ at external momentum
squared $p^2$. We do 
not include the contributions due to the transitions $h_i\to G/Z$,
since their contributions are negligible for light Higgs bosons. For
extremely heavy Higgs bosons, they can have some effect as shown in
\cite{Domingo:2020wiy}. \s

\begin{figure}[htbp]
   \centering
   \includegraphics[height=0.4\textheight, width=0.8\textwidth]{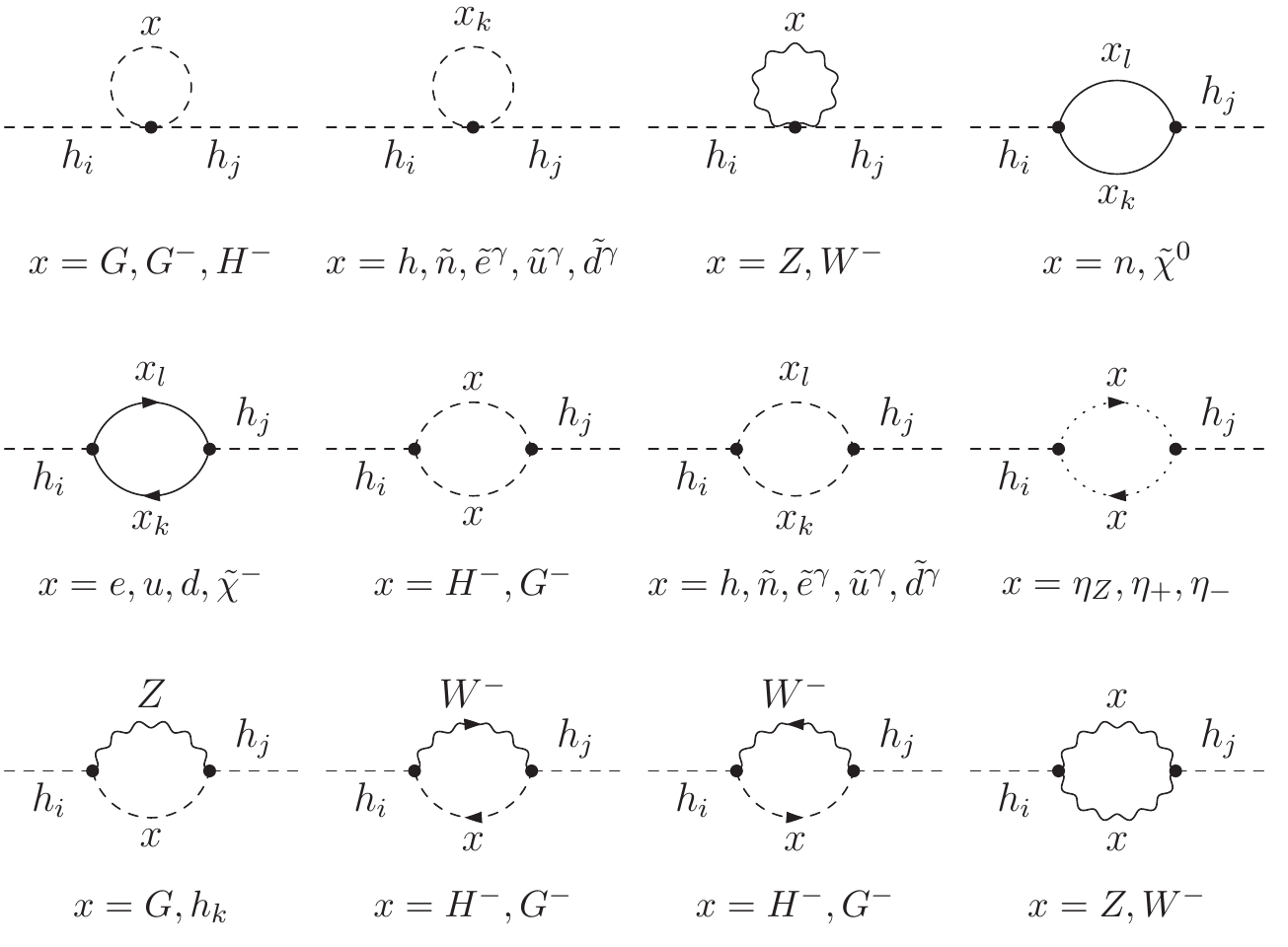} 
   \caption{Generic
    Feynman diagrams contributing to the one-loop neutral
     Higgs self-energies. The indices $k,l$ take several sets of values
depending on the fields that they go with: $k,l=1,\cdots,5$ for $x=h,\ti \chi^0$,
$k,l=1,\cdots,3$ for $x=e,u,d,\ti e,\ti u,\ti d$,\, $k,l=1,\cdots,18$ for $x=\ti n$
and $k,l=1,\cdots,9$ for $x=n$. The index $\gamma$ denotes the left- and
right-handed scalars. Color indices for the quarks and squarks are
suppressed.}
   \label{fig:FAHSdiags}
\end{figure}

The renormalized Higgs self-energies at one-loop level can be written in
terms of the unrenormalized self-energies $ \Sigma_{h_i h_j} (p^2)$ and
the counterterms as 
\begin{align}
\hat{\Sigma}_{h_i h_j} (p^2) =\ 
& \Sigma_{h_i h_j} (p^2) + \half p^2 \qty[\calR \qty(\delta {Z}_{hh}^\dagger + \delta {Z}_{hh}) \calR^T]_{ij} \notag\\
& - \half \qty[\calR\qty(\delta {Z}_{hh}^\dagger M_{hh} + M_{hh} \delta {Z}_{hh})\calR^T]_{ij} - \qty[\calR \delta M_{hh} (\calR)^T]_{ij}. \label{renself}
\end{align}
where the Higgs mass counterterm matrix is denoted by $\delta M_{hh}$
and the wave-function renormalization constant matrix
by $\delta {Z}_{hh}$ in the basis $(h_d,h_u,h_s,a,a_s)$. In
\figref{fig:FAHSdiags}, the one-loop Feynman diagrams contributing to
the unrenormalized self-energies $ \Sigma_{h_i h_j} (p^2)$ are shown.
In the following sections, we will discuss the counterterms and
renormalization conditions in detail. We furthermore include the
dominant two-loop corrections of order ${\cal O}(\alpha_s\alpha_t)$
\cite{Muhlleitner:2015dua} and ${\cal O}(\alpha_t^2)$
\cite{Dao:2019qaz}, which are available both for the real and for the
complex NMSSM, to increase the precision for the 
phenomenological analysis presented in Section \ref{sec:analysis}. \s

For the diagonalization of the loop-corrected mass matrix we apply the
iterative method presented in
\cite{Dao:2019qaz,Ender:2011qh,Graf:2012hh,Muhlleitner:2015dua}.\footnote{While
  this method includes contributions beyond the fixed-order
  renormalized self-energies, it can give rise to the gauge-parameter
  dependence of the loop-corrected masses due to the incomplete
  higher-order terms as studied in \cite{Dao:2019nxi,
    Baglio:2019nlc,Domingo:2020wiy}.  } For the loop-corrected $ n $-th Higgs
boson mass in the first iteration, the external momentum squared is
set equal to the tree-level Higgs boson mass, $ p^2 = m_{h_n}^2 $. The
obtained matrix is then diagonalized, yielding the $ n $-th diagonal
element. This value is then used as input momentum squared for the new
iteration. The process is repeated until the change in $ p^2 $ between
two consecutive iterations is less than $ 10^{-9} $. The $n$-th
loop-corrected mass squared $M_{h_n}^2$ is then defined as the real
part of the last iterative $ n $-th diagonal
element.\footnote{We use capital $M$ to
    denote loop-corrected masses in contrast to $m$ for tree-level
    masses. For masses that are renormalized on-shell ($M_{H^\pm},
    M_Z, M_W$, {\it cf.}~Subsec.~\ref{sec:renmwetc}) where the
    distinction need not be made, we use capital 
    $M$ as well.} The algorithm is 
repeated for all neutral Higgs boson masses. The loop-corrected masses
are then sorted in ascending order, $M_{h_1} \leq M_{h_2} \leq M_{h_3}
\leq M_{h_4} \leq M_{h_5}$. These loop-corrected masses obtained by
the iterative method will be the outputs used in the decay width
calculations and in the phenomenological study, if not stated otherwise.  \s  

We now define the loop-corrected mixing matrix that will be used to
compute the effective couplings of the Higgs bosons with gauge bosons,
fermions and among themselves.  We define the loop-corrected mixing
matrix $R^0$ to be the rotation of the loop-corrected mass matrix in the
approximation of vanishing external momentum, 
\be
\text{diag}(M_{0,H_1}^2,M_{0,H_2}^2,M_{0,H_3}^2,M_{0,H_4}^2,M_{0,H_5}^2)=R^0  \mathcal{M}_{hh}(0) (R^0)^T \;.
\label{eq:R0}
\ee
The corresponding loop-corrected mass eigenvalues are denoted by an
index 0 and sorted in ascending order, $M_{0,H_1} \le M_{0,H_2}\le
M_{0,H_3} \le M_{0,H_4}\le M_{0,H_5}$. In this approximation, the
mixing matrix $R^0$ is unitary but does not capture the proper OS
properties of the external loop-corrected states as momentum-dependent
effects are neglected. Using these thus defined mixing
  matrix elements, we obtain the Higgs effective couplings.
Following the strategy presented in the
\texttt{NMSSMCALC}~\cite{Baglio:2013iia}, these Higgs effective
couplings will be used to compute the Higgs decay widths, taking into
account also higher-order QCD corrections when available. 

\subsection{Counterterms of the Higgs Sector}
\label{ssec.CT}
 Closely following the renormalization procedure at one-loop level
 described in \cite{Ender:2011qh,Graf:2012hh}, we choose the following
 set of quantities as our independent input,  
\begin{align}
\qty{t_{h_d}, t_{h_u}, t_{h_s}, t_{a_d}, t_{a_s},e,M_W,M_Z,M_{H^\pm},
  \tan\beta, v_s, \absl, \absk, \reAk, \vp_\lambda, \vp_\kappa, \vp_u,
  \vp_s},  
\end{align}
where the five soft SUSY breaking parameters $\ti m_{ H_d}^2,\ti
m_{H_u}^2,\ti  m_{S}^2, \imAl, \imAk$ have been replaced by the five
independent tadpoles $t_{h_d}, t_{h_u}, t_{h_s}, t_{a_d}, t_{a_s}$
which vanish at tree level. The complex phases $\vp_\lambda,
\vp_\kappa, \vp_u, \vp_s$ do not need to be renormalized at one-loop
level. The remaining input parameters are replaced by the sum of the 
corresponding renormalized parameters and their counterterm as  
\begin{align}
t_{\phi } &\to t_{\phi} + \delta t_{\phi} \text{ with } \phi =
            \qty{h_d, h_u, h_s, a_d, a_s} \label{ct1} \\
M^2_{H^\pm } &\to M^2_{H^\pm} + \delta M^2_{H^\pm}\\
M^2_{W } &\to M^2_{W} + \delta M^2_{W}\\
M^2_{Z} &\to M^2_{Z} + \delta M^2_{Z}\\
e &\to e(1 + \delta Z_e)\\
\tb &\to \tb + \delta \tb\\
v_{s} &\to v_s + \delta v_s\\
\absl &\to \absl + \delta \absl\\
\absk &\to \absk + \delta \absk\\
\reAk &\to \reAk + \delta \reAk.
\end{align}
The neutral Higgs boson  counterterm matrix $ \delta M_{hh} $ in
\eqref{renself}  can be written in terms of the counterterms of the input
parameters. The analytical expression of $ \delta M_{hh} $ in terms of
these counterterms can be found in \appen{appen:CTHiggsMass}. In order
to determine the counterterms, we need renormalization conditions.  
In this study, we use a mixture of the \DRbt and the OS scheme specified as  
\begin{align}
\underbrace{t_{h_d}, t_{h_u}, t_{h_s}, t_{a_d}, t_{a_s}, e,M_W,M_Z,M_{H^\pm}}_{\text{OS scheme}},\underbrace{\tan\beta, v_s, \absl, \absk, \reAk}_{\text{\DRbt scheme}}.
\end{align}
In our code, there is also the possibility to chose $\reAl$ to be the
input parameter instead of the charged Higgs mass. In this case
$\reAl$ is renormalized in the \DRbt scheme, while $M_{H^\pm}$ is
computed at the same order as the one of the neutral
  Higgs boson masses. \s

The neutral Higgs wave function renormalization constants are
introduced for the neutral components of both doublets and the singlet as 
\begin{align}
H_{d} \to \qty(1 + \half \delta Z_{H_d}) H_d
\ ,\quad
H_{u}\to\qty(1 + \half \delta Z_{H_u}) H_u
\ ,\quad
S\to\qty(1 + \half \delta Z_S) S
\label{higgs field renormalization}.
\end{align}
Hence the wave-function renormalization constant
matrix introduced in \eqref{renself} in the basis $ \phi = (h_d, h_u, h_s, a, a_s)^T $ is given by
\begin{align}
\phi&\to \qty(1 + \half \delta {Z}_{hh}) \phi,\\
\delta {Z}_{hh} &= \diag\qty(\delta Z_{H_d},\delta Z_{H_u}, \delta Z_S,s_\beta^2 \delta Z_{H_d}+ c_\beta^2 \delta Z_{H_u}, 
\delta Z_S).
\end{align}
%
\subsubsection{The Neutral Wave Function Renormalization Constants}
We use the \DRbt scheme to define the Higgs wave function
renormalization constants.\footnote{While this renormalization is
  simple in practice, corrections arising from
  $ \hat{\Sigma}_{h_ih_i} (\partial\hat{\Sigma}_{h_i
      h_i}(p^2)/\partial p^2)$ and from
  $\hat{\Sigma}_{h_ih_j}\hat{\Sigma}_{h_jh_i}/(m_{h_i}^2-m_{h_j}^2)$
  enter the loop-corrected masses, which are of higher order compared to the
fixed order correction included in $\hat{\Sigma}_{h_ih_j}$. It is also
possible to choose the OS scheme for the wave function renormalization
constants, so that 
$(\partial\hat{\Sigma}_{h_i h_j}(p^2)/\partial p^2)$ and
$\hat{\Sigma}_{h_ih_j}$ vanish at the tree-level Higgs masses. This can be
done in the case where $m_{h_i}^2,m_{h_j}^2$ are very different in
magnitude so that mixing effects can be expected to be
small. In the other cases where the mixing terms are significant,
by using the \DRbt scheme they are absorbed into the loop-corrected
masses.} 
The \DRbt scheme requires that the divergent part of the first
derivative of the renormalized self-energies with respect to the
momentum squared vanishes, 
\begin{align}
\widetilde{\text{Re}} \eval{\dfrac{\partial\hat{\Sigma}_{h_i
  h_j}(p^2)}{\partial p^2}}_{\divt} = 0, \quad \forall i,j=1,\ldots,5 \;,
\end{align}
where the notation $ \widetilde{\text{Re}} $ means that only  the real
part of the loop integral is taken, and the superscript 'div' denotes
the divergent part. This equation should hold for any
external momentum. In practice we chose $p^2=0$. This renormalization
condition leads to the equation  
\begin{align}
\delta {Z}_{h_ih_i} = - \left[\calR^T \widetilde{\text{Re}} \eval{\dfrac{\partial\Sigma_{hh}}{\partial p^2}}_{\divt} \calR
\right]_{ii},
\end{align}
which is real by definition and the off-diagonal elements
vanish. Since  $ \delta {Z}_{hh} $ contains only three unknown 
variables $ \delta Z_{H_u}, \delta Z_{H_d} $ and $ \delta Z_S $, one
needs a set of three independent equations. Any chosen set must give
the same solution due to the $SU(2)_L$ symmetry.
%
\subsubsection{Tadpole Renormalization}
\begin{figure}[htbp]
   \centering
   \includegraphics[height=0.3\textheight, width=0.7\textwidth]{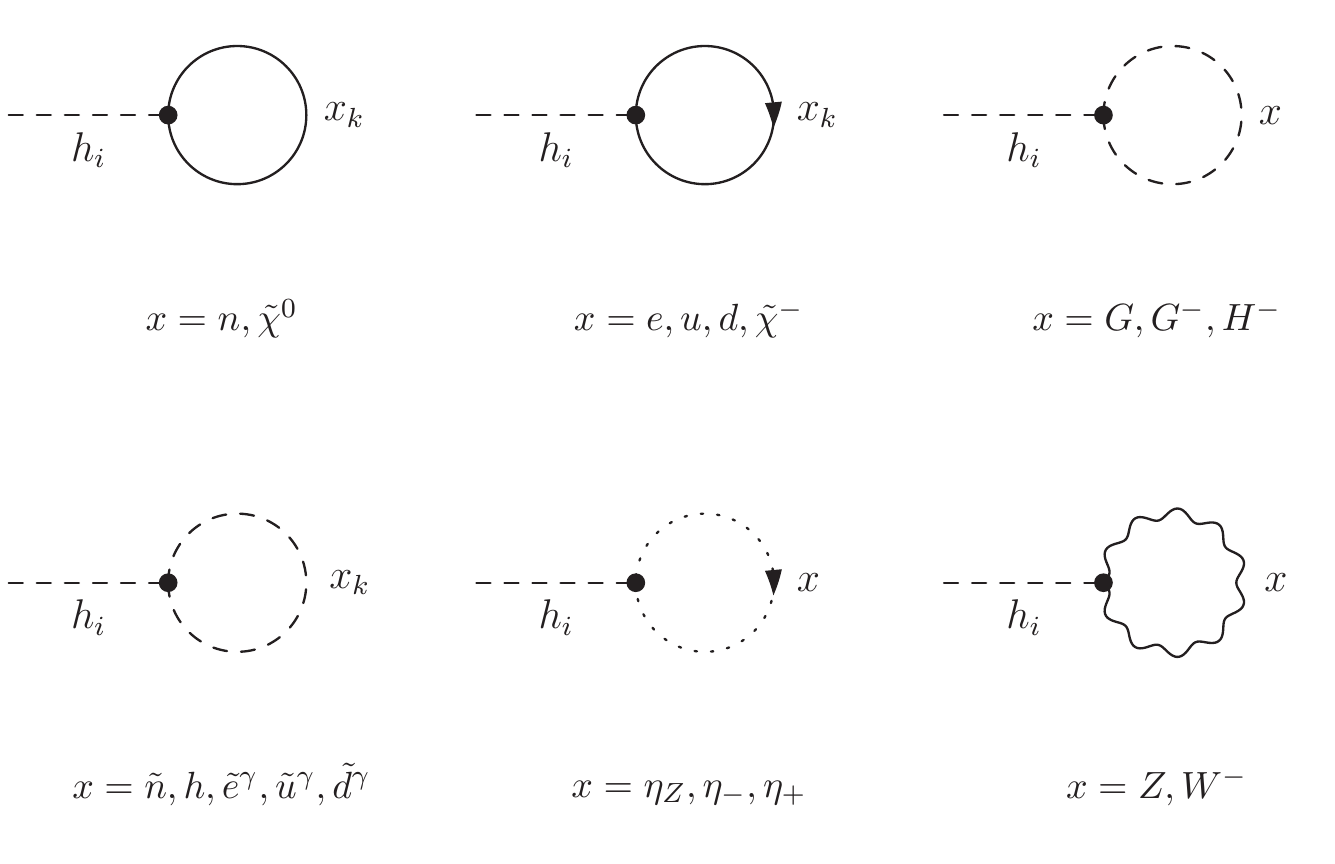} 
   \caption{Generic Feynman diagrams contributing to the one-loop tadpole
     counterterms. The index $k$ is given by $k=1,\dots,18$ for $x=\ti
     n$, $k=1,\dots,5$ for $x=h,\chi^0$, $k=1,\dots,9$ for $x=n$ and
     $k=1,\dots,3$ for $x=e,u,d,\ti e,\ti u,\ti d$. The index $\gamma$
     denotes the left- and right-handed scalars. Color indices for the
     quarks and squarks are suppressed.}
   \label{fig:FATaddiags}
\end{figure}
The tadpole counterterms are defined such that the minima of the Higgs
potential do not change at higher order. The tadpole counterterms
hence have to cancel any contribution from the diagrams at one-loop
level leading to the renormalization conditions  
\begin{align}
\delta t_\phi = t_\phi, \quad \phi = h_d, h_u, h_s, a_d, a_s \;,
\end{align}
with the  one-loop tadpole diagrams contributing to $ t_\phi$ 
depicted in \figref{fig:FATaddiags}.  
%
\subsubsection{Renormalization of \texorpdfstring{$M_W,M_Z,M_{H^\pm}$}{Lg} \label{sec:renmwetc}}
For the masses $M_W,M_Z$ of the massive gauge bosons as well as 
$M_{H^\pm}$ of the charged Higgs boson we apply OS
renormalization by requiring that the pole of the
corresponding two-point correlation function at one-loop level occurs
at the value of the input mass. In particular, the mass counterterms
are given by the unrenormalized self-energies as
\begin{align}
\delta M_{H^\pm}^2 &= \widetilde{\text{Re}} \Sigma_{H^+ H^-}
                     (M_{H^\pm}^2)\label{eq,CHct},\\ 
\delta M_W^2 &= \widetilde{\text{Re}} \Sigma^{T}_{WW} (M_W^2) \label{eq,Wct}, \\
\delta M_Z^2 &= \widetilde{\text{Re}} \Sigma^{T}_{ZZ}
               (M_Z^2) \;,\label{eq,Zct} %
\end{align}
where the superscript $T$ denotes the transverse parts of the
respective self-energies.  The wave function renormalization constants
$\delta Z_{H^+H^-}$ for the charged Higgs boson, $\delta Z_{H^+G^-}$,
$\delta Z_{H^+W^-}$ for the $H^+-G^+/W^+$ mixings, $\delta Z_{W^+W^-}$ for
the $W$ boson, $\delta Z_{Z^+Z^-}$ for the $Z$ boson and $\delta
Z_{ZG}, \delta Z_{Z\gamma}$ for the $Z-G/\gamma$ mixings are all
renormalized in the OS scheme, so that there is no additional
contribution to  Eqs.~(\ref{eq,CHct})-(\ref{eq,Zct}). 
%
\subsubsection{Renormalization of the Electric Charge}
The electric charge is renormalized in the OS scheme
\cite{Denner:1992vza,Denner:1991kt}. We take, however, the fine structure constant
at the $Z$ boson mass, $\alpha (M_Z^2)$, as input so that the
counterterm is given by
\begin{align}
\delta Z_e &=\half \left. \dfrac{\partial \Sigma^T_{\gamma\gamma}(k^2)}{\partial k^2} \right|_{k^2=0} + \dfrac{s_W}{c_W} \dfrac{\Sigma^T_{\gamma Z} (0)}{M_Z^2}-
\fr{1}{2} \Delta\alpha(M_Z),\\
\Delta\alpha(M_Z)&=\left. \dfrac{\partial \Sigma^{\text{light},T}_{\gamma\gamma}(k^2) }{\partial k^2} \right|_{k^2=0} - \fr{\widetilde{\text{Re}} \Sigma^{\text{light},T}_{\gamma\gamma}(M_Z^2) }{M_Z^2},
\end{align}
where the photon self-energy $\Sigma^{\text{light},T}_{\gamma\gamma}$
includes only the light SM fermion ($f$ with $m_f <m_t$) contributions. 
%
\subsubsection{Renormalization of \texorpdfstring{$ \tan\beta $}{Lg}}
The ratio of the two vacuum expectation values, $\tan\beta$, is
renormalized in the \DRbt scheme with the counterterm given by
\cite{Freitas:2002um,Dabelstein:1995js,Dabelstein:1994hb},  
\begin{align}
\delta \tan\beta = \left. \half \tan\beta \qty(\delta Z_{H_u} - \delta
  Z_{H_d})\right|_{\text{div}}.
\end{align}	
%
\subsubsection{Renormalization of the remaining \texorpdfstring{\DRbt}{Lg} quantities}
The renormalization of the remaining \DRbt quantities, $\delta v_S, \delta \absl, \delta \absk, \delta\reAk$ is defined such that
\begin{align}
\eval{\hat{\Sigma}_{h_i h_j}}_{\divt} = 0.
\end{align}
This system has more equations than the number of unknown
counterterms. We need only four independent equations to solve for the
four counterterms. Any set of four chosen equations resulted in the
same values for the counterterms, confirming that the renormalization procedure works.
 
\section{Constraints}
\label{sec:constraints}
In this section, we discuss all constraints that have been taken into account in our 
present study. Since we concentrate on the effects of the loop
corrections of the extended (s)neutrino sector on the loop-corrected Higgs masses and
their mixing, we consider here only the most relevant constraints
from the Higgs data, the active
	light neutrino oscillations, the electroweak precision observables, 
	 and the lepton flavor-violating radiative
decays $l_1 \to l_2 +\gamma$.  
\subsection{Higgs Data}
Our model, which contains five neutral and two charged Higgs bosons,
 must satisfy the experimental results on the $125\,\gev$ Higgs boson
 and the experimental constraints on new scalars. For a parameter
 point, we will calculate the Higgs boson masses including the
 available two-loop corrections at
 $\order{\alpha_t\alpha_s+\alpha_t^2}$ described in \sect{sec:HMass}
 and 
 the Higgs decay widths and branching ratios including the
 state-of-the-art higher-order QCD corrections which we take from the
 code \texttt{NMSSMCALC} \cite{Baglio:2013iia}. To check if a
 parameter point passes all the exclusion limits from searches at LEP,
 Tevatron and LHC we make use of the code  
\texttt{HiggsBounds-5} \cite{Bechtle:2020pkv}. We provide the Higgs
spectrum, decay widths, and the effective couplings as required by
\texttt{HiggsBounds} in an SLHA file \cite{Skands:2003cj}. If the
parameter point is allowed by \texttt{HiggsBounds}, it then will be
checked against the $125\,\gev$ Higgs boson data by using the code
\texttt{HiggsSignals} \cite{Bechtle:2013xfa}. We allow the uncertainty
of the SM-like Higgs boson mass to be $3\,\gev$ which means that at
least one Higgs boson must have a mass in the range $[122, 128]\,
\gev$. For the experimental data set, we use the "latestresults"
option. Using our input parameters, \texttt{HiggsSignals} computes the
$\chi^2$ from 107 observables  including the signal strength peak,
simplified template cross sections, the LHC Run-1 signal rates, and Higgs masses. We allow the total $\chi^2$ to vary within $2\sigma$ from the total $\chi^2$
obtained from the SM Higgs boson.  In \texttt{HiggsSignals-2.5.1}, the
SM $\chi^2$ is 84.44 and the $2\sigma$ for two degrees of freedom
corresponds to a $6.18$ $\chi^2$ difference.   
Therefore the NMSSM $\chi^2$ is allowed in the range $[78.26,90.62]$.

\subsection{The Active Light Neutrino Data}
\label{ssect.neutrino}
In our neutrino sector, there are three light neutrinos which correspond to the three types
of neutrinos observed in experiments. As discussed in \sect{sec:tree-levelspectrum},
for the neutrino sector, we use the Casas-Ibarra parameterization. This
means that we need three light mass values $m_{\nu_i}$, $i=1,2,3$, three angles and one complex phase of the PMNS matrix, three complex angles
of the orthogonal matrix $R$, defined in
  Eq.~(\ref{eq:rdefmat}), together with the matrices $\mu_X$ and 
$M_X$ to compute the mass matrix $M_D$ specified in
\eqref{eq:MDdefinition}. The obtained $M_D$ will be used to compute
the neutrino Yukawa couplings $y_\nu$ and the neutrino mass matrix in
\eqref{eq:neumassmatrix}. We then diagonalize this mass matrix using
quadruple precision. The obtained mass eigenvalues $m_{n_j}$
($j=1,...,9$) are the neutrino mass eigenvalues. 
From the $9\times 9$ neutrino mixing matrix we take the $3\times 3$
block which describes 
the mixing between the three light neutrinos and define
\be 
N_{ij} =  U^{\nu}_{ij}, \quad i,j=1,2,3.
\ee
The $N$ matrix is not unitary and can be written as
\be 
N= (I-\eta)U_{\text{PMNS}}.
\ee
We require that our input parameters, $m_{\nu_i}$, $i=1,2,3$, and
$U_{\text{PMNS}}$, satisfy the active light neutrino data. We take the best
fit points and the $3\sigma$ ranges  from the global fit, NuFIT 5.0
\cite{PhysRevD.95.096014}. For convenience, we list here the $3\sigma$
ranges for the mass differences and the mixing angles that are
obtained from the combined analysis including the latest neutrino
oscillation data presented at the ``Neutrino2020'' conference with the 
Super-Kamiokande atmospheric neutrino data. As usual, we define
\bea 
\Delta m_{ij}^2 &=&m_{\nu_i}^2 -m_{\nu_j}^2,\crn
\quad U_{\text{PMNS}}&=&\bpmatrix c_{12} c_{13} & s_{12}c_{13} & s_{13}e^{-i\delta_{CP}}\\
                          -s_{12}c_{23}- c_{12}s_{23}s_{13} e^{i\delta_{CP}}& c_{12}c_{23} - s_{12}s_{23}s_{13} e^{i\delta_{CP}} & s_{23}c_{13}\\
s_{12}s_{23}-c_{12}c_{23}s_{13} e^{i\delta_{CP}}& -c_{12}s_{23}- s_{12}c_{23}s_{13} e^{i\delta_{CP}} & c_{23}c_{13}\epmatrix,
\eea
where the short-hand notation $c_{ij}=\cos\theta_{ij}$ and
$s_{ij}=\sin\theta_{ij}$ has been used. We request that
\begin{align}
\Delta m_{21}^2\in [6.82,8.04] \times 10^{-5} \;\EV^2 , \quad s_{12}^2\in [0.269,0.343],
\end{align}
and for the normal ordering 
\begin{align}
 \Delta m_{31}^2 &\in [2.435,2.598]\times 10^{-3}\,\EV^2, & s_{23}^2&\in [0.415,0.616],\crn
 s_{13}^2&\in [0.02032,0.02410], &\delta_{CP} &\in [120,369],
\end{align}
and for the inverted ordering
\begin{align}
 \Delta m_{23}^2 &\in [2.414,2.581]\times 10^{-3}\,\EV^2, & s_{23}^2&\in [0.419,0.617],\crn
 s_{13}^2&\in [0.02052,0.02428],  & \delta_{CP} &\in [193,352].
\end{align}
In our analysis we use the constraint on the
non-unitary  $N$ matrix that arises from a combined analysis of short
and long-baseline neutrino oscillation data \cite{Parke:2015goa}. We
use the three most stringent bounds at the 3$\sigma$ CL expressed in a
parameterization-independent way, in particular 
\bea 
1 - \sum_{k=1}^3 N_{e k}N_{e k}^* &<&0.07,
\crn
1 - \sum_{k=1}^3 N_{\mu k}N_{\mu k}^* &<&0.07,
\crn
      \sum_{k=1}^3 N_{e k}N_{\mu k}^* &<& 0.03. 
\eea
This constraint will be denoted as non-unitary constraint in the
numerical section \ssect{sec:LFV}. \s

Furthermore, we use the Planck 2018 results for the upper limit of the
sum of the three light neutrino masses, 
\be 
\sum_{i=1}^3 m_{\nu_i} < 0.12 \, \text{eV}. 
\ee 

\subsection{The Oblique Parameters}
The presence of the supersymmetric particles, multiple Higgs boson
states and sterile neutrinos affects the masses and decay properties
of the electroweak bosons and the low-energy data. We use the three
well-known gauge self-energy parameters $S,T$ and $U$
\cite{Peskin:1991sw} at the one-loop level to describe 
the effects arising from new particles. Following \cite{Zyla:2020zbs}, we
also define the parameters $S,T,U$ from the transverse part of the
gauge boson self-energies as 
\begin{align}
T&= \fr{1
   }{\alpha}\braket{\fr{\Pi^{\text{new}}_{WW}(0)}{M_W^2}-\fr{\Pi^{\text{new}}_{ZZ}(0)}{M_Z^2}
   }, \\
S&= \fr{4 s^2_W c^2_W }{\alpha} \Bigg(\fr{\Pi^{\text{new}}_{ZZ}(M_Z^2)-\Pi^{\text{new}}_{ZZ}(0)}{M_Z^2} 
-\fr{c_W^2-s_W^2}{s_Wc_W}\fr{\Pi^{\text{new}}_{Z\gamma}(M_Z^2)}{M_Z^2}
   - \fr{\Pi^{\text{new}}_{\gamma\gamma}(M_Z^2)}{M_Z^2}\Bigg), \\
S+U&=\fr{4 s^2_W}{\alpha}
     \Bigg(\fr{\Pi^{\text{new}}_{WW}(M_W^2)-\Pi^{\text{new}}_{WW}(0)}{M_W^2}  
-\fr{c_W}{s_W}\fr{\Pi^{\text{new}}_{Z\gamma}(M_Z^2)}{M_Z^2} -
     \fr{\Pi^{\text{new}}_{\gamma\gamma}(M_Z^2)}{M_Z^2}\Bigg), 
\end{align}
where the fine structure constant $\alpha$ is given at the scale
$M_Z$. The superscript "{\text{new}}" means that we have subtracted the
SM contribution  
computed with a Higgs boson mass of $125\,\gev$ so that only new
physics contributions remain. Using data from physics at the $Z$ pole, 
 \cite{Zyla:2020zbs} has found the following best fit point and
 $1\sigma$ uncertainties for these parameters, 
\begin{align}
T&= 0.03\pm 0.12, \\
S&= -0.01\pm 0.10, \\
U&= 0.02 \pm 0.11.
\end{align} 
In our analysis, a valid parameter point satisfies constraints on new
physics if the $S,T,U$ values vary within the $1\sigma$ uncertainty ranges
around the best fit point. 

\subsection{The Radiative \texorpdfstring{$l_1 \to l_2 +\gamma$}{Lg} Decays}
\label{sec:LFVdecays}
We work in the NMSSM in which the soft SUSY breaking mass matrices
$\ti m_L^2, \ti m_E^2$ and trilinear couplings $A_{e}$ as well as the
Yukawa couplings $y_e$ are diagonal in any
basis. In general, if off-diagonal elements of these matrices exist,
they have large effects on charged lepton flavor-violating (LFV) processes
which have been severely constrained \cite{Patrignani:2016xqp}. 
Although in our setting these matrices are flavor conserving, the
presence of the low-scale sterile neutrinos and mixings with active
neutrinos can still induce large charged LFV processes. The most constraining
LFV processes are the radiative decays $\tau \to \mu \gamma$, $\tau
\to e \gamma$ and $\mu \to e \gamma$, which are calculated in this
section. The corresponding experimental bounds at 90\% confidence
level \cite{Zyla:2020zbs} are
\beq 
\Br (\mu^- \to e^-\gamma) \; &< \; 4.2 \times 10^{-13}, \\
\Br (\tau^- \to e^-\gamma) \; &< \; 3.3 \times 10^{-8}, \\
\Br (\tau^- \to \mu^-\gamma) \; &< \; 4.4 \times 10^{-8}.
\eeq
\begin{figure}[htbp]
   \centering
   \includegraphics[height=0.3\textheight, width=0.8\textwidth]{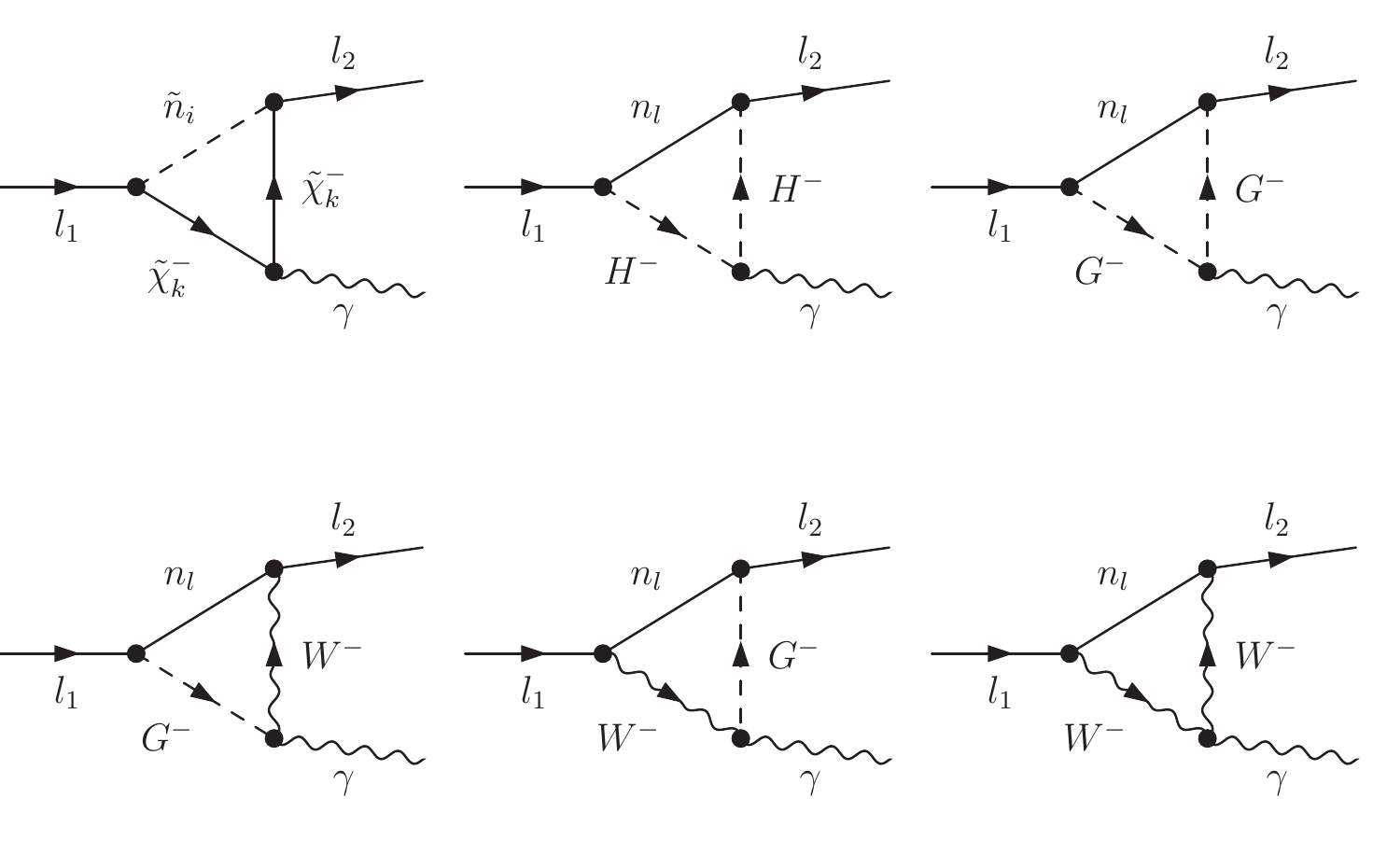} 
   \caption{Generic Feynman diagrams contributing to the charged lepton
     flavor-violating decays $l_1 \to l_2
     +\gamma$. The ranges of the indices 
     are $i=1,...,18, k=1,2$ and $  l=1,...,9$.}
   \label{fig:FAdiags}
\end{figure}
These processes have been widely studied in the literature for
non-supersymmetric models and the MSSM using the exact diagonalization
of the mass matrices or using the mass insertion approximation, for a
review see \cite{Calibbi:2017uvl} and references therein. In our
calculation we use the exact diagonalization of the relevant
(s)neutrino mass matrices. We have used the model file obtained from
\texttt{SARAH} to generate one-loop Feynman diagrams and amplitudes
using \texttt{FeynArts}, and further simplified the amplitudes with 
the help of the package \texttt{FeynCalc}. The one-loop Feynman
diagrams contributing to the decay processes  
\beq 
l_1 (p) \to l_2 (p-q) + \gamma(q)
\eeq
are depicted in \figref{fig:FAdiags}. The amplitude of this process is given by
\be  
\calA =  i \epsilon^*_\mu (q) \calM^\mu,
\ee
where $\epsilon_\mu$ denotes the polarization vector of the external photon.
Using gauge invariance, $ q_\mu \calM^\mu = 0 $, we can prove that 
$\calM^\mu $ must take the form 
\beq 
\calM^\mu = \baru_2  \sigma^{\mu\nu} q_\nu \qty(F_L P_L + F_R P_R)
u_1,
\eeq
where $ \sigma^{\mu\nu} = i/2 \comm{\gamma^{\mu}}{\gamma^\nu}$, $
P_{L/R} = (1 \mp \gamma_5)/2 $ and $F_{L/R}$ are left- and right-handed
form factors. The partial width is then given by 
\beq 
\Gamma(l_1\to l_2 \gamma) = \dfrac{m_1^3}{16\pi} \qty(\abs{F_L}^2 +
\abs{F_R}^2) , 
\eeq
where $m_1$ is the mass of $ l_{1} $ and we have neglected the mass
of $l_2$ since $m_2 \ll m_1$ for the processes
  considered here. Following the common procedure presented
in \cite{Lavoura:2003xp}, the branching ratios of the decays $l_1\to l_2
\gamma$ can be written in terms of the branching ratios of $l_1\to
l_2\bar \nu_2 \nu_1$ which are experimentally measured. 
In the NMSSM,
the tree-level decay width for the decays $l_1\to l_2\bar \nu_2 \nu_1 $ is
the same as that of the SM, \ie 
\be  
\Gamma(l_1\to l_2\bar \nu_2 \nu_1) =  \fr{G_F^2 m_1^5}{192 \pi^3};
\ee
therefore,
\be 
\Br(l_1\to l_2 \gamma) =\fr{12\pi^2}{G_F^2 m_1^2} \qty(\abs{F_L}^2 +
\abs{F_R}^2) \Br(l_1\to l_2\bar \nu_2 \nu_1). 
\ee
We use the following numerical values taken from
\cite{Patrignani:2016xqp} for the branching ratios,
\bea 
\Br(\mu\to e\bar \nu_e \nu_\mu) = 1, \quad \Br(\tau\to e\bar \nu_e
\nu_\tau) = 17.82 \%, \quad \Br(\tau\to \mu\bar \nu_\mu \nu_\tau) =
17.39 \%.
\eea
The contributions to the form factors in our model can be decomposed
into three parts as
\beq 
F_{L,R} = F^{W^\pm}_{L,R} + F_{L,R}^{H^\pm} + F_{L,R}^{\tilde{\chi}^\pm} \
, \label{eq:formfactor}
\eeq
where $ F^{W^\pm}_{L,R}, F_{L,R}^{H^\pm}$ and $ F_{L,R}^{\tilde{\chi}^\pm} $
denote the contributions from the one-loop diagrams  with $W^\pm$ and
charged Goldstone bosons, charged Higgs bosons and charginos,
respectively, on the internal lines. Their explicit expressions are
given in \appen{appen:formfactor}. Since in the numerical analysis, all the
\DRbt input parameters are given at the SUSY scale, we do not
consider contributions from the off-diagonal elements of $\ti m_L^2,
\ti m_E^2,A_{e}$ due to the renormalization group equations as discussed in
\cite{Casas:2001sr}. \s

It is also possible to keep the lepton masses of the external lines in
the three-point loop integrals. However, we explicitly checked that
the differences between the branching ratios obtained with the full lepton masses
using {\tt Mathematica} and the ones with zero lepton masses are below
the per-mille level. The same result has been found in
the Standard Model case \cite{Hue:2017lak}.  We, therefore, use the zero
lepton mass approximation implemented in the  code
\texttt{NMSSMCALC-nuSS} in the numerical analysis.
\section{Numerical analysis}
\label{sec:analysis}
In this section we will discuss the numerical impact of the neutrino
and sneutrino sectors on the loop-corrected neutral Higgs boson masses
and on the charged lepton flavor-violating
decays. We have performed a scan over the parameter space of our model to obtain
parameter points that satisfy all our constrains mentioned in
\sect{sec:constraints}. We chose SM input parameters as
\cite{Zyla:2020zbs,Dennerlhcnote}     
\begin{equation}
\begin{tabular}{lcllcl}
\quad $\alpha(M_Z)$ &=& 1/127.955, &\quad $\alpha^{\overline{\mbox{MS}}}_s(M_Z)$ &=&
0.1181\,, \\
\quad $M_Z$ &=& 91.1876~GeV\,, &\quad $M_W$ &=& 80.379~GeV \,, \\
\quad $m_t$ &=& 172.74~GeV\,, &\quad $m^{\overline{\mbox{MS}}}_b(m_b^{\overline{\mbox{MS}}})$ &=& 4.18~GeV\,, \\
\quad $m_c$ &=& 1.274~GeV\,, &\quad $m_s$ &=& 95.0~MeV\,,\\
\quad $m_u$ &=& 2.2~MeV\,, &\quad $m_d$ &=& 4.7~MeV\,, \\
\quad $m_\tau$ &=& 1.77682~GeV\,, &\quad $m_\mu$ &=& 105.6584~MeV\,,  \\
\quad $m_e$ &=& 510.9989~KeV\,, &\quad $G_F$ &=& $1.16637 \cdot 10^{-5}$~GeV$^{-2}$\,.
\end{tabular}
\end{equation} 
The light neutrino inputs are in the normal ordering and  according to the constraints
in \ssect{ssect.neutrino} are chosen randomly in the following ranges 
\bea m_{\nu_1} &\in& [0, 2.98152\times 10^{-11}] \,\gev\,, \crn
 m_{\nu_2} &\in&  [\sqrt{ m_{\nu_1}^2 + 6.82\times 10^{-23}},\sqrt{ m_{\nu_1}^2 + 8.04\times 10^{-23}} ]\, \gev\,,\crn
  m_{\nu_3} &\in&  [\sqrt{ m_{\nu_1}^2 + 2.435\times 10^{-21}},\sqrt{ m_{\nu_1}^2 + 2.598\times 10^{-21}} ]\, \gev\,,\crn
\theta_{12} &\in&  [\sqrt{\arcsin(0.269)},\sqrt{\arcsin(0.343)}]\,,\crn
\theta_{23} &\in&  [\sqrt{\arcsin(0.415)},\sqrt{\arcsin(0.617)}]\,,\crn
\theta_{13} &\in&  [\sqrt{\arcsin(0.02052)},\sqrt{\arcsin(0.02428)}]\,,\crn
\delta_{CP} &\in&  [120,369]\,.
\eea
Following the convention of the SUSY Les Houches Accord (SLHA)
format~\cite{Skands:2003cj}, the soft SUSY breaking masses and
trilinear couplings are understood as $\DRb$ parameters at 
the scale
\begin{eqnarray} 
\mu_R = M_{\text{SUSY}}= \sqrt{m_{\tilde{Q}_3} m_{\tilde{t}_R}} \;.  \label{eq:renscale}
\end{eqnarray}
This is also the renormalization scale that we use in all of our computations
of the higher-order corrections. In the Higgs sector we use per default the mixed
$\DRbar$-OS scheme specified in \sect{ssec.CT} and the OS charged Higgs boson mass
as input parameters. Furthermore, we choose OS renormalization for the
top/stop sector and include the two-loop corrections of order ${\cal
  O}(\alpha_s\alpha_t +\alpha_t^2)$ which are computed in
\cite{Muhlleitner:2014vsa,Dao:2019qaz} and are implemented in 
\texttt{NMSSMCALC}.  We perform the scan in the framework of the
CP-violating NMSSM where 
we chose the phase $\delta_{CP}$ in the neutrino sector as the only
non-vanishing complex phase. All other SUSY parameters are assumed to  
be real and are varied in the ranges specified in \tab{tab:scanranges}.
\begin{table}[h]
\centering
\begin{minipage}[t]{0.5\textwidth}
\centering
    \begin{tabular}[t]{ll}
        parameter              & scan range                 \\ \hline
$M_{H^\pm}$            & [0.6, 1]  TeV             \\
$M_1,M_2$              & [0.5, 1]  TeV                   \\
$\mueff$               & [0.2, 1]  TeV             \\
$m_{\tilde{Q}_3}, m_{\tilde{t}_R}$      & [1, 3] TeV               \\
$m_{\tilde{L}_3}, m_{\tilde{\tau}_R}$      & [1, 3] TeV               \\
$A_t$                  & [-4, 4]  TeV                 \\ 
$\Re A_\kappa$              & [-2,2] TeV               \\
$\tan\beta$            & [1, 10]                    \\
$\lambda$              & [0.0001, 0.7]                 \\
$\kappa$               & [0.0001, 0.7]          \\
\end{tabular}
\end{minipage}\hfill
\begin{minipage}[t]{0.5\textwidth}
\centering
    \begin{tabular}[t]{ll}
parameter              & scan range                 \\ \hline
$m_{\ti X}$            & [1,3] TeV \\
$m_{\ti N}$            & [1,3] TeV \\
$A_{\nu}$              & [-2, 2] TeV \\ 
$A_{X}$              & [-2, 2] TeV \\ 
$\mu_X$              & [1, 100] TeV \\ 
$B_{\mu_X}$              & [1, 1000] GeV \\ 
$\lambda_X$             & [$10^{-12},10^{-8}$]\\
$\theta_{1,2,3}$           & [0,2$\pi$]
\end{tabular}
\end{minipage}
\caption{Scan ranges for the random scan over the NMSSM parameter
  space. }
\label{tab:scanranges}
\end{table}
The remaining parameters are fixed as follows
\bea M_3= 1850\,\gev, \quad m_{\tilde{Q}_{1/2}}=m_{\tilde{L}_{1/2}}=m_{\tilde{x}_R}= 3\,\tev, A_{b,\tau}= 2\,\tev\eea
where $x=u,d,c,s,b,e,\mu$.
To ensure perturbativity below the GUT scale we omit points with
 \be 
\kappa^2 +\lambda^2 >(0.7)^2, 
\ee
and/or any element of the neutrino Yukawa matrix $y_\nu$ being larger
than $\sqrt{4\pi}$. Note that in our numerical
  analysis we take the various input parameters of the (s)neutrino sector to be the
  same for all three generations.

\subsection{Impact of the (S)Neutrinos on the Loop-corrected Higgs
  Boson Masses}
For the investigation of the impact of the (s)neutrino contributions
on the loop-corrected Higgs boson masses we choose a parameter point
from our generated scan sample satisfying all the described
constraints. We subsequently vary individual parameters of the
neutrino and/or sneutrino sectors to analyze their impact on the
loop-corrected Higgs boson masses. Our chosen parameter point is
called {\tt P1}. The light neutrino input parameters are set equal to
their best-fit values together with a fixed value for the lightest
neutrino mass, in particular,  
\begin{table}[h]
\centering
\begin{minipage}[t]{0.5\textwidth}
\centering
    \begin{tabular}[t]{ll}
 $m_{\nu_1}$ &= $10^{-11} \,\gev\,,$ \\
 $m_{\nu_2} $&=$  \sqrt{ m_{\nu_1}^2 + 7.37\times 10^{-23}}\, \gev\,,$\\
$ m_{\nu_3}$ &=$ \sqrt{ m_{\nu_1}^2 + 2.525\times 10^{-21}}\, \gev\,,$\\
\end{tabular}
\end{minipage}\hfill
\begin{minipage}[t]{0.5\textwidth}
\centering
    \begin{tabular}[t]{ll}
$\theta_{12} $&=$  \sqrt{\arcsin(0.297)}\,,$\\
$\theta_{23} $&=$  \sqrt{\arcsin(0.425)}\,,$\\
$\theta_{13} $&=$  \sqrt{\arcsin(0.0215)}\,,$\\
$\delta_{CP} $&=$  248.4^\circ\,.$\\
\end{tabular}
\end{minipage}
\label{bestfitp}
\end{table}

All other complex phases are set to zero and
the remaining input parameters are given by 
\begin{align}
M_{H^\pm}            &= 850   \, \gev \,,   & m_{\ti X}            & =1\, \tev \,,       \crn
M_1             & =660  \, \gev  \,,      &m_{\ti N}           & =1 \, \tev  \,,       \crn    
M_2              & =580  \, \gev   \,,   &A_{\nu}              & =1.2 \, \tev \,,       \crn             
\mueff              & =208 \, \gev    \,, &A_{X}              & =1 \, \tev  \,,       \crn      
m_{\tilde{Q}_3} & = 1300 \, \gev \,, & \mu_X              & =40 \, \tev \,,       \\
 m_{\tilde{t}_R}     & =1100 \, \gev \,,  & B_{\mu_X}              & =1 \, \tev  \,,       \crn            
m_{\tilde{\tau}_R}      &=1900 \, \gev \,, & \lambda_X             & =6.5 \times 10^{-10}\,,       \crn
A_t                  & =-1500  \, \gev \,,  &    \theta_{1}           & =2  \,,       \crn        
\mbox{Re} A_k             & =-791 \, \gev   \,,   &\theta_{2}
                                                              &
                                                                =3 \,,       \crn        
\tan\beta           & =4.4  \,,        &     \theta_{3}           & =4
                                                                    \,,       \crn     
\lambda             &= 0.30   \,,      &             
\kappa              &=0.30  \,.        &  \nonumber
\end{align}
In \tab{tab:massP1OS}, we present the Higgs mass spectrum  with and
without inverse seesaw mechanism at tree-level, one-loop, two-loop
$\order{\alpha_t\alpha_s}$ and two-loop $\order{\alpha_t \alpha_s +
  \alpha_t^2}$. The  main components of the Higgs mass eigenstates are
also shown in the last row. We have chosen the OS condition for
the top/stop sector. In order to quantify the  impact of the
(s)neutrino contributions on the Higgs boson masses we define the
relative correction $\Delta_i$ as 
\be 
\Delta_i= \left|\fr{M_{i} -M_{i}^{\text{no}} }{M_{i}^{\text{no}}}
\right|\,\label{eq:correction}
\ee
where $M_i$ is the loop-corrected  mass of the Higgs boson $i$
computed in the NMSSM with ISS and  $M_{i}^{\text{no}}$ the one in the
NMSSM without ISS. Note
that the NMSSM without ISS mechanism contains three massless neutrinos
and three complex sneutrinos which do not mix with each other. 
The massless neutrinos do not interact with the Higgs bosons while the sneutrinos
couple to the Higgs bosons through the $D$-terms with couplings
proportional to $g_1$ and $g_2$. For this parameter point the second
lightest Higgs boson is the $h_u$-like one and hence behaves SM-like. The
$h_s$-like Higgs boson is the lightest one with a mass of 
$90\,\gev$. The $h_d$- and $a$-like states have masses of about
$850\,\gev$ while the $a_s$-like Higgs boson mass is about $700\,\gev$. Although
we have a non-zero complex phase in the $U_{\text{PMNS}}$ matrix, the 
mixing between CP-even and CP-odd states is negligible. 
The $h_u$-like Higgs boson mass is affected the most by 
the inclusion of the ISS mechanism which raises $M_{h_u}$ by about $7
\,\gev$, $5\,\gev$ and $5.6\,\gev$ at one-loop level,
$\order{\alpha_t \alpha_s }$ and  $\order{\alpha_t^2}$,
respectively. If we quantify this change by using the relative
correction defined in \eqref{eq:correction} we see that $\Delta_{h_u}$
is about $5.3\%$ at one-loop level, then
decreases to $4.2\%$ at $\order{\alpha_t\alpha_s}$ and reaches $4.7\%$ 
at $\order{\alpha_t \alpha_s +  \alpha_t^2}$. We remind the reader
that we have used the iterative method to evaluate the loop-corrected
Higgs boson masses. This means that we have mixed orders of perturbation
theory. As common in supersymmetric theories, loop contributions
from particles and their superpartners are opposite in sign. This is
the case for the neutrinos and sneutrinos here as well. While the
sneutrinos give positive contributions to the mass $M_{h_u}$, those of
the neutrinos are negative. Soft-SUSY breaking terms
together with electroweak symmetry breaking prohibit the cancellation 
between the two contributions. We will elaborate this
further in the following by varying parameters related to the change of these two
contributions. We finish our comments on \tab{tab:massP1OS} by
remarking that the other Higgs boson masses are only slightly changed
by the ISS for this particular point. \s

\begin{table}[h]
\begin{center}
\begin{tabular}{|l|l|c|c|c|c|c|}
\hline
                              \multicolumn{2}{|l|}{}         & ${h_1}$   & ${h_2}$ & ${h_3}$ & ${h_4}$ & ${h_5}$ \\ \hline \hline
\multicolumn{2}{|l|}{tree-level}                                          & 84.96 & 85.38  & 705.10 & 847.86  & 850.00 \\ \hline

\multirow{2}{*}{one-loop}&without ISS                 &90.31 & 129.77 & 700.63 & 847.76 & 849.95  \\
                         &with ISS                      &89.43 & 136.62 & 701.07 & 848.06 & 850.74 \\ \hline

\multirow{2}{*}{two-loop  ${\cal O}(\alpha_t \alpha_s)$}&  without ISS           &90.21 & 114.69 & 700.65 & 847.77 & 849.88   \\
                         & with ISS                      &89.17 & 119.52 & 701.08 & 848.06 & 850.62 \\ \hline
\multirow{2}{*}{two-loop ${\cal O}(\alpha_t \alpha_s+ \alpha_t^2)$} & without ISS &90.24 & 120.72 & 700.65 & 847.77 & 849.91  \\
             & with ISS                      &89.28 & 126.37 & 701.08 & 848.06 & 850.67   \\ \hline
\multicolumn{2}{|l|}{main component}                                    & $h_s$     & $h_u$   & $a_s$   & $a$   & $h_d$   \\ \hline
\end{tabular}
\caption{Parameter point {\tt P1}:  Mass values in GeV and main
  components of the neutral Higgs 
  bosons at tree-level, one-loop, two-loop $\order{\alpha_t\alpha_s}$ and
  two-loop $\order{\alpha_t \alpha_s +  \alpha_t^2}$ obtained for the 
  NMSSM without or with the inverse seesaw mechanism together with $\OS$
  conditions in the top/stop sector. }
\label{tab:massP1OS}
\end{center}
\end{table}
%
\begin{figure}[h]
    \centering
    \subfloat[]{
    \begin{tabular}[b]{r}
        \includegraphics[width=0.495\textwidth]{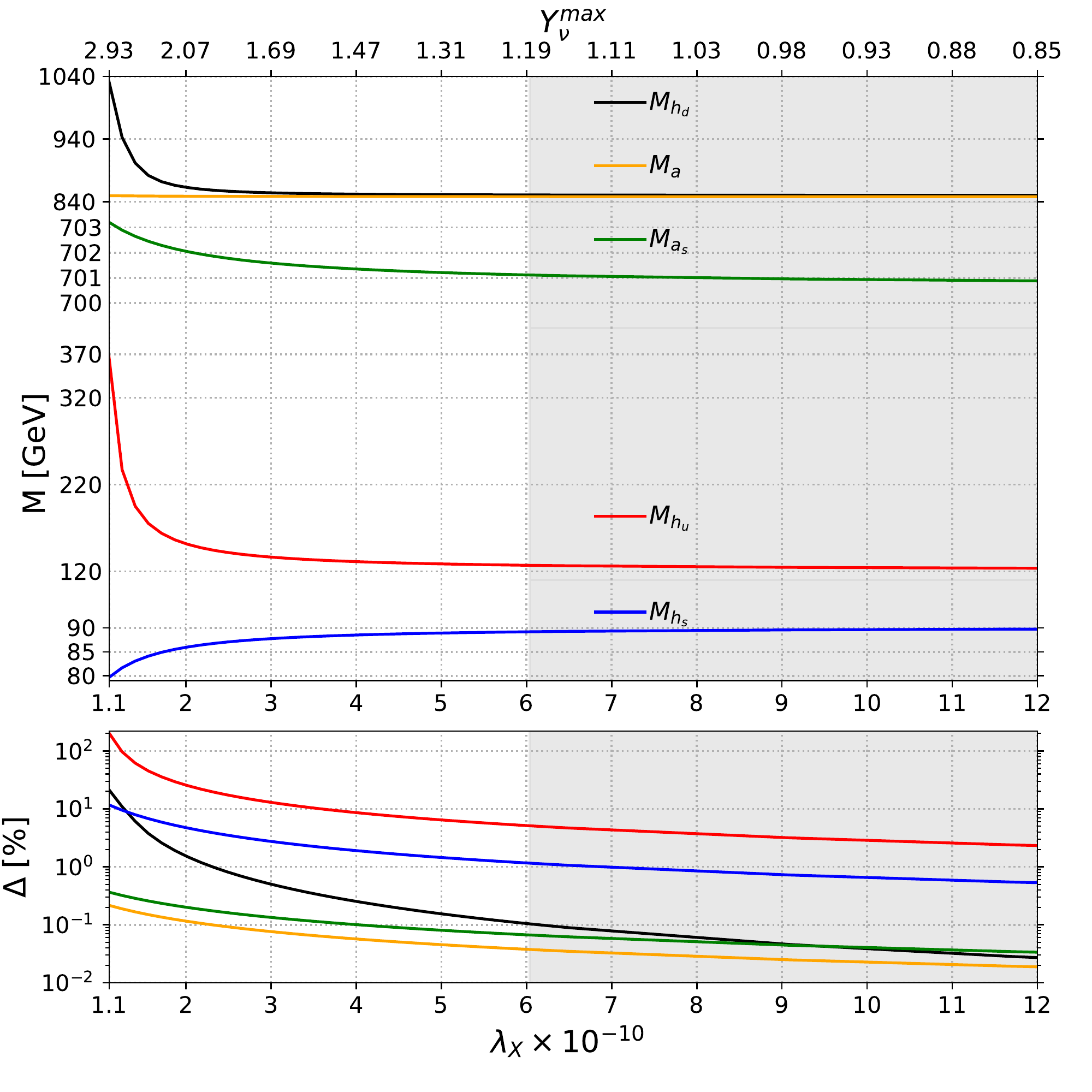}
    \end{tabular}
    }
    \subfloat[]{
    \begin{tabular}[b]{r}
        \includegraphics[width=0.495\textwidth]{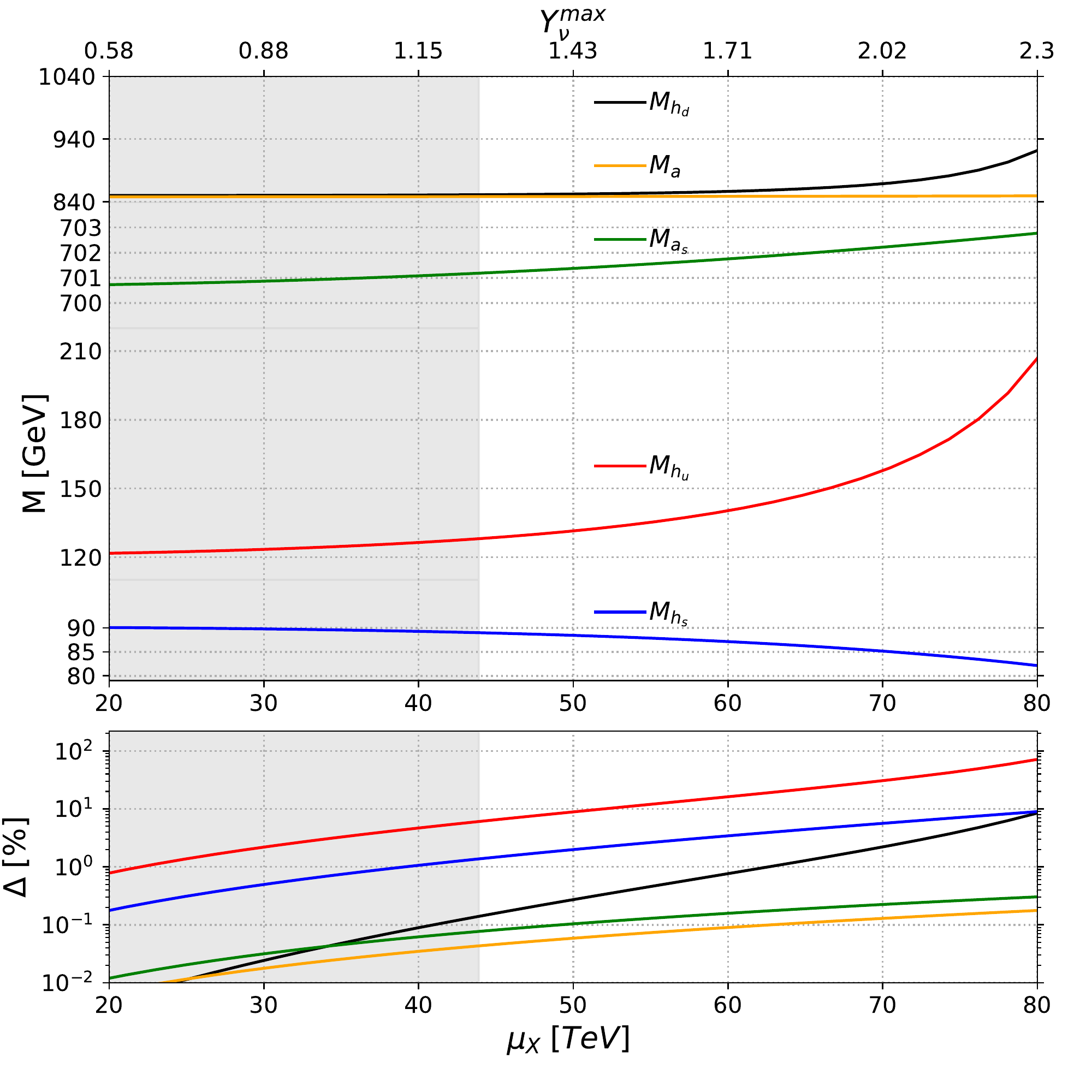}
    \end{tabular}
    }
    \caption{Parameter point {\tt P1}: The loop-corrected Higgs boson masses in GeV at order $\order{\alpha_t \alpha_s +  \alpha_t^2}$ as function of the 
parameter $\lambda_X$ (upper left) and of $\mu_X$ (upper right) The
relative corrections $\Delta$ defined in
\eqref{eq:correction} are shown in the lower panels. The five Higgs
boson masses are denoted by 
their main components ($h_u,h_d,h_s,a,a_s$) corresponding to the five
colors (red, black, blue, yellow, green). The gray area on the
plots denotes the points satisfying our
constraints. $Y_{\nu}^{\text{max}}$ denotes the maximum element of the
neutrino Yukawa matrix.  
  } 
    \label{fig:lambdaMUX}
\end{figure}

We continue by investigating the effects of the various neutrino parameters on
the loop-corrected Higgs boson masses. From now on we present only the
masses at two-loop $\order{\alpha_t \alpha_s +  \alpha_t^2}$ which is
the highest precision of our numerical code that includes the
ISS.\footnote{We very recently completed the two-loop contributions at
  $\order{(\alpha_t+\alpha_\lambda+\alpha_\kappa)^2}$ \cite{Dao:2021khm}. They
  are included in {\tt NMSSMCALC} and {\tt NMSSMCALCEW} and will soon
  be included in \texttt{NMSSMCALC-nuSS} as well.}
In \figref{fig:lambdaMUX} we show the dependence of the loop-corrected Higgs boson masses on the coupling $\lambda_X$
in the left panel and on the mass $\mu_X$ in the right panel and the dependence of the correction $\Delta$ defined in \eqref{eq:correction} in the lower panels. We remind the reader that both $\lambda_X$ and $\mu_X$ appear in the neutrino mixing matrix as shown in \eqref{eq:neumassmatrix} and \eqref{eq:lambdaxinmasses}. 
 In the Casas-Ibarra parameterization which is used in our
 computation, both $\lambda_X$ and $\mu_X$ enter the evaluation of the
 neutrino Dirac mass matrix $M_D$, see \eqref{eq:MDdefinition},
 therefore directly affect the neutrino Yukawa matrix $Y_\nu=\sqrt{2}
 M_D/(v_u e^{i\varphi_u})$ that can be written as 
\be 
Y_\nu \propto \fr{m_\nu\sqrt{\mu_X}}{v_u\sqrt{\lambda_X v_s}}. 
\ee
We denote the maximum element of the neutrino Yukawa matrix by
$Y_{\nu}^{\text{max}}$. We chose the range of variation for
$\lambda_X$ and $\mu_X$ such that $Y_{\nu}^{\text{max}}$ is smaller
than the perturbativity limit $\sqrt{4\pi}$ applied in our analysis. On
the $x$-axis on top of each plot  in \figref{fig:lambdaMUX}, we see
the variation of $Y_{\nu}^{\text{max}}$ corresponding to the range of
variation for $\lambda_X$ and $\mu_X$. For our chosen parameter point
{\tt P1}, the value of $Y_{\nu}^{\text{max}}$ is equal to 1.15.  As
can be inferred from the plots, the impact from the (s)neutrino sector
is less than 2.3\% on $M_{h_u}$ if $Y_{\nu}^{\text{max}}$ is smaller
than 0.85 (0.89) corresponding to $\lambda_X> 1.2\times 10^{-9}$
($\mu_X<31\, \tev$) in the left (right) plot. With increasing value of
$Y_{\nu}^{\text{max}}$ ($\lambda_X$ becomes smaller or $\mu_X$ gets
larger) the effect increases significantly. The relative correction
$\Delta_{h_u}$ can even go up to 100\% for $\lambda_X=1.24\times
10^{-10}$ ($Y_{\nu}^{\text{max}}=2.63$) or $\mu_X=83\,\tev$
($Y_{\nu}^{\text{max}}=2.39$). This makes us question the
perturbativity limit of $\sqrt{4\pi}$ applied on $Y_{\nu}^{\text{max}}$
at the SUSY scale. A more stringent constraint that demands
$Y_{\nu}^{\text{max}}<\sqrt{4\pi}$ up to the Planck scale may imply a
much smaller value for $Y_{\nu}^{\text{max}}$ at the SUSY scale. A
recent study in \cite{Mandal:2020lhl} found  that
$Y_{\nu}^{\text{max}}<0.8$ at the TeV scale for the SM with inverse
seesaw mechanism. We expect a similar value for the model in our study.  
The large value of $\mu_X$ results in large values for the sterile
neutrinos and additional sneutrinos. One then has to
worry about the validity of 
the fixed-order calculation applied in the computation of the Higgs
mass corrections. We can choose, however, a much smaller value for
$\mu_X$ and $\lambda_X$. As long as $Y_{\nu}^{\text{max}}$ is large we
still get a large value for $\Delta_{h_u}$. With a low mass spectrum
of sterile neutrinos and sneutrinos one can get large branching ratios
for the charged lepton flavor-violating processes
$l_1 \to l_2 +\gamma$ that will be discussed in \ssect{sec:LFV}. \s 

From \figref{fig:lambdaMUX}, we see that not only the $h_u$-like Higgs
boson is strongly affected by  large $Y_{\nu}^{\text{max}}$ but also
the $h_d$- and $h_s$-like states. While both the $h_{u}$- and
$h_d$-like Higgs boson mass get positive corrections, the $h_{s}$-like
Higgs boson mass receives negative corrections. On the left plot
$\Delta_{h_{d}}$ can go up to $21\%$ at $Y_{\nu}^{\text{max}}=2.93$
while $\Delta_{h_{s}}$ goes to 12\%. Note that neutrinos interact with
$h_u$  interaction states through the interaction term $ h_u
\bar{\nu}_i(Y_\nu P_L+Y_\nu^\dagger P_R )\nu_j $ and  with $h_s$
through the interaction term $ h_s\bar{\nu}_i(\lambda_X P_L +
\lambda_X^\dagger P_R)\nu_j $ where $\lambda_X$ is a very small
number. Neutrinos do not interact with $h_d$, but sneutrino do
interact through $D$-terms. For this particular parameter point, the
dominantly $h_d$- and $h_s$-like Higgs mass eigenstates have a
significant admixture of the $h_u$ component. The impact of the
neutrinos on the other Higgs bosons depends on their mixtures with
$h_u$.  The states $a$ and $a_s$ are less affected. 

\begin{figure}[h]
    \centering
    \subfloat[]{
    \begin{tabular}[b]{r}
        \includegraphics[width=0.49\textwidth]{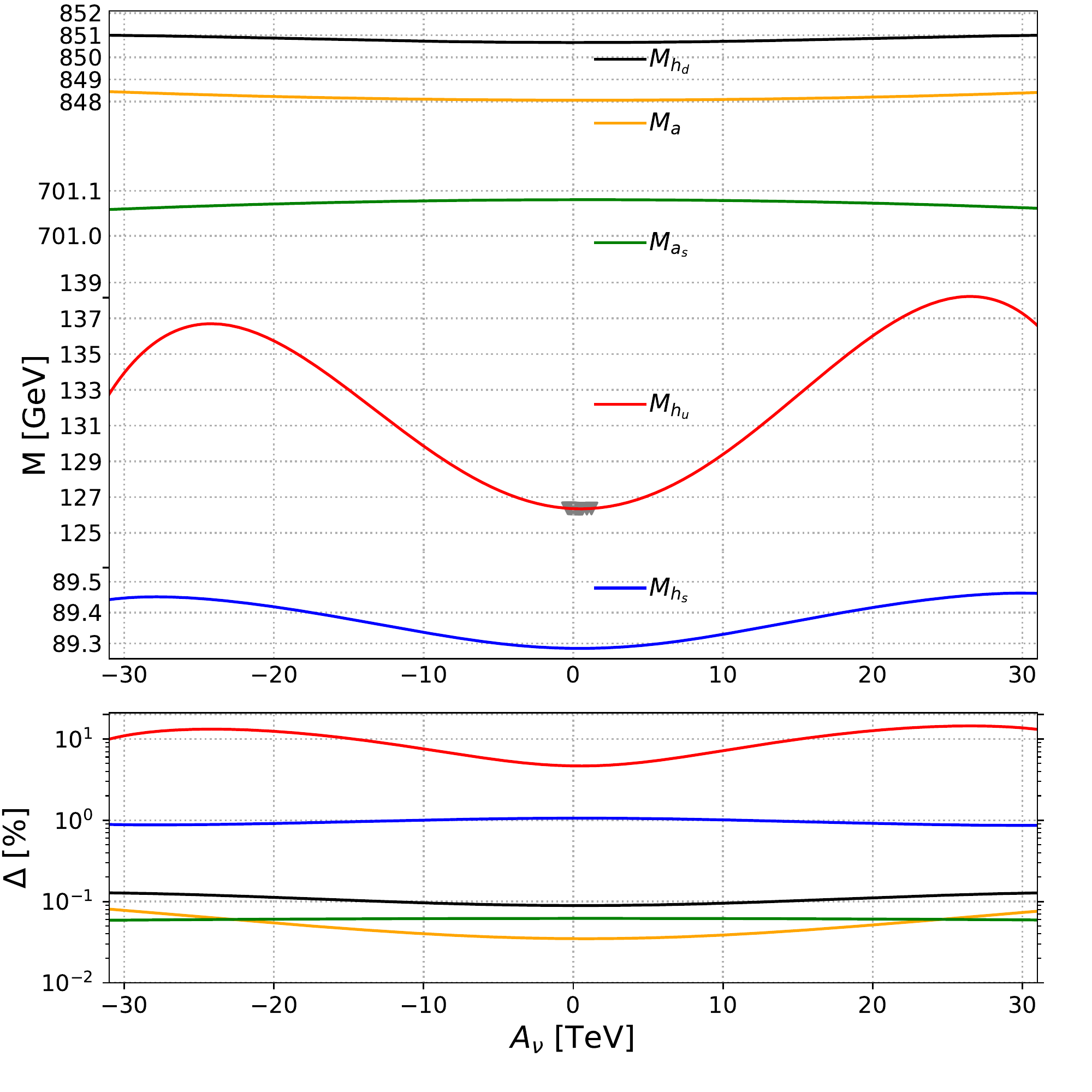}
    \end{tabular}
    }
    \subfloat[]{
    \begin{tabular}[b]{r}
        \includegraphics[width=0.49\textwidth]{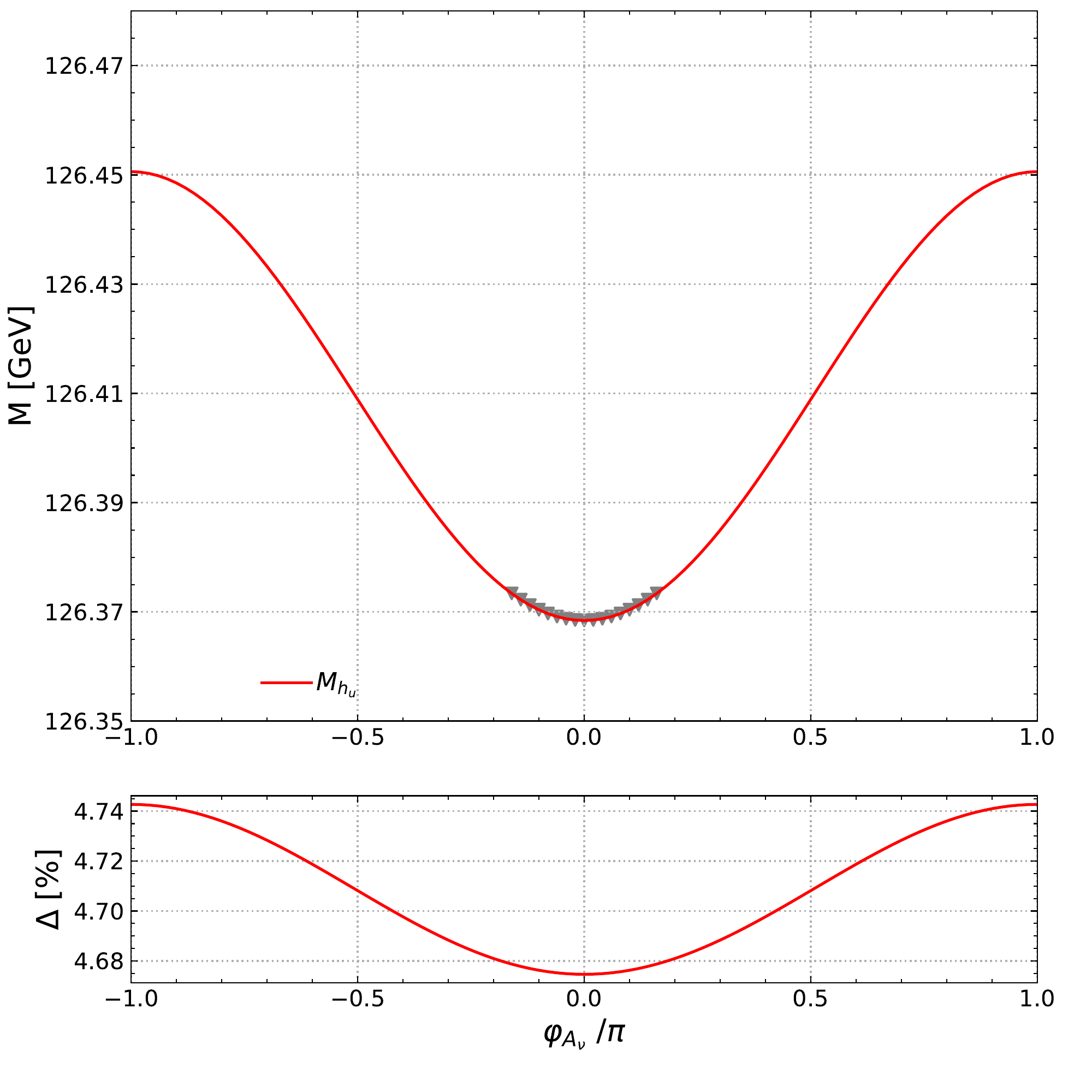}
    \end{tabular}
    }
    \caption{The same as \figref{fig:lambdaMUX} but now we show the
      dependence on: (a) the neutrino trilinear
      coupling parameter $A_\nu$ and (b) the complex phase of $A_\nu$. 
  The gray triangles denote the points satisfying our
constraints.} 
    \label{fig:Av}
\end{figure}

We now move on to the discussion of the dependence of the
loop-corrected Higgs boson masses on the neutrino trilinear
coupling parameter $A_\nu$ that affects only the
sneutrino sector and leaves the neutrino sector unchanged. In
\figref{fig:Av} we vary the value of $A_\nu$ in the left plots and the
complex phase of $A_\nu$ in the right plots. The color code and
notation of the left plots are the same as in
  \figref{fig:lambdaMUX}. In the right plot, we show only the
loop-corrected mass of the $h_u$-like Higgs boson, since the impact of
the complex phase on the other Higgs boson masses is negligible.
As can be inferred from the plots, the loop-corrected $h_u$-like Higgs
boson mass is strongly affected by $A_\nu$. This dependence looks like
the dependence on the top trilinear coupling $A_t$
and its complex phase, see for example
\cite{Muhlleitner:2014vsa,Dao:2019qaz}, but the
  relative size of the corrections can cover a larger range if  
$Y_{\nu}^{\text{max}}$ is large. 
The correction $\Delta_{h_u}$ is about 4.7\% at $A_\nu= 0$ and maximal
(14.5\%) at $A_\nu\approx \pm 26\,\tev$. The $h_s$-like Higgs boson
mass depends slightly on $A_\nu$ while the other Higgs boson masses
are barely affected by the variation of $A_\nu$. 
We remind the reader that the maximum value of $Y_\nu$ that is
obtained during the variation of $A_\nu$ is given by $Y_{\nu}^{\text{max}}= 1.15$.
In our analysis we also reduced $Y_{\nu}^{\text{max}}$
to 0.8 by setting $\lambda_X=1.33\times 10^{-9}$ and varying
$A_\nu$. We then obtained the variation of $\Delta_{h_u}$ in the range [2, 3.6]\%. 
In the right panel of \figref{fig:Av}, we
observe a change of $0.08\%$ for $\Delta_{h_u}$ when the complex phase
$\varphi_{A_\nu}$ is varied in the range $[-\pi,\pi]$. Other complex
phases of the neutrino sector like $\delta_{CP}$, of the phases of
$\lambda_X,$ and $\mu_X$ have an insignificant impact on the
loop-corrected Higgs masses. \s

Note that in \figref{fig:lambdaMUX} and \figref{fig:Av} we present
parameter points that satisfy all constraints by the gray area,
respectively, the gray triangles. The other points violate charged
flavor-violating lepton decays, the $S,$ $T,$ $U$
parameters and/or the Higgs data.  \s

\begin{figure}[h]
    \centering
        \includegraphics[width=0.49\textwidth]{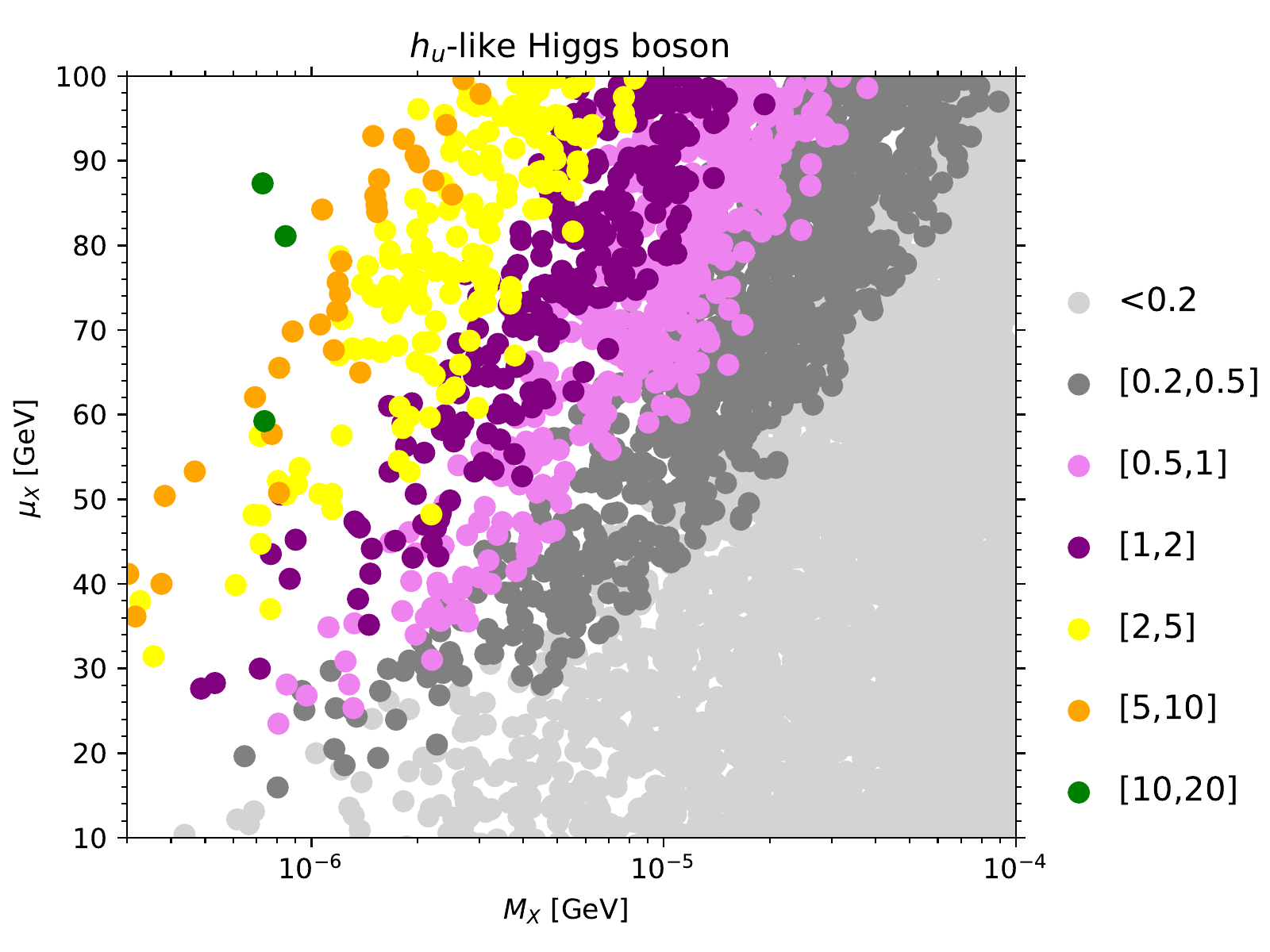}
        \includegraphics[width=0.49\textwidth]{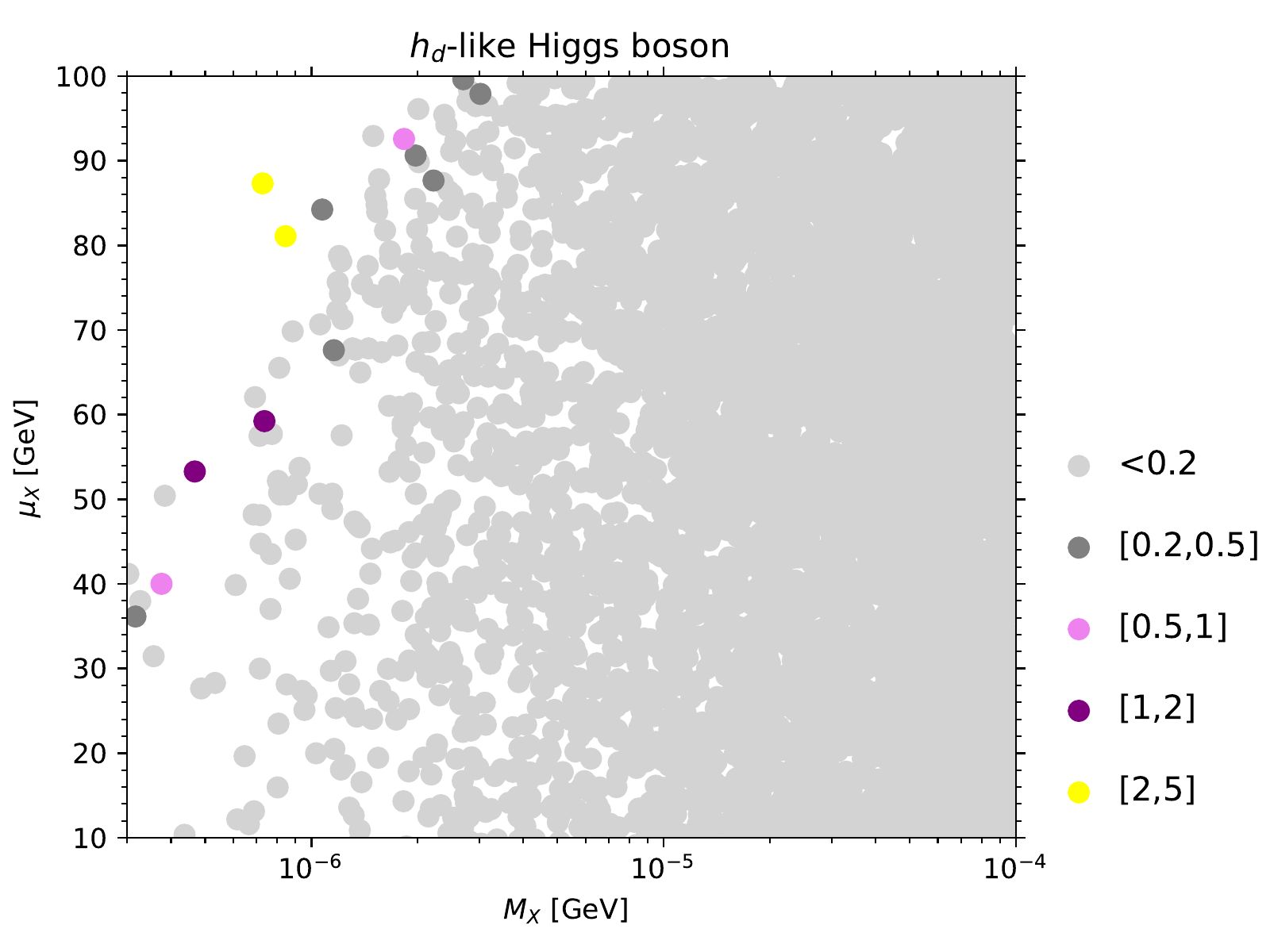}
    \caption{Scatter plots in the plain of two variables of the
      neutrino sector, $(M_X,\mu_X)$ for the $h_u$-like (left) and the
      $h_d$-like (right) Higgs boson. The color code indicates the
      relative correction $\Delta$ (defined in \eqref{eq:correction})
      for each Higgs state in percent.  
  } 
    \label{fig:scatter1}
\end{figure}

In the remainder of this section, we present scatter plots in
\figref{fig:scatter1} which we obtained from
our scan keeping only parameter points that satisfy all our mentioned
constraints. The points are depicted in two-dimensional planes with
$M_X$ on the $x$-axis and $\mu_X$ on the $y$-axis. Note that $M_X$ is
related to $\lambda_X$ and $v_s$ as given in
\eqref{eq:lambdaxinmasses}. Since $\lambda_X$ and $v_s$ are both
varied in the scan, $M_X$ may be a more appropriate parameter
than $\lambda_X$ for the scatter plots. 
The color code in the plot quantifies the size of the relative
corrections $\Delta$ for the respective Higgs boson in the individual plots. 
The light gray points denote $\Delta \le 0.2\%$, gray $0.2\%<\Delta \le
0.5\%$, violet $0.5\%<\Delta \le 1\%$, purple $1\%<\Delta \le
2\%$, yellow $2\%<\Delta \le 5\%$, orange $5\%<\Delta \le 10\%$ and
green $10\%<\Delta \le 20\%$. The left plot of
the \figref{fig:scatter1} presents the relative corrections for the
$h_u$-like Higgs boson, while the right plot  for the $h_d$-like state.
The $\Delta$ for the other Higgs bosons are less significant and
therefore we do not present them here. \s
 
Most of the points obtained in our scan have small relative
corrections $\Delta \le 0.2\%$. Larger relative corrections are
realized for $\mu_X$-$M_X$ towards the top-left corner of each plot,
corresponding to increasing values of $Y_{\nu}^{\text{max}}$. 
In the  top-left corner, there are no points because they either
violate the perturbativity constraint of
$Y_{\nu}^{\text{max}}<\sqrt{4\pi}$ or they lead to
  unstable numerical results due to large corrections. 
The color pattern is rather clear for the $h_u$-like Higgs boson,
which confirms our conclusion on the strong dependence of
$\Delta_{h_u}$ on  $Y_{\nu}^{\text{max}}$. There are some outliers
which do not lie in their color bands, since  $\Delta_{h_u}$ depends
not only on   $Y_{\nu}^{\text{max}}$ but also on the sneutrino soft SUSY breaking 
parameters. 

\subsection{Impact of the (S)Neutrinos on the LFV decays}
\label{sec:LFV}
In this section, we investigate the impact of the neutrinos and
sneutrinos in the NMSSM with ISS on the radiative $l_1\to l_2\gamma$
decays. As explained in \ssect{sec:LFVdecays}, we consider three decay
processes, namely $\mu\to e\gamma$, $\tau\to e\gamma$, and $\tau\to
\mu\gamma$ among which the most stringent constraint
exists for the decay $\mu\to e\gamma$. For all
processes, the dominant contributions arise from the right-handed form
factors $F_R$. The ratio between the contributions
from the right- and left-handed form factors is approximately
proportional to the ratio $m_{l_1}^2/m_{l_2}^2$ as can be inferred
from the analytic expressions in \appen{appen:formfactor}.
One can therefore safely neglect the contribution from the left-handed form
factors. We have divided the contributions to the form factors into three
parts given by $F_{L,R}^{W^\pm}$, $F_{L,R}^{H^\pm}$,  and
$F_{L,R}^{\ti\chi^\pm}$. \s

\begin{figure}[h]
    \centering
        \includegraphics[width=0.49\textwidth]{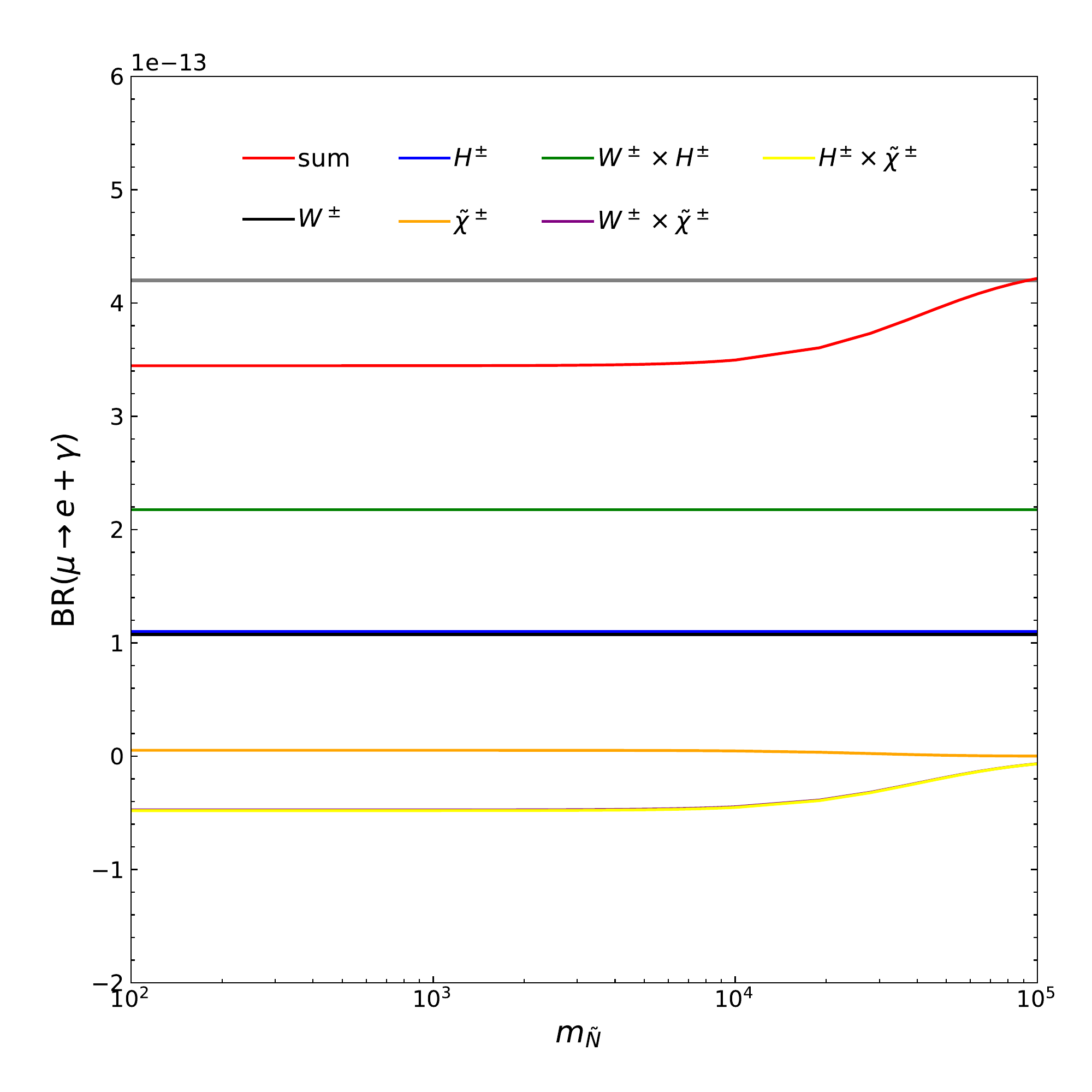}
        \includegraphics[width=0.49\textwidth]{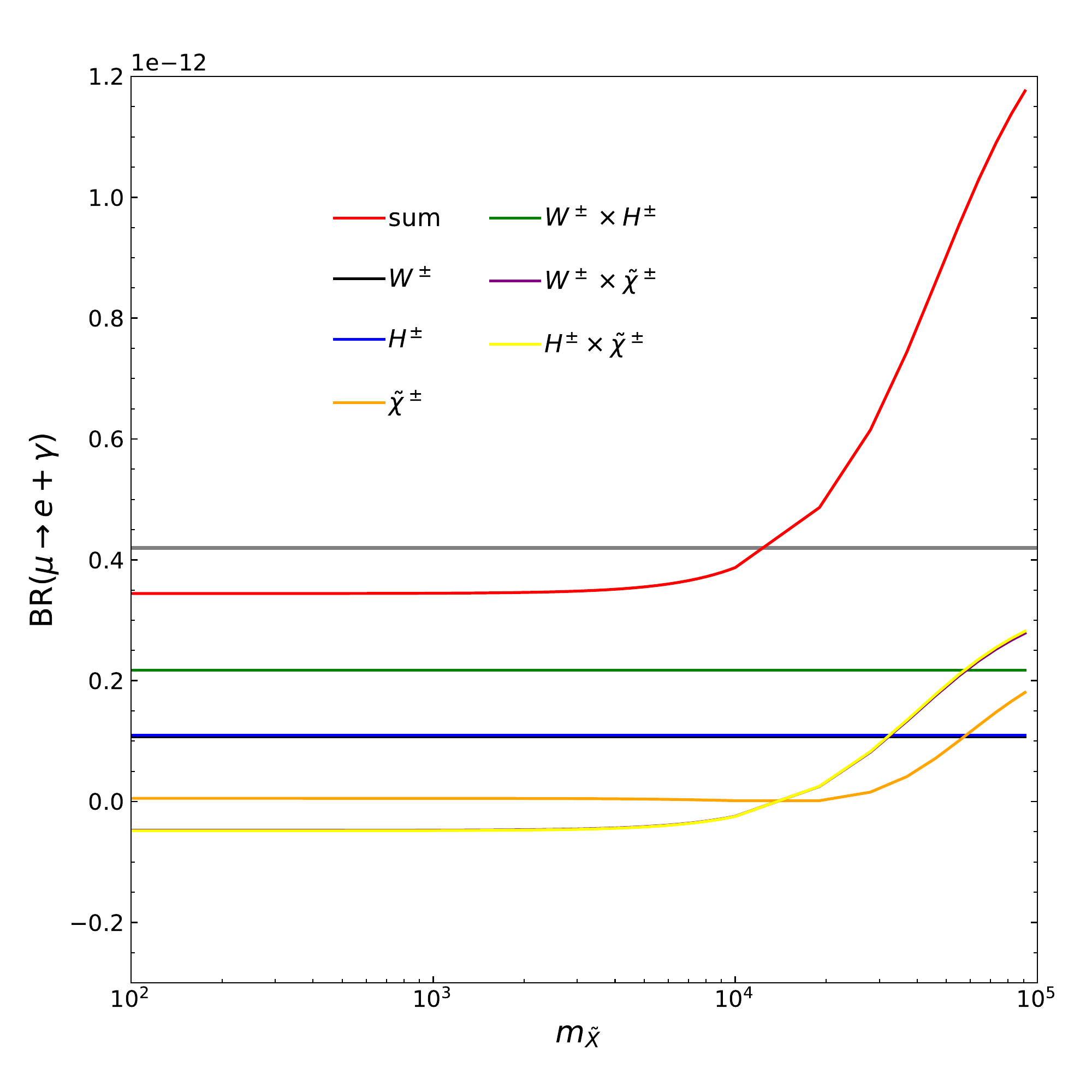}
    \caption{ Same as the left plot of \figref{fig:LFV3}, but $m_{\ti
        N}$ (left) and $m_{\ti X}$ (right) are varied instead.} 
    \label{fig:LFV3}
\end{figure}

In our scan, the constraint on $\mu\to e\gamma$ is very important and
rules out many points. The branching ratios are very sensitive to 
the spectrum of the neutrino and sneutrino sectors. In particular, they
increase when the mixings between sterile neutrinos and active
neutrinos increase. Both $\mu_X$ and  $\lambda_X$ have strong impacts
on these mixings. 
We investigate this dependence by taking the parameter point {\tt P1}
of the previous section and varying again the parameters  $\mu_X$ and $\lambda_X$. 
\begin{figure}[h]
    \centering
        \includegraphics[width=0.49\textwidth]{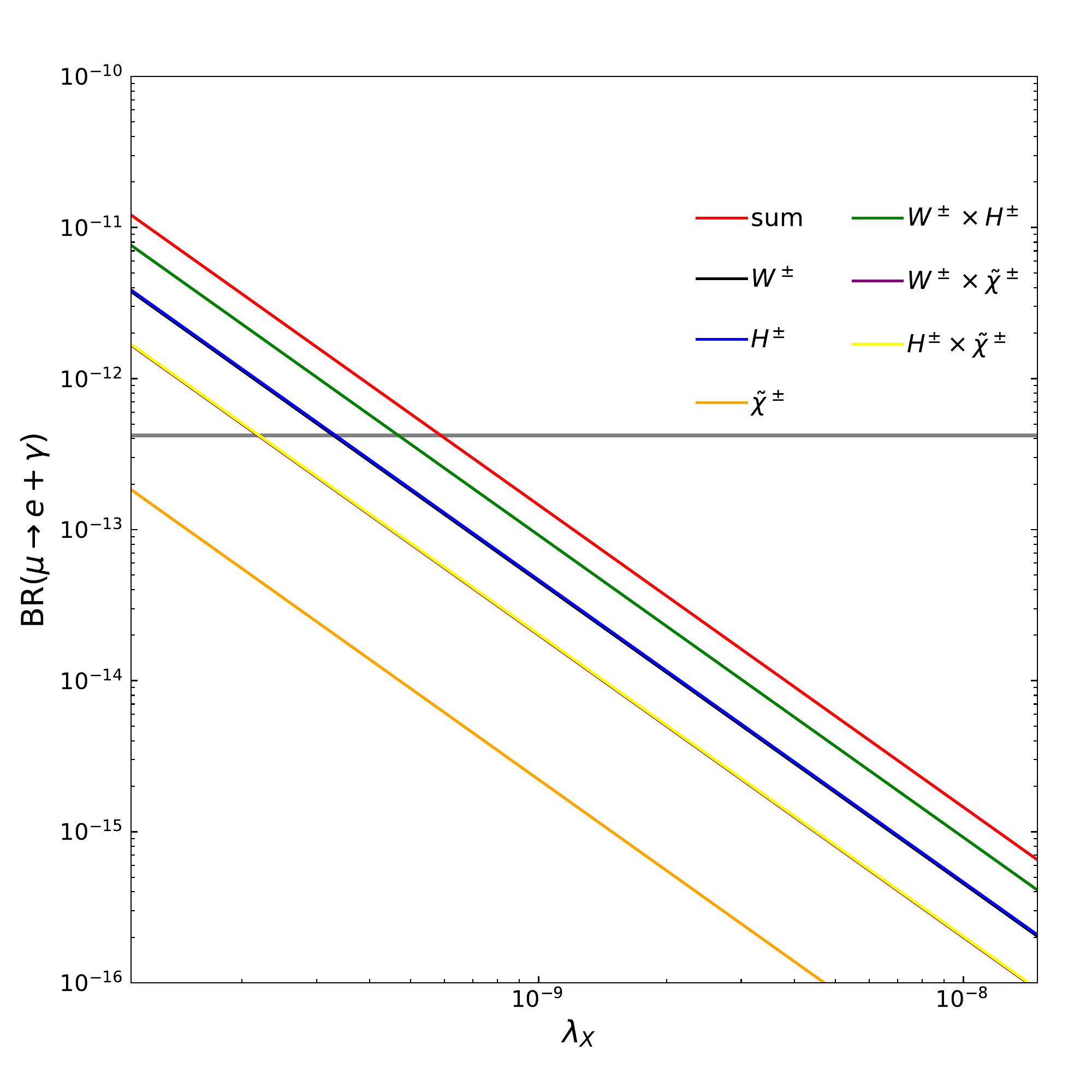}
        \includegraphics[width=0.49\textwidth]{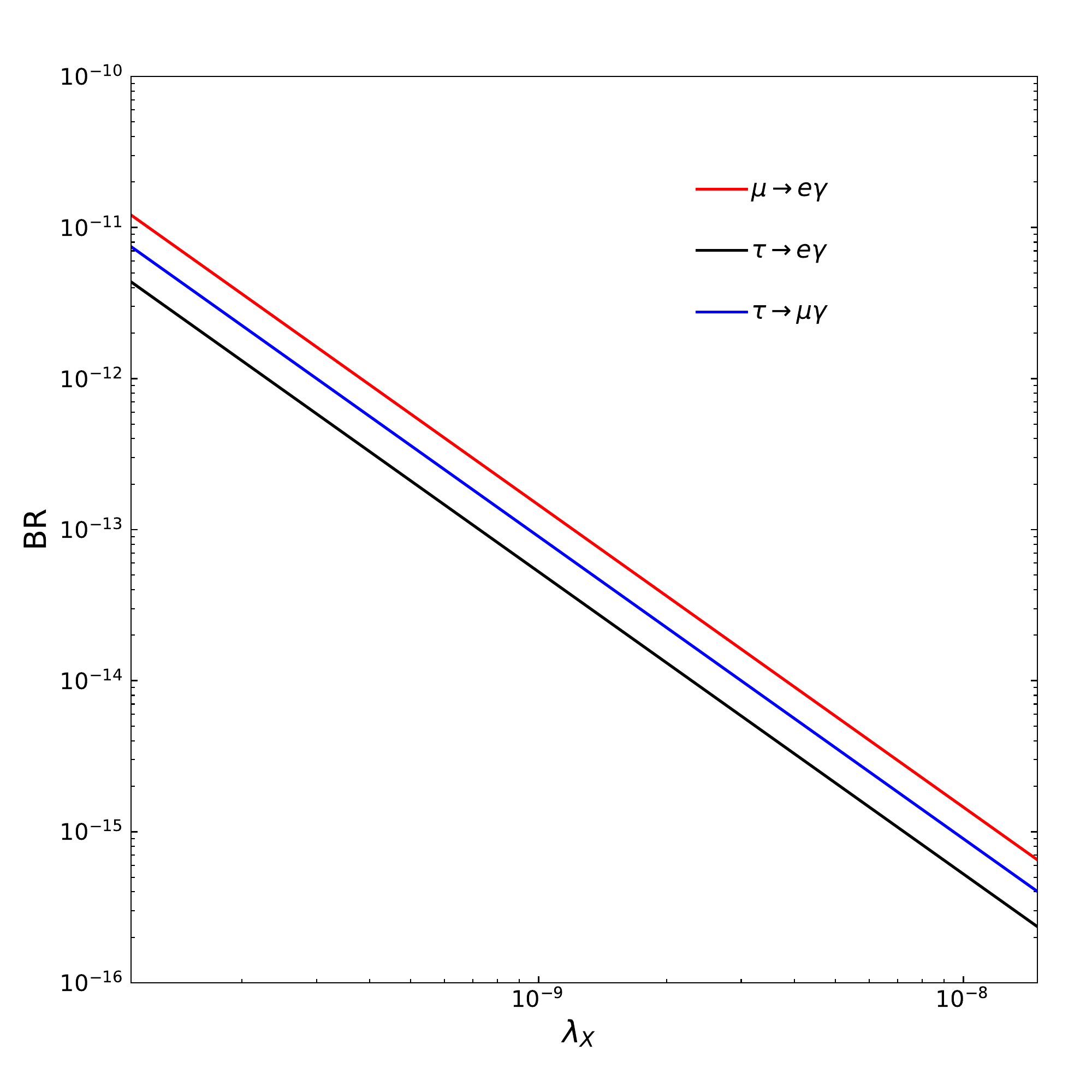}
    \caption{Left: Branching ratio and its individual contributions
      for the decay $\mu\to e\gamma$ as a function of $\lambda_X$. 
      The interference contributions
      $W^\pm \times \ti\chi^\pm$ between the chargino and the $W^\pm$ form
      factors (violet), and $H^\pm \times \ti\chi^\pm$ between the chargino
      and charged Higgs form factors (yellow) are negative. In order
      to present them in one plot we have changed the sign. The black
      line for $W^\pm$ is hidden under the blue line and the purple line is under
      the yellow one. The horizontal black line shows the experimental
      upper limit on the branching ratio. Right:
      Branching ratios for the three decays $\mu\to e\gamma$ (red),
      $\tau\to e\gamma$ (black), and $\tau\to \mu\gamma$ (blue) as
      function of $\lambda_X$.} 
    \label{fig:LFV1}
\end{figure}
In the left plot of \figref{fig:LFV1}, we show the dependence of
the branching ratio (red line) for the decay $\mu\to e\gamma$ on
$\lambda_X$. We also depict the individual contributions from the squared form factors
$(F_{L,R}^{x})^2$ with $x=W^\pm$ (black), $H^\pm$ (blue), and
$\ti\chi^\pm$ (orange) lines, as well as the interference terms
$2(F_{L,R}^{x})(F_{L,R}^{y})$ with $(x,y)=(W^\pm,H^\pm)$ (green)
$(W^\pm,\ti\chi^\pm)$ (purple), and $(H^\pm,\ti\chi^\pm)$ (yellow).
For this parameter point scenarios, we find that the
contributions from the $W$ boson and from the charged Higgs boson are dominant, and
the interference term between  
the $W$ boson and charged Higgs form factors adds a significant
contribution to the sum. We observed that the form factors
$F_{L,R}^{W^\pm}$ and $F_{L,R}^{H^\pm}$ have the same sign
while they are opposite in sign compared to
$F_{L,R}^{\ti\chi^\pm}$. The chargino contributions hence suppress the total 
decay widths. In the right plot of \figref{fig:LFV1}, 
we present the branching ratios for the three considered decay
processes, $\mu\to e\gamma$, $\tau\to e\gamma$, and $\tau\to
\mu\gamma$. We observe that the branching ratio for the process
$\mu\to e\gamma$ is larger than for the other two processes. 
While the contributions from the $W$ boson and charged Higgs form
factors depend only on the spectrum and mixing of the neutrinos, the
chargino contributions depend on those of the sneutrinos. One can vary
the parameters of the sneutrino sector such as $m_{\ti N}$ and $m_{\ti
  X}$ to change the sign and the magnitude of the chargino
contributions. We show the dependence of the branching ratio (red
line) for the decay $\mu\to e\gamma$ on $m_{\ti N}$ and $m_{\ti X}$ in
the left and right plot of \figref{fig:LFV3}, respectively. \s 

\begin{figure}[h]
    \centering
        \includegraphics[width=0.49\textwidth]{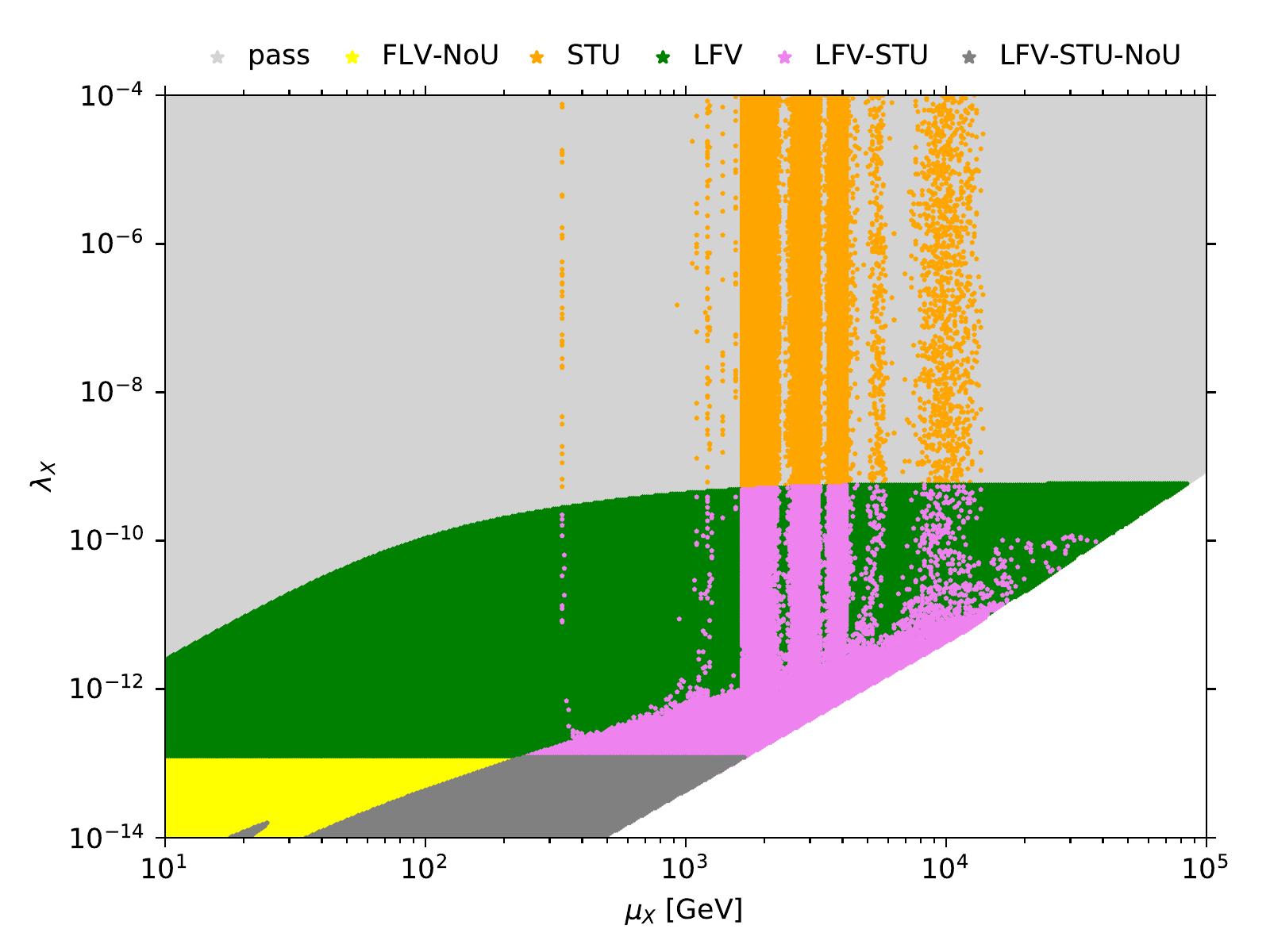}
    \caption{Scatter plot in the $\mu_X$ and $\lambda_X$ plane
      starting from the parameter point {\tt P1}. The
      color code is used to distinguish between points that pass all three
      constraints LFV, STU, NoU (light gray), that do not pass any of
      the three constraints (dark gray) and that violate individual
      constraints or combinations of two constraints: LFV (green), STU (orange), LFV
      and STU (pink), LFV and NoU (yellow). See text for details. 
  } 
    \label{fig:LFV2}
\end{figure}

We now investigate the impact of the parameters in the neutrino sector on
the $S$, $T$, $U$ parameters (STU), on the LFV decays and on the
non-unitary $3\times 3$ neutrino mixing matrix (NoU) discussed in
\ssect{ssect.neutrino}. We started from the parameter point {\tt P1}
and we changed only the following parameters in the corresponding
ranges, 
\beq 
&&\mu_X \in [10,10^5] \gev,\quad \lambda_X\in [10^{-14}, 10^{-4}].
\eeq
All remaining parameters are kept fixed. We show in \figref{fig:LFV2}
a scatter plot in the plane of $\mu_X $ and $\lambda_X$. We do not
consider the {\tt HiggsBounds} and the {\tt HiggsSignals} constraints 
in this particular plot since we want to focus on the three mentioned
constraints. The gray color denotes points which pass all three
constraints while dark gray reflects points that violate all three
constraints. The orange and green colors are for points violating STU
and LFV constraints, respectively. The pink and yellow points,
respectively, violate combinations of two constraints, namely LFV-STU and LFV-NoU.
The white area on the plot is not accessible since we encounter either
negative values for one of the Higgs boson masses or
$Y_{\nu}^{\text{max}}>\sqrt{4\pi}$. As can be inferred from the plot,
small values of $\lambda_X$ ($\lsim 10^{-12}$) are
not preferred, independent  of $\mu_X$, since this parameter region
is very sensitive to the constraints from LFV and NoU. For
$\lambda_X>10^{-12}$, the STU constraint is important in the region
$\mu_X\in [10^3,10^4]\gev$ and for $\mu_X=320$~GeV,
independent of $\lambda_X$. There are also regions such as
$\mu_X<300\,\gev$ or $\mu_X>20\,\tev$ with 
$\lambda_X>8\times 10^{-10}$ where the three mentioned constraints do
not play a role. This plot demonstrates the importance of
taking into account the STU and LFV constraints when performing
numerical analysis.
Although the magnitude of $ \lambda_X $ signifies the
  magnitude of charged lepton flavor violation, {\it i.e.}~the larger
  $ \lambda_X $ is the more violation we would expect, the reverse
  happens here. The reason is as follows. When we fix light neutrino
  masses $ m_\nu $, $ \lambda_X $ becomes inversely proportional to $
  y_\nu $. Smaller $ \lambda_X $, or larger $ y_\nu$, yields a larger
  mixing between sterile and active neutrinos, or a weaker GIM
  mechanism, and thus a larger LFV decay rate. This is also why small
  values of $ \lambda_X $ violate unitarity.

\section{Conclusions}
\label{sec:conclusions}
In this paper, we studied the impact of an extended neutrino sector on
the NMSSM Higgs sector. We considered the framework of both the
CP-conserving and CP-violating NMSSM extended by six singlet leptonic
superfields. Their mixing with the three doublet leptonic superfields
allows for the explanation of the tiny non-zero neutrino masses through the
inverse seesaw mechanism. While $R$-parity is conserved in this model
lepton number is explicitly violated through the interaction between
two singlet neutrino superfields and a singlet Higgs superfield. \s

We quantified the indirect neutrino effects on the NMSSM Higgs sector
by computing the complete one-loop corrections to the Higgs boson
masses at non-vanishing external momentum. For the renormalization, we
applied a mixed OS-$\overline{\mbox{DR}}$ scheme and consistently
combined our one-loop result with the two-loop ${\cal O}(\alpha_t
(\alpha_s + \alpha_t))$ results computed previously by our group. In
the numerical analysis, we performed a parameter scan of the model and
kept only those points for our study that respect the constraints from the Higgs
data, the neutrino oscillation data, the charged lepton
flavor-violating decays $l_i \to l_j + \gamma$, and the new physics
constraints from the oblique parameters $S,T,U$. We presented 
 the explicit calculation of the one-loop decay width for
$l_i \to l_j + \gamma$. Our one-loop results have been implemented in
the Fortran code {\tt NMSSMCALC-nuSS}, that has been made publicly
available and is based on the code {\tt NMSSMCALC} that also computes
the Higgs decays widths and branching ratios. \s

We found for our investigated benchmark point that the impact of the neutrinos on the
one-loop corrections is largest for the $h_u$-like, and hence SM-like,
Higgs boson mass with about 5\%, decreasing slightly when the two-loop
corrections are
included. While the neutrino and sneutrino contributions come with
opposite signs, their complete cancellation is prohibited by the
soft-SUSY breaking terms in combination with electroweak symmetry
breaking. We furthermore showed that a large neutrino Yukawa coupling
has a significant impact on the loop corrections. The same is true for
the soft-SUSY breaking trilinear coupling $A_\nu$ and its complex
phase that affects the sneutrino sector while leaving the neutrino
sector unchanged. Our findings on the dependence on the neutrino
Yukawa coupling are confirmed by the presented scatter plots taking
into account all of the parameter points passing the constraints. \s

Our investigation of the one-loop corrected LFV decay $l_1 \to l_2 \gamma$
shows that the constraints from this decay are relevant and need to be
taken into account. This result is underlined by an analysis of the
impact of all considered constraints that shows the importance of the
$S, T, U$ and LFV constraints on the validity of the parameter
scenarios. \s

In summary, the one-loop analysis of the impact of an extended
(s)neutrino sector on the NMSSM demonstrates the importance of taking
into account these indirect effects through loop
contributions. Together with the usual constraints from Higgs data and
new physics as well as from LFV decays, they constrain the valid
parameter space of the model.

\section*{Acknowledgements}
The research of MM was supported by the Deutsche
Forschungsgemeinschaft (DFG, German Research Foundation) under grant
396021762 - TRR 257. T.N.D is funded by the Vietnam National Foundation for Science
and Technology Development (NAFOSTED) under grant number 103.01-2020.17. 

\begin{appendix}
\section{Form Factors} \label{appen:formfactor}
In this appendix we give the left- and right-handed form factors for
the LFV radiative decays $l_1 \to l_2 + \gamma$ discussed in \sect{sec:LFVdecays}.
For simplicity, we introduce the following abbreviations for the
one-loop three-point integrals  
\begin{align}
    C_{i...}^{W} &= C_{i...} \qty(m_1^2,0,m_2^2, m_{\nu_i}^2,M_W^2,M_W^2)\\
	C_{i...}^{H} &= C_{i...} \qty(m_1^2,0,m_2^2,
                       m_{\nu_i}^2,M_{H^\pm}^2,M_{H^\pm}^2) \\
	C_{i...}^{\chi} &= C_{i...} \qty(m_1^2,0,m_2^2,m_{\ti
                          n_i}^2,m_{\ti\chi_j^\pm}^2,m_{\ti\chi_j^\pm}^2). 
\end{align}
We use the following conventions, 
\begin{align}
C_0^W &= \fr{(\mu_R^2 \pi)^{(4-D)/2}}{i\pi^2}\int dq^D\fr{1}{D_x},\\
C_1^W k_1^\mu+ C_2^W k_2^\mu&=  \fr{(\mu_R^2 \pi)^{(4-D)/2}}{i\pi^2}\int dq^D\fr{q^\mu}{D_x},\\
C_{00}^W g^{\mu\nu} + C_{11}^W k_1^\mu k_1^\nu+ C_{12}^W (k_1^\mu k_2^\nu + k_1^\nu k_2^\mu)+ C_{22}^W k_2^\mu  k_2^\nu&=  \fr{(\mu_R^2 \pi)^{(4-D)/2}}{i\pi^2}\int dq^D\fr{q^\mu q^\nu}{D_x},
\end{align}
where the denominator $D_x$ is given by 
\be 
D_x=(q^2- m_{n_i}^2)((q-k_1)^2- M_W^2)((q-k_2)^2- M_W^2),
\ee
with $k_1^2= m_1^2, k_2^2=m_2^2$ and $(k_1-k_2)^2=0$
and $m_1$ ($m_2$) denoting the $l_1$ ($l_2$) mass,
  $m_{n_i}$ the mass of all the neutrinos, the active and the sterile
  ones. \s

In the 't Hooft-Feynman gauge, the left- and right-handed form factors, 
$ F^{W^\pm, H^\pm, \chi^\pm}_{L,R}$, defined in
\eqref{eq:formfactor} are given by 
\begin{align} 
 F^{W^\pm}_{L} = & \fr{e^3m_{2}}{16 \pi^2M_W^2s_W^2}\sum_{i=1}^{9} U_{\nu\;il_1}U_{\nu\;il_2}^*\bigg[m_{n_i}^2\left(
C_0^{W} + C_1^W + 2 C_2^W +  C_{12}^W +  C_{22}^W \right) \crn
&+M_W^2 \left( -2 C_1^W + 2 C_{12}^W + 2
  C_{22}^W\right) + m_{1}^2 \left( C_1^W +  C_{11}^W+C_{12}^W \right)
  \bigg] \\
 F^{W^\pm}_{R} = &  \fr{e^3m_{1}}{16 \pi^2M_W^2s_W^2}\sum_{i=1}^{9} U_{\nu\;il_1}U_{\nu\;il_2}^*\bigg[m_{n_i}^2\left(
C_0^{W} + 2C_1^W +  C_2^W +  C_{11}^W +  C_{12}^W \right) \crn
&+M_W^2 \left( -2 C_2^W + 2 C_{11}^W + 2 C_{12}^W\right) + m_{2}^2 \left( C_2^W +  C_{12}^W+C_{22}^W \right) \bigg],
\end{align}
for the $W^\pm$ and charged Goldstone boson triangle diagrams,
\begin{align} 
 F^{H^\pm}_{L}=& \fr{e^3m_{2}}{16 \pi^2M_W^2s_W^2}\sum_{i=1}^{9} U_{\nu\;il_1}U_{\nu\;il_2}^*\bigg[m_{n_i}^2\left(
-(C_0^{H^\pm} + C_1^{H^\pm} +C_2^{H^\pm})+ \fr{1}{t_\beta^2}( C_2^{H^\pm} +  C_{12}^{H^\pm} +  C_{22}^{H^\pm}) \right) \crn
& + m_{1}^2 t_\beta^2 \left( C_1^{H^\pm} +
  C_{11}^{H^\pm}+C_{12}^{H^\pm} \right) \bigg] \\
 F^{H^\pm}_{R}=&\fr{e^3m_{1}}{16 \pi^2M_W^2s_W^2}\sum_{i=1}^{9} U_{\nu\;il_1}U_{\nu\;il_2}^*\bigg[m_{n_i}^2\left(
-(C_0^{H^\pm} + C_1^{H^\pm} +  C_2^{H^\pm} )+ \fr{1}{t_\beta^2}(C_1^{H^\pm}+  C_{11}^{H^\pm} +  C_{12}^{H^\pm}) \right) \crn
&+ m_{2}^2 t_\beta^2 \left( C_2^{H^\pm} +  C_{12}^{H^\pm}+C_{22}^{H^\pm} \right) \bigg],
 \end{align}
for  the charged Higgs triangle diagram, and
\begin{align}
F_{L}^{\tilde{\chi}^\pm}&=-\fr{e}{8 \pi^2} \sum_{i=1}^{18}\sum_{j=1}^{2} \bigg[
  m_1 g^{L*}_{e^+_{l_1}\ti\chi_j \ti\nu_i}  g^L_{e^+_{l_2}\ti\chi_j \ti\nu_i}
   \left(C_1^{\ti \chi^\pm_j}+C_{12}^{\ti \chi^\pm_j}+C_{11}^{\ti \chi^\pm_j}\right) +m_{\ti \chi^\pm_j} g^{R*}_{e^+_{l_1}\ti\chi_j \ti\nu_i} g^L_{e^+_{l_2}\ti\chi_j \ti\nu_i}  \left(C_1^{\ti \chi^\pm_j}+C_2^{\ti \chi^\pm_j}\right)  \crn
&+ m_2 
g^{R*}_{e^+_{l_1}\ti\chi_j \ti\nu_i}   g^R_{e^+_{l_2}\ti\chi_j \ti\nu_i} \left(C_2^{\ti \chi^\pm_j}+C_{22}^{\ti \chi^\pm_j} +C_{12}^{\ti \chi^\pm_j}\right)  \bigg]\\
F_{R}^{\tilde{\chi}^\pm}&=-\fr{e}{8 \pi^2}\sum_{i=1}^{18}\sum_{j=1}^{2} \bigg[
 m_2g^{L*}_{e^+_{l_1}\ti\chi_j \ti\nu_i}  
   g^L_{e^+_{l_2}\ti\chi_j \ti\nu_i}\left(C_2^{\ti \chi^\pm_j}+C_{12}^{\ti \chi^\pm_j}+C_{22}^{\ti \chi^\pm_j}\right) + m_{\ti \chi^\pm_j} g^{L*}_{e^+_{l_1}\ti\chi_j \ti\nu_i}  g^R_{e^+_{l_2}\ti\chi_j \ti\nu_i} \left(C_1^{\ti \chi^\pm_j}+C_2^{\ti \chi^\pm_j}\right)\crn
&+
m_1 g^{R*}_{e^+_{l_1}\ti\chi_j \ti\nu_i}  g^R_{e^+_{l_2}\ti\chi_j \ti\nu_i}\left(C_1^{\ti \chi^\pm_j}+C_{11}^{\ti \chi^\pm_j}+C_{12}^{\ti \chi^\pm_j}\right) 
\bigg]\,,
\end{align}
for the triangle diagrams with sneutrinos and charginos in the
internal lines. The left- and right-handed couplings between the
leptons, charginos and sneutrinos are defined in the interaction Lagrangian,
\be 
\bar{e}_k ( i g^L_{e^+_{k}\ti\chi_j \ti\nu_i} P_L + i
g^R_{e^+_{k}\ti\chi_j \ti\nu_i} P_R ) \chi_j^- \ti\nu_i ,
\ee
where 
\begin{align}
	g^L_{e^+_{k}\ti\chi_j \ti\nu_i} &= \fr{m_k}{v
                                          c_\beta}\braket{U^{\ti\nu}_{ik}
                                          + i U^{\ti\nu}_{i(k+9)}} U^*_{j2}\\
	g^R_{e^+_{k}\ti\chi_j \ti\nu_i} &=
                                          -\fr{g_2V_{j1}}{\sqrt{2}}\braket{U^{\ti\nu}_{ik}
                                          + i U^{\ti\nu}_{i(k+9)}
                                          }+\fr{1}{\sqrt 2}V_{j2} 
		\sum_{n=1}^3 \braket{U^{\ti\nu}_{i(n+3)} + iU^{\ti\nu}_{i(n+12)}} Y^*_{\nu,kn} .
\end{align}
In the numerical analysis we used the
massless limit for the external lines. In this limit, one has the following simple
expressions for the one-loop three-point integrals
\cite{Lavoura:2003xp} 
\begin{align}
C_0(0,0,0,x,y,y)&= \fr{1}{x}\braket{-\fr{1}{t-1} + \fr{\ln(t)}{(t-1)^2}}\\
C_1(0,0,0,x,y,y)&= \fr{1}{x} \braket{ \fr{(t - 3)}{4(t - 1)^2} + \fr{\ln(t)}{2(t - 1)^3} }\\
C_{11}(0,0,0,x,y,y)&=\fr{1}{x} \braket{\fr{(-2 t^2 + 7 t - 11)}{18(t - 1)^3} + 
      \fr{\ln(t)}{3(t - 1)^4} }\\
C_2(0,0,0,x,y,y)&= C_1(0,0,0,x,y,y)\\
C_{22}(0,0,0,x,y,y)&=2 C_{12}(0,0,0,x,y,y)=C_{11}(0,0,0,x,y,y),
\end{align}
where $t= y/x$.
\section{Sneutrino mass matrix} \label{appen:sneumass}
The mass matrix of the sneutrinos written in each $ 3\times 3 $ block
is given by 
\begin{align}
(M_\tn)_{\tn_+ \tn_+} &= \half I_3 M_z^2 \cos 2\beta + \half \qty(\ti m_L^2 + \ti m_L^{2T}) + \dfrac{1}{2} v_u^2 \Re \qty(y_\nu y_\nu^\dagger)\\
(M_\tn)_{\tn_+ \tN_+} &= \dfrac{1}{\sqrt{2}} v_u \Re\qty(e^{i\vp_u} y_\nu A_\nu) - \dfrac{1}{2} v_d v_s \Re\qty(e^{i\vp_s} \lambda y_\nu^* )\\
(M_\tn)_{\tn_+ \tX_+} &= \dfrac{1}{\sqrt{2}} v_u \Re\qty( e^{i\vp_u} y_\nu \mu_X^* )\\
(M_\tn)_{\tn_+ \tn_-} &= \dfrac{i}{2} \qty(\ti m_L^2 - \ti m_L^{2T}) + \dfrac{1}{2} v_u^2 \Im \qty(y_\nu y_\nu^\dagger)\\
(M_\tn)_{\tn_+ \tN_-} &= \dfrac{1}{\sqrt{2}} v_u \Im\qty(e^{i\vp_u} y_\nu A_\nu) - \dfrac{1}{2} v_d v_s \Im\qty(e^{i\vp_s} \lambda y_\nu^* )\\
(M_\tn)_{\tn_+ \tX_-} &= \dfrac{1}{\sqrt{2}} v_u \Im\qty( e^{i\vp_u} y_\nu \mu_X^* )\\
(M_\tn)_{\tN_+ \tN_+} &= \half\qty(\ti m^2_N + \ti m^{2T}_N) + \Re\qty(\mu_X \mu_X^\dagger) + \dfrac{1}{2} v_u^2 \Re \qty(y_\nu^T y_\nu^*)\\
(M_\tn)_{\tN_+ \tX_+} &= \Re\qty(\mu_X B_{\mu_X}) + \dfrac{1}{\sqrt{2}} v_s \Re\qty[ e^{-i\vp_s}\mu_X\qty(\lambda_X^\dagger + \lambda_X^*) ]\\
(M_\tn)_{\tN_+ \tn_-} &= -\dfrac{1}{\sqrt{2}} v_u \Im\qty(e^{i\vp_u}A_\nu^T y_\nu^T) - \dfrac{1}{2} v_d v_s \Im\qty(e^{i\vp_s} \lambda y_\nu^\dagger )\\
(M_\tn)_{\tN_+ \tN_-} &= \dfrac{i}{2}\qty(\ti m^2_N - \ti m^{2T}_N) - \Im\qty(\mu_X \mu_X^\dagger) - \dfrac{1}{2} v_u^2 \Im \qty(y_\nu^T y_\nu^*)\\
(M_\tn)_{\tN_+ \tX_-} &= -\Im\qty(\mu_X B_{\mu_X}) + \dfrac{1}{\sqrt{2}} v_s \Im\qty[ e^{-i\vp_s}\mu_X\qty(\lambda_X^\dagger + \lambda_X^*) ]\\
(M_\tn)_{\tX_+ \tX_+} &= \half \qty(\ti m_X^2 + \ti m_X^{2T}) + \Re\qty(\mu_X^T \mu_X^*) + \half \Re\qty[ \qty(e^{2i\vp_s}v_s^2 \kappa - e^{i\vp_u}v_d v_u \lambda) \qty(\lambda_X^* + \lambda_X^\dagger)] \notag\\
&\quad+ \half v_s^2 \Re\qty[ \qty(\lambda_X + \lambda_X^T )\qty(\lambda_X^\dagger + \lambda_X^*) ] + \dfrac{1}{\sqrt{2}} v_s \Re \qty[e^{i\vp_s}\qty(\lambda_X A_X + A_X^T\lambda_X^T)] \\
(M_\tn)_{\tX_+ \tn_-} &= -\dfrac{1}{\sqrt{2}} v_u \Im\qty( e^{i\vp_u} \mu_X^\dagger y_\nu^T )\\
(M_\tn)_{\tX_+ \tN_-} &= \Im\qty(B_{\mu_X}^T \mu_X^T) + \dfrac{1}{\sqrt{2}} v_s \Im\qty[ e^{-i\vp_s} \qty(\lambda_X^\dagger + \lambda_X^*) \mu^T_X ]\\
(M_\tn)_{\tX_+ \tX_-} &= \dfrac{i}{2} \qty(\ti m_X^2 - \ti m_X^{2T}) + \Im\qty(\mu_X^T \mu_X^*) + \half \Im\qty[ \qty(e^{2i\vp_s}v_s^2 \kappa - e^{i\vp_u}v_d v_u \lambda) \qty(\lambda_X^* + \lambda_X^\dagger)] \notag\\
&\quad+ \half v_s^2 \Im\qty[ \qty(\lambda_X + \lambda_X^T )\qty(\lambda_X^\dagger + \lambda_X^*) ] - \dfrac{1}{\sqrt{2}} v_s \Im \qty[e^{i\vp_s}\qty(\lambda_X A_X + A_X^T\lambda_X^T)] \\
(M_\tn)_{\tn_- \tn_-} &= \half I_3 M_z^2 \cos 2\beta + \half \qty(\ti m_L^2 + \ti m_L^{2T}) + \dfrac{1}{2} v_u^2 \Re \qty(y_\nu y_\nu^\dagger)\\
(M_\tn)_{\tn_- \tN_-} &= \dfrac{1}{\sqrt{2}} v_u \Re\qty(e^{i\vp_u} y_\nu A_\nu) - \dfrac{1}{2} v_d v_s \Re\qty(e^{i\vp_s} \lambda y_\nu^* )\\
(M_\tn)_{\tn_- \tX_-} &= \dfrac{1}{\sqrt{2}} v_u \Re\qty( e^{i\vp_u} y_\nu \mu_X^* )\\	
(M_\tn)_{\tN_- \tN_-} &= \half\qty(\ti m^2_N + \ti m^{2T}_N) + \Re\qty(\mu_X \mu_X^\dagger) + \dfrac{1}{2} v_u^2 \Re \qty(y_\nu^T y_\nu^*)\\
(M_\tn)_{\tN_- \tX_-} &= \Re\qty(\mu_X B_{\mu_X}) - \dfrac{1}{\sqrt{2}} v_s \Re\qty[ e^{-i\vp_s}\mu_X\qty(\lambda_X^\dagger + \lambda_X^*) ]\\
(M_\tn)_{\tX_- \tX_-} &= \half \qty(\ti m_X^2 + \ti m_X^{2T}) + \Re\qty(\mu_X^T \mu_X^*) \textcolor{red}{-} \half \Re\qty[ \qty(e^{2i\vp_s}v_s^2 \kappa - e^{i\vp_u}v_d v_u \lambda) \qty(\lambda_X^* + \lambda_X^\dagger)] \notag\\
&\quad+ \half v_s^2 \Re\qty[ \qty(\lambda_X + \lambda_X^T )\qty(\lambda_X^\dagger + \lambda_X^*) ] \textcolor{red}{-} \dfrac{1}{\sqrt{2}} v_s \Re \qty[e^{i\vp_s}\qty(\lambda_X A_X + A_X^T\lambda_X^T)].
\end{align}

\section{Neutral Higgs mass matrix counterterm} \label{appen:CTHiggsMass}
In this section we present the counterterm mass matrix for the neutral
Higgs bosons in the basis $ \qty(h_d,h_u,h_s,a,a_s)^T
$. We use the convention $ \varphi_y =
  \varphi_\kappa-\varphi_\lambda+2\varphi_s - \varphi_u $, $
  \varphi_\omega = \varphi_\kappa + 3 \varphi_s $ and the short-hand
notation $s_x \equiv \sin x, c_x \equiv \cos x$.
\begin{align}
	\qty(\delta^{(1)}M_{hh})_{h_d h_d} &= c_{\beta }^3 s_{\beta } \delta^{(1)}{t_{\beta }} \left(v^2 \left| \lambda \right| ^2-2 M_W^2-2 M_Z^2+2 M^2_{H^\pm}\right)+v^2 \left| \lambda \right|  s_{\beta }^2 \delta^{(1)}{\left| \lambda \right| }+v \left| \lambda \right| ^2 s_{\beta }^2 \delta^{(1)} v\notag\\
	&\quad-\frac{\left(c_{2 \beta }-3\right) c_{\beta } \delta^{(1)}{t_{h_d}}}{2 v}-\frac{c_{\beta }^2 s_{\beta } \delta^{(1)}{t_{h_u}}}{v}+c_{\beta }^2 \delta^{(1)}{M_Z^2}-s_{\beta }^2 \delta^{(1)}{M_W^2}+\delta^{(1)}{M^2_{H^\pm}} s_{\beta }^2\\
	\qty(\delta^{(1)}M_{hh})_{h_d h_u} &= \frac{1}{2} c_{2 \beta } c_{\beta }^2 \delta^{(1)}{t_{\beta }} \left(v^2 \left| \lambda \right| ^2+2 M_W^2-2 M_Z^2-2 M^2_{H^\pm}\right)+v^2 \left| \lambda \right|  c_{\beta } s_{\beta } \delta^{(1)}{\left| \lambda \right| }\notag\\
	&\quad+v \left| \lambda \right| ^2 c_{\beta } s_{\beta } \delta^{(1)} v+\frac{c_{\beta }^3 \delta^{(1)}{t_{h_u}}}{v}+c_{\beta } s_{\beta } \delta^{(1)}{M_W^2}-c_{\beta } s_{\beta } \delta^{(1)}{M_Z^2}\notag\\
	&\quad-c_{\beta } \delta^{(1)}{M^2_{H^\pm}} s_{\beta }+\frac{s_{\beta }^3 \delta^{(1)}{t_{h_d}}}{v}\\
	\qty(\delta^{(1)}M_{hh})_{h_d h_s} &= -\frac{v c_{\beta }^2 \delta^{(1)}{t_{\beta }}}{2 v_s} \left(2 c_{\beta }^2 s_{\beta } \left(v^2 \left| \lambda \right| ^2-2 M_W^2+2 M^2_{H^\pm}\right)+\left| \kappa \right|  \left| \lambda \right|  c_{\beta } v_s^2 c_{\varphi _y}\right.\notag\\
	&\quad\left.+s_{\beta }^3 \left(-\left(v^2 \left| \lambda \right| ^2-2 M_W^2+2 M^2_{H^\pm}\right)\right)+2 \left| \lambda \right| ^2 s_{\beta } v_s^2\right)\notag\\
	&\quad+\frac{v \delta^{(1)}{v_s} \left(c_{\beta } \left(s_{\beta }^2 \left(v^2 \left| \lambda \right| ^2-2 M_W^2\right)+2 \left| \lambda \right| ^2 v_s^2\right)+s_{\beta } \left(M^2_{H^\pm} s_{2 \beta }-\left| \kappa \right|  \left| \lambda \right|  v_s^2 c_{\varphi _y}\right)\right)}{2 v_s^2}\notag\\
	&\quad+\frac{\delta^{(1)} v \left(s_{\beta } \left(M_W^2 s_{2 \beta }-\left| \kappa \right|  \left| \lambda \right|  v_s^2 c_{\varphi _y}\right)+c_{\beta } \left(2 \left| \lambda \right| ^2 v_s^2-s_{\beta }^2 \left(3 v^2 \left| \lambda \right| ^2+2 M^2_{H^\pm}\right)\right)\right)}{2 v_s}\notag\\
	&\quad+\delta^{(1)}{\left| \lambda \right| } \left(c_{\beta } \left(2 v \left| \lambda \right|  v_s-\frac{v^3 \left| \lambda \right|  s_{\beta }^2}{v_s}\right)-\frac{1}{2} v \left| \kappa \right|  s_{\beta } v_s c_{\varphi _y}\right)\notag\\
	&\quad-\frac{1}{2} v \left| \lambda \right|  s_{\beta } v_s \delta^{(1)}{\left| \kappa \right| } c_{\varphi _y}+\frac{1}{2} v \left| \kappa \right|  \left| \lambda \right|  s_{\beta } v_s s_{\varphi _y} \delta^{(1)}{\varphi _y}+\frac{c_{\beta }^3 s_{\beta } \delta^{(1)}{t_{h_u}}}{v_s}\notag\\
	&\quad+\frac{v c_{\beta } s_{\beta }^2 \delta^{(1)}{M_W^2}}{v_s}-\frac{v c_{\beta } \delta^{(1)}{M^2_{H^\pm}} s_{\beta }^2}{v_s}+\frac{s_{\beta }^4 \delta^{(1)}{t_{h_d}}}{v_s}\\
	\qty(\delta^{(1)}M_{hh})_{h_d a} &= \frac{\delta^{(1)}{t_{a_d}}\cot\beta }{v}\\
	\qty(\delta^{(1)}M_{hh})_{h_d a_s} &= \frac{\delta^{(1)}{t_{a_d}}}{v_s}+\frac{3}{2} v \left| \kappa \right|  \left| \lambda \right|  c_{\beta }^3 v_s s_{\varphi _y} \delta^{(1)}{t_{\beta }}+\frac{3}{2} v \left| \kappa \right|  \left| \lambda \right|  s_{\beta } v_s c_{\varphi _y} \delta^{(1)}{\varphi _y}\notag\\
	&\quad+\frac{3}{2} v \left| \lambda \right|  s_{\beta } v_s \delta^{(1)}{\left| \kappa \right| } s_{\varphi _y}+\frac{3}{2} v \left| \kappa \right|  s_{\beta } v_s \delta^{(1)}{\left| \lambda \right| } s_{\varphi _y}+\frac{3}{2} v \left| \kappa \right|  \left| \lambda \right|  s_{\beta } \delta^{(1)}{v_s} s_{\varphi _y}\notag\\
	&\quad+\frac{3}{2} \left| \kappa \right|  \left| \lambda \right|  s_{\beta } v_s \delta^{(1)} v s_{\varphi _y}\\
	\qty(\delta^{(1)}M_{hh})_{h_u h_u} &= c_{\beta }^3 s_{\beta } \delta^{(1)}{t_{\beta }} \left(-v^2 \left| \lambda \right| ^2+2 M_W^2+2 M_Z^2-2 M^2_{H^\pm}\right)+v^2 \left| \lambda \right|  c_{\beta }^2 \delta^{(1)}{\left| \lambda \right| }\notag\\
	&\quad+v \left| \lambda \right| ^2 c_{\beta }^2 \delta^{(1)} v-\frac{c_{\beta } s_{\beta }^2 \delta^{(1)}{t_{h_d}}}{v}+\frac{\left(c_{2 \beta }+3\right) s_{\beta } \delta^{(1)}{t_{h_u}}}{2 v}-c_{\beta }^2 \delta^{(1)}{M_W^2}\notag\\
	&\quad+c_{\beta }^2 \delta^{(1)}{M^2_{H^\pm}}+s_{\beta }^2 \delta^{(1)}{M_Z^2}\\
	\qty(\delta^{(1)}M_{hh})_{h_u h_s} &= \frac{v c_{\beta }^2 \delta^{(1)}{t_{\beta }}}{2 v_s} \left(2 c_{\beta } \left(\left| \lambda \right| ^2 v_s^2+2 s_{\beta }^2 \left(M^2_{H^\pm}-M_W^2\right)\right)\right.\notag\\
	&\quad\left.+c_{\beta }^3 \left(-\left(v^2 \left| \lambda \right| ^2-2 M_W^2+2 M^2_{H^\pm}\right)\right)+\left| \lambda \right|  s_{\beta } \left(\left| \kappa \right|  v_s^2 c_{\varphi _y}+v^2 \left| \lambda \right|  s_{2 \beta }\right)\right)\notag\\
	&\quad+\frac{1}{2} v \delta^{(1)}{v_s} \left(\frac{c_{\beta }^2 s_{\beta } \left(v^2 \left| \lambda \right| ^2-2 M_W^2+2 M^2_{H^\pm}\right)}{v_s^2}-\left| \kappa \right|  \left| \lambda \right|  c_{\beta } c_{\varphi _y}+2 \left| \lambda \right| ^2 s_{\beta }\right)\notag\\
	&\quad+\delta^{(1)} v \left(-\frac{c_{\beta }^2 s_{\beta } \left(3 v^2 \left| \lambda \right| ^2-2 M_W^2+2 M^2_{H^\pm}\right)}{2 v_s}-\frac{1}{2} \left| \kappa \right|  \left| \lambda \right|  c_{\beta } v_s c_{\varphi _y}+\left| \lambda \right| ^2 s_{\beta } v_s\right)\notag\\
	&\quad+\delta^{(1)}{\left| \lambda \right| } \left(-\frac{v^3 \left| \lambda \right|  c_{\beta }^2 s_{\beta }}{v_s}-\frac{1}{2} v \left| \kappa \right|  c_{\beta } v_s c_{\varphi _y}+2 v \left| \lambda \right|  s_{\beta } v_s\right)\notag\\
	&\quad-\frac{1}{2} v \left| \lambda \right|  c_{\beta } v_s \delta^{(1)}{\left| \kappa \right| } c_{\varphi _y}+\frac{1}{2} v \left| \kappa \right|  \left| \lambda \right|  c_{\beta } v_s s_{\varphi _y} \delta^{(1)}{\varphi _y}+\frac{c_{\beta } s_{\beta }^3 \delta^{(1)}{t_{h_d}}}{v_s}\notag\\
	&\quad+\frac{c_{\beta }^4 \delta^{(1)}{t_{h_u}}}{v_s}+\frac{v c_{\beta }^2 s_{\beta } \delta^{(1)}{M_W^2}}{v_s}-\frac{v c_{\beta }^2 \delta^{(1)}{M^2_{H^\pm}} s_{\beta }}{v_s}\\
	\qty(\delta^{(1)}M_{hh})_{h_u a} &= \frac{\delta^{(1)}{t_{a_d}}}{v}\\
	\qty(\delta^{(1)}M_{hh})_{h_u a_s} &= \frac{\cot\beta \delta^{(1)}{t_{a_d}}}{v_s}-\frac{3}{2} v \left| \kappa \right|  \left| \lambda \right|  c_{\beta }^2 s_{\beta } v_s s_{\varphi _y} \delta^{(1)}{t_{\beta }}+\frac{3}{2} v \left| \lambda \right|  c_{\beta } v_s \delta^{(1)}{\left| \kappa \right| } s_{\varphi _y}\notag\\
	&\quad+\frac{3}{2} v \left| \kappa \right|  c_{\beta } v_s \delta^{(1)}{\left| \lambda \right| } s_{\varphi _y}+\frac{3}{2} v \left| \kappa \right|  \left| \lambda \right|  c_{\beta } \delta^{(1)}{v_s} s_{\varphi _y}+\frac{3}{2} v \left| \kappa \right|  \left| \lambda \right|  c_{\beta } v_s c_{\varphi _y} \delta^{(1)}{\varphi _y}\notag\\
	&\quad+\frac{3}{2} \left| \kappa \right|  \left| \lambda \right|  c_{\beta } v_s \delta^{(1)} v s_{\varphi _y}\\
	\qty(\delta^{(1)}M_{hh})_{h_s h_s} &= -\frac{v s_{\beta } \delta^{(1)}{t_{h_u}} c_{\beta }^4}{v_s^2}+\frac{v^2 s_{\beta }^2 \delta^{(1)}{M^2_{H^\pm}} c_{\beta }^2}{v_s^2}-\frac{v^2 s_{\beta }^2 \delta^{(1)}{M_W^2} c_{\beta }^2}{v_s^2}\notag\\
	&\quad+\frac{1}{2} v^2 c_{2 \beta } \left(-\left| \kappa \right|  \left| \lambda \right|  c_{\varphi _y}+\frac{3 i \left(-1+e^{2 i \varphi _{\omega }}\right) \left| \kappa \right|  \left| \lambda \right|  s_{\varphi _y}}{1+e^{2 i \varphi _{\omega }}}\right.\notag\\
	&\quad\left.+\frac{\left(v^2 \left| \lambda \right| ^2-2 M_W^2+2 M^2_{H^\pm}\right) s_{2 \beta }}{v_s^2}\right) \delta^{(1)}{t_{\beta }} c_{\beta }^2+\frac{i \left(-1+e^{2 i \varphi _{\omega }}\right) v \delta^{(1)}{t_{a_d}} c_{\beta }}{\left(1+e^{2 i \varphi _{\omega }}\right) v_s^2}\notag\\
	&\quad-\frac{v s_{\beta }^4 \delta^{(1)}{t_{h_d}} c_{\beta }}{v_s^2}+\frac{v^2 \left| \kappa \right|  \left| \lambda \right|  s_{\beta } \left(3 i \left(-1+e^{2 i \varphi _{\omega }}\right) c_{\varphi _y}+\left(1+e^{2 i \varphi _{\omega }}\right) s_{\varphi _y}\right) \delta^{(1)}{\varphi _y} c_{\beta }}{2 \left(1+e^{2 i \varphi _{\omega }}\right)}\notag\\
	&\quad+\frac{1}{2} v \delta^{(1)} v s_{2 \beta } \left(-\left| \kappa \right|  \left| \lambda \right|  c_{\varphi _y}+\frac{3 i \left(-1+e^{2 i \varphi _{\omega }}\right) \left| \kappa \right|  \left| \lambda \right|  s_{\varphi _y}}{1+e^{2 i \varphi _{\omega }}}\right.\notag\\
	&\quad\left.+\frac{\left(v^2 \left| \lambda \right| ^2-M_W^2+M^2_{H^\pm}\right) s_{2 \beta }}{v_s^2}\right)+\left(\frac{1}{2} \left| \lambda \right|  c_{\beta } s_{\beta } \left(\frac{3 i \left(-1+e^{2 i \varphi _{\omega }}\right) s_{\varphi _y}}{1+e^{2 i \varphi _{\omega }}}-c_{\varphi _y}\right) v^2\right.\notag\\
	&\quad\left.+v_s \left(\frac{\sqrt{2} e^{i \varphi _{\omega }} 	\text{Re}A_{\kappa}}{1+e^{2 i \varphi _{\omega }}}+4 \left| \kappa \right|  v_s\right)\right) \delta^{(1)}{\left| \kappa \right| }+\frac{1}{4} v^2 s_{2 \beta } \left(\frac{\left| \lambda \right|  s_{2 \beta } v^2}{v_s^2}-\left| \kappa \right|  c_{\varphi _y}\right.\notag\\
	&\quad\left.+\frac{3 i \left(-1+e^{2 i \varphi _{\omega }}\right) \left| \kappa \right|  s_{\varphi _y}}{1+e^{2 i \varphi _{\omega }}}\right) \delta^{(1)}{\left| \lambda \right| }+\frac{\sqrt{2} e^{i \varphi _{\omega }} \left| \kappa \right|  v_s \delta^{(1)}{	\text{Re}A_{\kappa}}}{1+e^{2 i \varphi _{\omega }}}+\frac{\left(i-i e^{2 i \varphi _{\omega }}\right) \delta^{(1)}{t_{a_s}}}{e^{2 i \varphi _{\omega }} v_s+v_s}\notag\\
	&\quad+\frac{\delta^{(1)}{t_{h_s}}}{v_s}+\left(-\frac{\left| \lambda \right| ^2 c_{\beta }^2 s_{\beta }^2 v^4}{v_s^3}-\frac{\left(M^2_{H^\pm}-M_W^2\right) s_{2 \beta }^2 v^2}{2 v_s^3}\right.\notag\\
	&\quad\left.+\left| \kappa \right|  \left(\frac{\sqrt{2} e^{i \varphi _{\omega }} 	\text{Re}A_{\kappa}}{1+e^{2 i \varphi _{\omega }}}+4 \left| \kappa \right|  v_s\right)\right) \delta^{(1)}{v_s}\notag\\
	&\quad+\frac{\left(-6 e^{2 i \varphi _{\omega }} \left| \kappa \right|  \left| \lambda \right|  c_{\beta } s_{\beta } s_{\varphi _y} v^2-i \sqrt{2} e^{i \varphi _{\omega }} \left(-1+e^{2 i \varphi _{\omega }}\right) \left| \kappa \right|  	\text{Re}A_{\kappa} v_s\right) \delta^{(1)}{\varphi _{\omega }}}{\left(1+e^{2 i \varphi _{\omega }}\right){}^2}\\
	\qty(\delta^{(1)}M_{hh})_{h_s a} &= \frac{\delta^{(1)}{t_{a_d}}}{s_{\beta } v_s}-\frac{1}{2} v \left| \kappa \right|  \left| \lambda \right|  v_s c_{\varphi _y} \delta^{(1)}{\varphi _y}-\frac{1}{2} v \left| \lambda \right|  v_s \delta^{(1)}{\left| \kappa \right| } s_{\varphi _y}-\frac{1}{2} v \left| \kappa \right|  v_s \delta^{(1)}{\left| \lambda \right| } s_{\varphi _y}\notag\\
	&\quad-\frac{1}{2} v \left| \kappa \right|  \left| \lambda \right|  \delta^{(1)}{v_s} s_{\varphi _y}-\frac{1}{2} \left| \kappa \right|  \left| \lambda \right|  v_s \delta^{(1)} v s_{\varphi _y}\\
	\qty(\delta^{(1)}M_{hh})_{h_s a_s} &= -\frac{2 v c_{\beta } \delta^{(1)}{t_{a_d}}}{v_s^2}+\frac{2 \delta^{(1)}{t_{a_s}}}{v_s}-2 v^2 \left| \kappa \right|  \left| \lambda \right|  c_{\beta }^2 c_{2 \beta } s_{\varphi _y} \delta^{(1)}{t_{\beta }}-2 v^2 \left| \lambda \right|  c_{\beta } s_{\beta } \delta^{(1)}{\left| \kappa \right| } s_{\varphi _y}\notag\\
	&\quad-2 v^2 \left| \kappa \right|  c_{\beta } s_{\beta } \delta^{(1)}{\left| \lambda \right| } s_{\varphi _y}-2 v^2 \left| \kappa \right|  \left| \lambda \right|  c_{\beta } s_{\beta } c_{\varphi _y} \delta^{(1)}{\varphi _y}-4 v \left| \kappa \right|  \left| \lambda \right|  c_{\beta } s_{\beta } \delta^{(1)} v s_{\varphi _y}\\
	\qty(\delta^{(1)}M_{hh})_{a a} &= v^2 \left| \lambda \right|  \delta^{(1)}{\left| \lambda \right| }+v \left| \lambda \right| ^2 \delta^{(1)} v-\delta^{(1)}{M_W^2}+\delta^{(1)}{M^2_{H^\pm}}\\
	\qty(\delta^{(1)}M_{hh})_{a a_s} &= \frac{v c_{2 \beta } c_{\beta }^2 \delta^{(1)}{t_{\beta }} \left(v^2 \left| \lambda \right| ^2-2 M_W^2+2 M^2_{H^\pm}\right)}{2 v_s} \notag\\
	&\quad-\frac{v \delta^{(1)}{v_s} \left(6 \left| \kappa \right|  \left| \lambda \right|  v_s^2 c_{\varphi _y}+s_{2 \beta } \left(v^2 \left| \lambda \right| ^2-2 M_W^2+2 M^2_{H^\pm}\right)\right)}{4 v_s^2} \notag\\
	&\quad+\delta^{(1)} v \left(\frac{s_{2 \beta } \left(3 v^2 \left| \lambda \right| ^2-2 M_W^2+2 M^2_{H^\pm}\right)}{4 v_s}-\frac{3}{2} \left| \kappa \right|  \left| \lambda \right|  v_s c_{\varphi _y}\right)\notag\\
	&\quad+\delta^{(1)}{\left| \lambda \right| } \left(\frac{v^3 \left| \lambda \right|  c_{\beta } s_{\beta }}{v_s}-\frac{3}{2} v \left| \kappa \right|  v_s c_{\varphi _y}\right)-\frac{3}{2} v \left| \lambda \right|  v_s \delta^{(1)}{\left| \kappa \right| } c_{\varphi _y} \notag\\
	&\quad+\frac{3}{2} v \left| \kappa \right|  \left| \lambda \right|  v_s s_{\varphi _y} \delta^{(1)}{\varphi _y}-\frac{c_{\beta }^3 \delta^{(1)}{t_{h_u}}}{v_s}-\frac{v c_{\beta } s_{\beta } \delta^{(1)}{M_W^2}}{v_s}+\frac{v c_{\beta } \delta^{(1)}{M^2_{H^\pm}} s_{\beta }}{v_s}\notag\\
	&\quad-\frac{s_{\beta }^3 \delta^{(1)}{t_{h_d}}}{v_s}\\
	\qty(\delta^{(1)}M_{hh})_{a_s a_s} &= -\frac{v s_{\beta } \delta^{(1)}{t_{h_u}} c_{\beta }^4}{v_s^2}+\frac{v^2 s_{\beta }^2 \delta^{(1)}{M^2_{H^\pm}} c_{\beta }^2}{v_s^2}-\frac{v^2 s_{\beta }^2 \delta^{(1)}{M_W^2} c_{\beta }^2}{v_s^2}\notag\\
	&\quad+\frac{1}{2} v^2 c_{2 \beta } \left(3 \left| \kappa \right|  \left| \lambda \right|  c_{\varphi _y}-\frac{9 i \left(-1+e^{2 i \varphi _{\omega }}\right) \left| \kappa \right|  \left| \lambda \right|  s_{\varphi _y}}{1+e^{2 i \varphi _{\omega }}}\right.\notag\\
	&\quad\left.+\frac{\left(v^2 \left| \lambda \right| ^2-2 M_W^2+2 M^2_{H^\pm}\right) s_{2 \beta }}{v_s^2}\right) \delta^{(1)}{t_{\beta }} c_{\beta }^2-\frac{3 i \left(-1+e^{2 i \varphi _{\omega }}\right) v \delta^{(1)}{t_{a_d}} c_{\beta }}{\left(1+e^{2 i \varphi _{\omega }}\right) v_s^2}\notag\\
	&\quad-\frac{v s_{\beta }^4 \delta^{(1)}{t_{h_d}} c_{\beta }}{v_s^2}+\frac{3}{2} v^2 \left| \kappa \right|  \left| \lambda \right|  s_{\beta } \left(-\frac{3 i \left(-1+e^{2 i \varphi _{\omega }}\right) c_{\varphi _y}}{1+e^{2 i \varphi _{\omega }}}-s_{\varphi _y}\right) \delta^{(1)}{\varphi _y} c_{\beta }\notag\\
	&\quad+\delta^{(1)} v \left(\frac{2 \left| \lambda \right| ^2 c_{\beta }^2 s_{\beta }^2 v^3}{v_s^2}+3 \left| \kappa \right|  \left| \lambda \right|  c_{\beta } s_{\beta } \left(c_{\varphi _y}-\frac{3 i \left(-1+e^{2 i \varphi _{\omega }}\right) s_{\varphi _y}}{1+e^{2 i \varphi _{\omega }}}\right) v\right.\notag\\
	&\quad\left.+\frac{\left(M^2_{H^\pm}-M_W^2\right) s_{2 \beta }^2 v}{2 v_s^2}\right)+\left(\frac{3}{2} v^2 \left| \lambda \right|  c_{\beta } s_{\beta } \left(c_{\varphi _y}-\frac{3 i \left(-1+e^{2 i \varphi _{\omega }}\right) s_{\varphi _y}}{1+e^{2 i \varphi _{\omega }}}\right)\right.\notag\\
	&\quad\left.-\frac{3 \sqrt{2} e^{i \varphi _{\omega }} 	\text{Re}A_{\kappa} v_s}{1+e^{2 i \varphi _{\omega }}}\right) \delta^{(1)}{\left| \kappa \right| }+\frac{1}{4} v^2 s_{2 \beta } \left(\frac{\left| \lambda \right|  s_{2 \beta } v^2}{v_s^2}+3 \left| \kappa \right|  c_{\varphi _y}\right.\notag\\
	&\quad\left.-\frac{9 i \left(-1+e^{2 i \varphi _{\omega }}\right) \left| \kappa \right|  s_{\varphi _y}}{1+e^{2 i \varphi _{\omega }}}\right) \delta^{(1)}{\left| \lambda \right| }-\frac{3 \sqrt{2} e^{i \varphi _{\omega }} \left| \kappa \right|  v_s \delta^{(1)}{	\text{Re}A_{\kappa}}}{1+e^{2 i \varphi _{\omega }}}\notag\\
	&\quad+\frac{3 i \left(-1+e^{2 i \varphi _{\omega }}\right) \delta^{(1)}{t_{a_s}}}{\left(1+e^{2 i \varphi _{\omega }}\right) v_s}+\frac{\delta^{(1)}{t_{h_s}}}{v_s}\notag\\
	&\quad+\left(-\frac{v^2 \left(v^2 \left| \lambda \right| ^2-2 M_W^2+2 M^2_{H^\pm}\right) s_{2 \beta }^2}{4 v_s^3}-\frac{3 \sqrt{2} e^{i \varphi _{\omega }} \left| \kappa \right|  	\text{Re}A_{\kappa}}{1+e^{2 i \varphi _{\omega }}}\right) \delta^{(1)}{v_s}\notag\\
	&\quad+\frac{3 e^{i \varphi _{\omega }} \left| \kappa \right|  \left(6 e^{i \varphi _{\omega }} \left| \lambda \right|  c_{\beta } s_{\beta } s_{\varphi _y} v^2+i \sqrt{2} \left(-1+e^{2 i \varphi _{\omega }}\right) 	\text{Re}A_{\kappa} v_s\right) \delta^{(1)}{\varphi _{\omega }}}{\left(1+e^{2 i \varphi _{\omega }}\right){}^2}.
\end{align}

\end{appendix}

\bibliographystyle{ieeetr}

\begin{thebibliography}{100}
\bibitem{Fayet:1974pd}
P.~Fayet, ``{Supergauge Invariant Extension of the Higgs Mechanism and a Model
  for the electron and Its Neutrino},'' {\em Nucl.Phys.}, vol.~B90,
  pp.~104--124, 1975.

\bibitem{Barbieri:1982eh}
R.~Barbieri, S.~Ferrara, and C.~A. Savoy, ``{Gauge Models with Spontaneously
  Broken Local Supersymmetry},'' {\em Phys.Lett.}, vol.~B119, p.~343, 1982.

\bibitem{Dine:1981rt}
M.~Dine, W.~Fischler, and M.~Srednicki, ``{A Simple Solution to the Strong CP
  Problem with a Harmless Axion},'' {\em Phys.Lett.}, vol.~B104, p.~199, 1981.

\bibitem{Nilles:1982dy}
H.~P. Nilles, M.~Srednicki, and D.~Wyler, ``{Weak Interaction Breakdown Induced
  by Supergravity},'' {\em Phys.Lett.}, vol.~B120, p.~346, 1983.

\bibitem{Frere:1983ag}
J.~Frere, D.~Jones, and S.~Raby, ``{Fermion Masses and Induction of the Weak
  Scale by Supergravity},'' {\em Nucl.Phys.}, vol.~B222, p.~11, 1983.

\bibitem{Derendinger:1983bz}
J.~Derendinger and C.~A. Savoy, ``{Quantum Effects and SU(2) x U(1) Breaking in
  Supergravity Gauge Theories},'' {\em Nucl.Phys.}, vol.~B237, p.~307, 1984.

\bibitem{Ellis:1988er}
J.~R. Ellis, J.~Gunion, H.~E. Haber, L.~Roszkowski, and F.~Zwirner, ``{Higgs
  Bosons in a Nonminimal Supersymmetric Model},'' {\em Phys.Rev.}, vol.~D39,
  p.~844, 1989.

\bibitem{Drees:1988fc}
M.~Drees, ``{Supersymmetric Models with Extended Higgs Sector},'' {\em
  Int.J.Mod.Phys.}, vol.~A4, p.~3635, 1989.

\bibitem{Ellwanger:1993xa}
U.~Ellwanger, M.~Rausch~de Traubenberg, and C.~A. Savoy, ``{Particle spectrum
  in supersymmetric models with a gauge singlet},'' {\em Phys.Lett.},
  vol.~B315, pp.~331--337, 1993.

\bibitem{Ellwanger:1995ru}
U.~Ellwanger, M.~Rausch~de Traubenberg, and C.~A. Savoy, ``{Higgs phenomenology
  of the supersymmetric model with a gauge singlet},'' {\em Z.Phys.}, vol.~C67,
  pp.~665--670, 1995.

\bibitem{Ellwanger:1996gw}
U.~Ellwanger, M.~Rausch~de Traubenberg, and C.~A. Savoy, ``{Phenomenology of
  supersymmetric models with a singlet},'' {\em Nucl.Phys.}, vol.~B492,
  pp.~21--50, 1997.

\bibitem{Elliott:1994ht}
T.~Elliott, S.~King, and P.~White, ``{Unification constraints in the
  next-to-minimal supersymmetric standard model},'' {\em Phys.Lett.},
  vol.~B351, pp.~213--219, 1995.

\bibitem{King:1995vk}
S.~King and P.~White, ``{Resolving the constrained minimal and next-to-minimal
  supersymmetric standard models},'' {\em Phys.Rev.}, vol.~D52, pp.~4183--4216,
  1995.

\bibitem{Franke:1995tc}
F.~Franke and H.~Fraas, ``{Neutralinos and Higgs bosons in the next-to-minimal
  supersymmetric standard model},'' {\em Int.J.Mod.Phys.}, vol.~A12,
  pp.~479--534, 1997.

\bibitem{Maniatis:2009re}
M.~Maniatis, ``{The Next-to-Minimal Supersymmetric extension of the Standard
  Model reviewed},'' {\em Int.J.Mod.Phys.}, vol.~A25, pp.~3505--3602, 2010.

\bibitem{Ellwanger:2009dp}
U.~Ellwanger, C.~Hugonie, and A.~M. Teixeira, ``{The Next-to-Minimal
  Supersymmetric Standard Model},'' {\em Phys.Rept.}, vol.~496, pp.~1--77,
  2010.

\bibitem{Aad:2012tfa}
G.~Aad {\em et~al.}, ``{Observation of a new particle in the search for the
  Standard Model Higgs boson with the ATLAS detector at the LHC},'' {\em
  Phys.Lett.}, vol.~B716, pp.~1--29, 2012.

\bibitem{Chatrchyan:2012ufa}
S.~Chatrchyan {\em et~al.}, ``{Observation of a new boson at a mass of 125 GeV
  with the CMS experiment at the LHC},'' {\em Phys.Lett.}, vol.~B716,
  pp.~30--61, 2012.

\bibitem{Gogoladze:2008wz}
I.~Gogoladze, N.~Okada, and Q.~Shafi, ``{NMSSM and Seesaw Physics at LHC},''
  {\em Phys. Lett.}, vol.~B672, pp.~235--239, 2009.

\bibitem{Mohapatra:1986aw}
R.~Mohapatra, ``{Mechanism for Understanding Small Neutrino Mass in Superstring
  Theories},'' {\em Phys. Rev. Lett.}, vol.~56, pp.~561--563, 1986.

\bibitem{PhysRevD.34.1642}
R.~N. Mohapatra and J.~W.~F. Valle, ``Neutrino mass and baryon-number
  nonconservation in superstring models,'' {\em Phys. Rev. D}, vol.~34,
  pp.~1642--1645, Sep 1986.

\bibitem{Bernabeu:1987gr}
J.~Bernabeu, A.~Santamaria, J.~Vidal, A.~Mendez, and J.~Valle, ``{Lepton Flavor
  Nonconservation at High-Energies in a Superstring Inspired Standard Model},''
  {\em Phys. Lett. B}, vol.~187, pp.~303--308, 1987.

\bibitem{Gogoladze:2012jp}
I.~Gogoladze, B.~He, and Q.~Shafi, ``{Inverse Seesaw in NMSSM and 126 GeV Higgs
  Boson},'' {\em Phys. Lett.}, vol.~B718, pp.~1008--1013, 2013.

\bibitem{Wang:2013jya}
W.~Wang, J.~M. Yang, and L.~L. You, ``{Higgs boson mass in NMSSM with
  right-handed neutrino},'' {\em JHEP}, vol.~07, p.~158, 2013.

\bibitem{Biekotter:2017xmf}
T.~Biek{\"o}tter, S.~Heinemeyer, and C.~Mu\~noz, ``{Precise prediction for the
  Higgs-boson masses in the $\mu \nu $ SSM},'' {\em Eur. Phys. J. C}, vol.~78,
  no.~6, p.~504, 2018.

\bibitem{Biekotter:2019gtq}
T.~Biek{\"o}tter, S.~Heinemeyer, and C.~Mu{\~no}z, ``{Precise prediction for
  the Higgs-Boson masses in the $\mu\nu $ SSM with three right-handed neutrino
  superfields},'' {\em Eur. Phys. J. C}, vol.~79, no.~8, p.~667, 2019.

\bibitem{Heinemeyer:2010eg}
S.~Heinemeyer, M.~J. Herrero, S.~Penaranda, and A.~M. Rodriguez-Sanchez,
  ``{Higgs Boson Masses in the MSSM with Heavy Majorana Neutrinos},'' {\em
  JHEP}, vol.~05, p.~063, 2011.

\bibitem{Draper:2013ava}
P.~Draper and H.~E. Haber, ``{Decoupling of the Right-handed Neutrino
  Contribution to the Higgs Mass in Supersymmetric Models},'' {\em Eur. Phys.
  J. C}, vol.~73, p.~2522, 2013.

\bibitem{Guo:2013sna}
J.~Guo, Z.~Kang, T.~Li, and Y.~Liu, ``{Higgs boson mass and complex sneutrino
  dark matter in the supersymmetric inverse seesaw models},'' {\em JHEP},
  vol.~02, p.~080, 2014.

\bibitem{Chun:2014tfa}
E.~J. Chun, V.~S. Mummidi, and S.~K. Vempati, ``{Anatomy of Higgs mass in
  Supersymmetric Inverse Seesaw Models},'' {\em Phys. Lett. B}, vol.~736,
  pp.~470--477, 2014.

\bibitem{Muhlleitner:2014vsa}
M.~Muhlleitner, D.~T. Nhung, H.~Rzehak, and K.~Walz, ``{Two-loop contributions
  of the order $ \mathcal{O}\left({\alpha}_t{\alpha}_s\right) $ to the masses
  of the Higgs bosons in the CP-violating NMSSM},'' {\em JHEP}, vol.~1505,
  p.~128, 2015.

\bibitem{Dao:2019qaz}
T.~Dao, R.~Gr\"ober, M.~Krause, M.~M\"uhlleitner, and H.~Rzehak, ``{Two-loop $
  \mathcal{O} $ ( $ {\alpha}_t^2 $ ) corrections to the neutral Higgs boson
  masses in the CP-violating NMSSM},'' {\em JHEP}, vol.~08, p.~114, 2019.

\bibitem{Baglio:2013vya}
J.~Baglio, T.~N. Dao, R.~Gr{\"o}ber, M.~M. M{\"u}hlleitner, H.~Rzehak,
  M.~Spira, J.~Streicher, and K.~Walz, ``{A new implementation of the NMSSM
  Higgs boson decays},'' {\em EPJ Web Conf.}, vol.~49, p.~12001, 2013.

\bibitem{Baglio:2019nlc}
J.~Baglio, T.~N. Dao, and M.~M\"uhlleitner, ``{One-Loop Corrections to the
  Two-Body Decays of the Neutral Higgs Bosons in the Complex NMSSM},'' 
  {\em Eur. Phys. J. C} vol.~80, no.~10, p.~960, 2020.

\bibitem{Dao:2020dfb}
T.~N. Dao, M.~Muhlleitner, S.~Patel, and K.~Sakurai, ``{One-loop Corrections to
  the Two-Body Decays of the Charged Higgs Bosons in the Real and Complex
  NMSSM},'' {\em Eur. Phys. J. C}, vol.~81, no.~4, p.~340, 2021.

\bibitem{Gonzalez-Garcia:1988okv}
M.~C. Gonzalez-Garcia and J.~W.~F. Valle, ``{Fast Decaying Neutrinos and
  Observable Flavor Violation in a New Class of Majoron Models},'' {\em Phys.
  Lett. B}, vol.~216, pp.~360--366, 1989.

\bibitem{Grimus:2000vj}
W.~Grimus and L.~Lavoura, ``{The Seesaw mechanism at arbitrary order:
  Disentangling the small scale from the large scale},'' {\em JHEP}, vol.~11,
  p.~042, 2000.

\bibitem{Casas:2001sr}
J.~A. Casas and A.~Ibarra, ``{Oscillating neutrinos and $\mu \rightarrow e, \gamma$},''
  {\em Nucl. Phys.}, vol.~B618, pp.~171--204, 2001.

\bibitem{Siegel:1979wq}
W.~Siegel, ``{Supersymmetric Dimensional Regularization via Dimensional
  Reduction},'' {\em Phys. Lett. B}, vol.~84, pp.~193--196, 1979.

\bibitem{Stockinger:2005gx}
D.~Stockinger, ``{Regularization by dimensional reduction: consistency, quantum
  action principle, and supersymmetry},'' {\em JHEP}, vol.~03, p.~076, 2005.

\bibitem{Kublbeck:1990xc}
J.~Kublbeck, M.~Bohm, and A.~Denner, ``{FeynArts: Computer Algebraic
  Generation of Feynman Graphs and Amplitudes},'' {\em Comput.Phys.Commun.},
  vol.~60, pp.~165--180, 1990.

\bibitem{Hahn:2000kx}
T.~Hahn, ``{Generating Feynman diagrams and amplitudes with FeynArts 3},'' {\em
  Comput.Phys.Commun.}, vol.~140, pp.~418--431, 2001.

\bibitem{Staub:2009bi}
F.~Staub, ``{From Superpotential to Model Files for FeynArts and
  CalcHep/CompHep},'' {\em Comput.Phys.Commun.}, vol.~181, pp.~1077--1086,
  2010.

\bibitem{Staub:2010jh}
F.~Staub, ``{Automatic Calculation of supersymmetric Renormalization Group
  Equations and Self Energies},'' {\em Comput.Phys.Commun.}, vol.~182,
  pp.~808--833, 2011.

\bibitem{Staub:2012pb}
F.~Staub, ``{SARAH 3.2: Dirac Gauginos, UFO output, and more},'' {\em Computer
  Physics Communications}, vol.~184, pp.~pp. 1792--1809, 2013.

\bibitem{Staub:2013tta}
F.~Staub, ``{SARAH 4: A tool for (not only SUSY) model builders},'' {\em
  Comput.Phys.Commun.}, vol.~185, pp.~1773--1790, 2014.

\bibitem{FeynCalc}
R.~Mertig, M.~B{\"o}hm, and A.~Denner, ``Feyncalc - computer-algebraic
  calculation of feynman amplitudes,'' {\em Computer Physics Communications},
  vol.~64, no.~3, pp.~345 -- 359, 1991.

\bibitem{Shtabovenko:2016sxi}
V.~Shtabovenko, R.~Mertig, and F.~Orellana, ``{New Developments in FeynCalc
  9.0},'' {\em Comput. Phys. Commun.}, vol.~207, pp.~432--444, 2016.

\bibitem{Baglio:2013iia}
J.~Baglio, R.~Grober, M.~Muhlleitner, D.~Nhung, H.~Rzehak, {\em et~al.},
  ``{NMSSMCALC: A Program Package for the Calculation of Loop-Corrected Higgs
  Boson Masses and Decay Widths in the (Complex) NMSSM},'' {\em
  Comput.Phys.Commun.}, vol.~185, no.~12, pp.~3372--3391, 2014.

\bibitem{Domingo:2020wiy}
F.~Domingo and S.~Pa\ss{}ehr, ``{Towards Higgs masses and decay widths
  satisfying the symmetries in the (N)MSSM},'' {\em Eur. Phys. J. C} 
  vol.~80, no.~12, pp.~1124, 2020.

\bibitem{Muhlleitner:2015dua}
M.~M{\"u}hlleitner, D.~T. Nhung, and H.~Ziesche, ``{The order $
  \mathcal{O}\left({\alpha}_t{\alpha}_s\right) $ corrections to the trilinear
  Higgs self-couplings in the complex NMSSM},'' {\em JHEP}, vol.~12, p.~034,
  2015.

\bibitem{Ender:2011qh}
K.~Ender, T.~Graf, M.~Muhlleitner, and H.~Rzehak, ``{Analysis of the NMSSM
  Higgs Boson Masses at One-Loop Level},'' {\em Phys.Rev.}, vol.~D85,
  p.~075024, 2012.

\bibitem{Graf:2012hh}
T.~Graf, R.~Grober, M.~Muhlleitner, H.~Rzehak, and K.~Walz, ``{Higgs Boson
  Masses in the Complex NMSSM at One-Loop Level},'' {\em JHEP}, vol.~10,
  p.~122, 2012.

\bibitem{Dao:2019nxi}
T.~N. Dao, L.~Fritz, M.~Krause, M.~M\"uhlleitner, and S.~Patel, ``{Gauge
  dependences of higher-order corrections to NMSSM Higgs boson masses and the
  charged Higgs Decay ${H^{\pm } \rightarrow W^\pm h_{i}}$},'' {\em Eur. Phys.
  J. C}, vol.~80, no.~3, p.~292, 2020.

\bibitem{Denner:1992vza}
A.~Denner, H.~Eck, O.~Hahn, and J.~Kublbeck, ``{Feynman rules for fermion
  number violating interactions},'' {\em Nucl. Phys.}, vol.~B387, pp.~467--481,
  1992.

\bibitem{Denner:1991kt}
A.~Denner, ``{Techniques for calculation of electroweak radiative corrections
  at the one loop level and results for W physics at LEP-200},'' {\em Fortsch.
  Phys.}, vol.~41, pp.~307--420, 1993.

\bibitem{Freitas:2002um}
A.~Freitas and D.~Stockinger, ``{Gauge dependence and renormalization of tan
  beta in the MSSM},'' {\em Phys. Rev.}, vol.~D66, p.~095014, 2002.

\bibitem{Dabelstein:1995js}
A.~Dabelstein, ``{Fermionic decays of neutral MSSM Higgs bosons at the one loop
  level},'' {\em Nucl. Phys.}, vol.~B456, pp.~25--56, 1995.

\bibitem{Dabelstein:1994hb}
A.~Dabelstein, ``{The One loop renormalization of the MSSM Higgs sector and its
  application to the neutral scalar Higgs masses},'' {\em Z. Phys.}, vol.~C67,
  pp.~495--512, 1995.

\bibitem{Bechtle:2020pkv}
P.~Bechtle, D.~Dercks, S.~Heinemeyer, T.~Klingl, T.~Stefaniak, G.~Weiglein, and
  J.~Wittbrodt, ``{HiggsBounds-5: Testing Higgs Sectors in the LHC 13 TeV
  Era},'' 6 2020.

\bibitem{Skands:2003cj}
P.~Z. Skands {\em et~al.}, ``{SUSY Les Houches accord: Interfacing SUSY
  spectrum calculators, decay packages, and event generators},'' {\em JHEP},
  vol.~07, p.~036, 2004.

\bibitem{Bechtle:2013xfa}
P.~Bechtle, S.~Heinemeyer, O.~Stål, T.~Stefaniak, and G.~Weiglein,
  ``{$HiggsSignals$: Confronting arbitrary Higgs sectors with measurements at
  the Tevatron and the LHC},'' {\em Eur. Phys. J.}, vol.~C74, no.~2, p.~2711,
  2014.

\bibitem{PhysRevD.95.096014}
F.~Capozzi, E.~Di~Valentino, E.~Lisi, A.~Marrone, A.~Melchiorri, and
  A.~Palazzo, ``Global constraints on absolute neutrino masses and their
  ordering,'' {\em Phys. Rev. D}, vol.~95, p.~096014, May 2017.

\bibitem{Parke:2015goa}
S.~Parke and M.~Ross-Lonergan, ``{Unitarity and the three flavor neutrino
  mixing matrix},'' {\em Phys. Rev. D}, vol.~93, no.~11, p.~113009, 2016.

\bibitem{Peskin:1991sw}
M.~E. Peskin and T.~Takeuchi, ``{Estimation of oblique electroweak
  corrections},'' {\em Phys. Rev. D}, vol.~46, pp.~381--409, 1992.

\bibitem{Zyla:2020zbs}
P.~Zyla {\em et~al.}, ``{Review of Particle Physics},'' {\em PTEP}, vol.~2020,
  no.~8, p.~083C01, 2020.

\bibitem{Patrignani:2016xqp}
C.~Patrignani {\em et~al.}, ``{Review of Particle Physics},'' {\em Chin.
  Phys.}, vol.~C40, no.~10, p.~100001, 2016.

\bibitem{Calibbi:2017uvl}
L.~Calibbi and G.~Signorelli, ``{Charged Lepton Flavour Violation: An
  Experimental and Theoretical Introduction},'' {\em Riv. Nuovo Cim.}, vol.~41,
  no.~2, pp.~71--174, 2018.

\bibitem{Lavoura:2003xp}
L.~Lavoura, ``{General formulae for f(1) $\to$ f(2) $\gamma$},'' {\em Eur. Phys.
  J.}, vol.~C29, pp.~191--195, 2003.

\bibitem{Hue:2017lak}
L.~Hue, L.~Ninh, T.~Thuc, and N.~Dat, ``{Exact one-loop results for $l_i \to
  l_j\gamma$ in 3-3-1 models},'' {\em Eur. Phys. J. C}, vol.~78, no.~2, p.~128,
  2018.

\bibitem{Dennerlhcnote}
A.~Denner {\em et~al.}, ``{Standard Model input parameters for Higgs physics},'' LHCHXSWG-INT-2015-006, 2015.

\bibitem{Dao:2021khm}
T.~N. Dao, M.~Gabelmann, M.~M\"uhlleitner, and H.~Rzehak, ``{Two-Loop ${\cal
  O}((\alpha_t+\alpha_\lambda+\alpha_\kappa)^2)$ Corrections to the Higgs Boson
  Masses in the CP-Violating NMSSM},'' {\em  2106.06990}.

\bibitem{Mandal:2020lhl}
S.~Mandal, R.~Srivastava, and J.~W.~F. Valle, ``{Electroweak symmetry breaking
  in the inverse seesaw mechanism},'' {\em JHEP}, vol.~03, p.~212, 2021.

\end{thebibliography}

\end{document}